\documentclass[preprint2,twocolappendix,appendixfloats]{aastex6}
\usepackage{graphicx}	
\usepackage{amsmath}	
\usepackage{amssymb}	
\usepackage{longtable}

\defcitealias{lehner13}{L13}
\defcitealias{wotta16}{W16}
\defcitealias{fumagalli16}{FOP16}

\def\nodata{ ~$\cdots$~ }
\newcommand{\xh}{{[\rm X/H]}}

\newcommand{\lyaf}{Ly$\alpha$ forest}

\newcommand{\lya}{Ly$\alpha$}

\newcommand{\lyb}{Ly$\beta$}

\newcommand{\nhi}{$N_{\rm HI}$}
\newcommand{\nh}{$N_{\rm H}$}

\newcommand{\novi}{$N_{\rm OVI}$}
\newcommand{\nhii}{$N_{\rm HII}$}
\newcommand{\mnhi}{N_{\rm HI}}
\newcommand{\mnhii}{N_{\rm HII}}

\newcommand{\mlnovi}{\log N_{\rm OVI}}

\newcommand{\mlnhi}{\log N_{\rm HI}}

\newcommand{\km}{${\rm km\,s}^{-1}$}

\newcommand{\hst}{{\em HST}}

\newcommand{\hi}{H$\;${\small\rm I}\relax}

\newcommand{\alii}{Al$\;${\small\rm II}\relax}

\newcommand{\cii}{C$\;${\small\rm II}\relax}
\newcommand{\ciii}{C$\;${\small\rm III}\relax}
\newcommand{\civ}{C$\;${\small\rm IV}\relax}

\newcommand{\nv}{N$\;${\small\rm V}\relax}
\newcommand{\oi}{O$\;${\small\rm I}\relax}

\newcommand{\ovi}{O$\;${\small\rm VI}\relax}

\newcommand{\siii}{Si$\;${\small\rm II}\relax}
\newcommand{\siiii}{Si$\;${\small\rm III}\relax}

\newcommand{\siiv}{Si$\;${\small\rm IV}\relax}

\newcommand{\mgii}{Mg$\;${\small\rm II}\relax}

\newcommand{\feii}{Fe$\;${\small\rm II}\relax}
\newcommand{\feiii}{Fe$\;${\small\rm III}\relax}

\newcommand{\hit}{H$\;${\scriptsize\rm I}\relax}

\newcommand{\ciit}{C$\;${\scriptsize\rm II}\relax}

\newcommand{\ciiit}{C$\;${\scriptsize\rm III}\relax}
\newcommand{\civt}{C$\;${\scriptsize\rm IV}\relax}

\newcommand{\oit}{O$\;${\scriptsize\rm I}\relax}

\newcommand{\ovit}{O$\;${\scriptsize\rm VI}\relax}

\newcommand{\siiit}{Si$\;${\scriptsize\rm II}\relax}

\newcommand{\siiiit}{Si$\;${\scriptsize\rm III}\relax}

\newcommand{\aliit}{Al$\;${\scriptsize\rm II}\relax}

\newcommand{\siivt}{Si$\;${\scriptsize\rm IV}\relax}

\newcommand{\feiit}{Fe$\;${\scriptsize\rm II}\relax}

\newcommand{\feiiit}{Fe$\;${\scriptsize\rm III}\relax}

\slugcomment{Accepted for publication in the ApJ -- 10/18/2016}
\shortauthors{Lehner et al.}
\shorttitle{Cosmic Evolution of the Metallicity of the Dense Ionized Gas}

\begin{document}

\title{The Cosmic Evolution of the Metallicity Distribution of  Ionized Gas Traced by Lyman Limit Systems}

\author{
Nicolas Lehner\altaffilmark{1},
John M. O'Meara\altaffilmark{2},
J. Christopher Howk\altaffilmark{1},
J. Xavier Prochaska\altaffilmark{3,4}, and
Michele Fumagalli\altaffilmark{5}
}

\altaffiltext{1}{Center of Astrophysics, Department of Physics, University of Notre Dame, 225 Nieuwland Science Hall, Notre Dame, IN 46556}
\altaffiltext{2}{Department of Chemistry and Physics, Saint Michael's College, One Winooski Park, Colchester, VT 05439}
\altaffiltext{3}{Department of Astronomy and Astrophysics, University of California, 1156 High Street, Santa Cruz, CA 95064}
\altaffiltext{4}{University of California Observatories, Lick Observatory 1156 High Street, Santa Cruz, CA 95064}
\altaffiltext{5}{Institute for Computational Cosmology and Centre for Extragalactic Astronomy, Department of Physics, Durham University, South Road, Durham, DH1 3LE, UK}

\begin{abstract}
We present the first results from our KODIAQ Z survey aimed to determine the metallicity distribution and physical properties of the   $z\ga 2$ partial and full Lyman limit systems (pLLSs and LLSs; $16.2 \le \mlnhi<19$), which are probed of the interface regions between the intergalactic medium (IGM) and galaxies. We study 31 \hi-selected pLLSs and LLSs at $2.3<z<3.3$ observed with Keck/HIRES in absorption against background QSOs. We compare the column densities of metal-ions  to \hi\ and use photoionization models to assess the metallicity. The metallicity distribution  of the pLLSs/LLSs at $2.3<z<3.3$ is consistent with a unimodal distribution peaking at $\xh \simeq -2$. The metallicity distribution of these absorbers therefore evolves markedly with $z$ since at $z\la 1$ it is bimodal with peaks at $\xh \simeq -1.8$ and $-0.3$. There is a substantial fraction (25--41\%) of pLLSs/LLSs with  metallicities well below those of damped \lya\ absorbers (DLAs) at any studied $z$ from $z\la 1$ to $z\sim 2$--4, implying reservoirs of metal-poor cool, dense gas in the IGM/galaxy interface at all $z$. However, the gas probed by pLLSs and LLSs is rarely pristine, with a fraction 3--18\% for pLLSs/LLSs with $\xh \le -3 $. We find C/$\alpha$ enhancement in several pLLSs and LLSs in the metallicity range $-2 \la \xh \la -0.5$, where C/$\alpha$ is 2–5 times larger than observed in Galactic metal-poor stars or high redshift DLAs at similar metallicities. This is likely caused by preferential ejection of carbon from metal-poor galaxies into their surroundings.
\end{abstract}


\section{Introduction}\label{s-intro}
Modern theory and simulations agree that the star formation of galaxies and the properties of their circumgalactic medium (CGM, defined here as the gas between the inner regions of galaxies and the diffuse intergalactic medium, IGM) should be intimately connected. This is especially true for the dense flows through the CGM: feedback from star formation is understood to drive outflows that carry mass and metals away from galaxies, while infall from the IGM is thought to bring in fresh gas to fuel on-going star formation. In fact, each of these is a necessary component for our current understanding of galaxy evolution. Without significant feedback, most baryons would cool into the centers of halos to form prodigious quantities of stars \citep[e.g.,][]{white78,keres09a}, but with feedback, the baryon content of stars and cold gas in galaxies can be matched \citep[$<20$\% of their cosmic baryons; e.g.,][]{fukugita98,conroy09} by driving matter into the CGM and beyond. Similarly, without continued infall of IGM material, star-forming galaxies would consume their interstellar gas in $\sim$1 Gyr \citep[e.g.,][]{genzel10,prochaska05}. The absence of star formation in some galaxies may be explained by the strangulation of IGM infall, wherein the hot ambient coronal matter in high-mass galaxies is sufficient to heat the infalling gas to temperatures that make it unavailable for immediate star formation (\citealt{dekel06,keres09b}).

These exchanges of matter, both in and out, through the CGM thus play critical roles in the evolution of galaxies. The competition between these large-scale inflows and outflows and its behavior with galactic mass is thought to shape such disparate properties of galaxies as the galactic mass-metallicity relation, the galaxy color bimodality, the maintenance of star formation in galaxies over billions of years, and the (stellar) baryonic mass fraction of galaxies \citep[e.g.,][]{keres05,dekel06,faucher-giguere11}. It has, however, been difficult to verify these predictions. There is good reason to believe feedback-driven outflows are important carriers of mass and metals through the CGM since ubiquitous outflows are observed toward galaxy centers \citep[e.g.,][]{pettini01,shapley03,steidel04,steidel10,weiner09,rubin14}. The COS-Halos and COS-Dwarfs surveys have demonstrated that the CGM is a massive reservoir of galactic metals, with galaxies having ejected at least as much metal mass as they have retained (\citealt{tumlinson11a,werk14,peeples14,bordoloi14}, and see also, e.g., \citealt{stocke13,liang14,lehner15} for other works). Similarly, characterizing the infall of matter requires  that the accreting gas is first found. It is not often seen in absorption against the galaxies themselves \citep[e.g.,][]{martin12,rubin12} and has been difficult to observe directly in the CGM.

To study the relationship between galaxy and CGM properties requires the development of methods for identifying gas infall, outflows, or other phenomena. Our team has approached this problem by using absorption lines toward background QSOs, searching for CGM gas with an \hi\ selection technique and determining the gas metallicity as a ``tracer" of the origin(s) of the gas \citep{ribaudo11,fumagalli11a,fumagalli11b,lehner13}. The selection based only on its \hi\ column density avoids biases that can be present with metal-line selection (e.g., via \mgii\ absorption).  We target absorbers with a detectable break at the Lyman limit and/or with the Lyman series so that the \hi\ column density is in the interval $16\la \mlnhi \la 19$. These are known as the partial Lyman limit systems (pLLS, defined in this work as $16\le \mlnhi < 17.2$) and LLSs (defined in this work as $17.2\le \mlnhi < 19$). The reasons for targeting these absorbers are twofold. First, in cosmological simulations, the LLSs have been shown to be good tracers of cold flows at $z\sim 2$--3 \citep[e.g.,][]{fumagalli11a,fumagalli14,faucher-giguere11,faucher-giguere15,vandevoort12b}. Second, Empirically, at $z\la 1$, the pLLSs and LLSs have been associated with the dense CGM  \citep{lanzetta95,penton02,bowen02,chen05}, and in particular for each specific pLLS and LLS with some galaxy information, they have been found well within the virial radius of galaxies (typically at impact parameter $<130$ kpc, (\citealt{lehner13}, hereafter \citetalias{lehner13}). Higher redshift studies can only observe the most luminous galaxies, but notably the Keck Baryonic Structure  Survey (KBSS) shows that at $z\sim 2$--3 there is a strong incidence of absorbers with $\mlnhi >14.5$ with galaxies at transverse physical distance $\le 300$ kpc and velocity separation between the absorber and galaxy redshifts $\le 300 $ \km, but not for the lower \nhi\ absorbers \citep{rudie12}. The same survey also found that nearly half of the absorbers with $\mlnhi >15.5$ are found in the CGM of (massive) galaxies, which also implies that some of the absorbers (especially the pLLSs) may probe more diffuse gas or the CGM of less massive galaxies at high $z$. In any case,  at all $z$, by definition of their \hi\ column densities, the pLLSs/LLSs are at the interface between the IGM probed by \lyaf\ (LYAF) absorbers with $\mlnhi \la 15.5$ and virialized structures  traced by super-LLSs (SLLS; $19\le \mlnhi <20.3$) and damped \lya\ absorbers (DLAs; $ \mlnhi \ge 20.3$). 

Recently, we have shown that the dense CGM of $z<1$ galaxies traced by pLLSs and LLSs has a bimodal metallicity distribution function (MDF) with two well-separated peaks at $Z\simeq 0.02 Z_{{\sun}}$ and $0.5 Z_{{\sun}}$ and with about equal proportions in each branch (\citetalias{lehner13}). We have now doubled the initial sample of pLLSs and LLSs  at $z<1$ and found the same MDF  (\citealt{wotta16}, hereafter \citetalias{wotta16}). However, as shown in \citetalias{wotta16}, the bimodal nature of the MDF is dominated by the pLLS population and may start to transition to a unimodal distribution in the LLS regime. As argued in these papers, the metal-rich branch must trace expelled matter: galactic winds, recycled outflows, and tidally-stripped gas, i.e., it traces gas that has been in a galaxy previously in view of the relatively large metal enrichment of the gas. On the other hand, the metallicities of pLLSs and LLSs in the  metal-poor branch are extremely low for the $z<1$ universe, lower than the metallicities of dwarf galaxies accreting onto central massive galaxies \citep[e.g.,][]{skillman89,tremonti04,nicholls14,jinmy15} and much lower than the lowest metallicities observed for the typical DLAs at similar redshift (\citetalias{lehner13}; \citetalias{wotta16}). These metal-poor LLSs appear to  have all the properties of those expected for infalling matter, including the temperature, ionization structure, kinematic properties, and metallicity \citep{fumagalli11a,vandevoort12b,shen13}. 

Having identified low-metallicity gas in the halos of galaxies at low redshift, we now want to determine how the metallicity of the pLLSs and LLSs evolves with $z$ and \nhi\ at $z>2$ using the same selection criteria and method to derive the metallicity. This program directly builds on our Keck Observatory Database of Ionized Absorbers towards Quasars (KODIAQ) survey \citep{lehner14,omeara15}, which has used the NASA Keck Observatory Archive (KOA) to characterize the properties of the highly ionized gas associated with pLLSs and LLSs.  With our new KODIAQ Z program, we will expand this effort to now determine the MDF and physical properties of the pLLSs and LLSs at $z\ga 2$ in an unprecedently large sample.

In this paper, we present the results from a pilot study from a subset of the KODIAQ Z sample with the goal to assemble a sample of pLLSs and LLSs at $2.3 < z < 3.3$ with a similar size as in \citetalias{lehner13} at $z<1$. The total sample consists of 32  \hi\ selected pLLSs and LLSs (19 pLLSs and 13 LLSs); the statistical sample for the metallicity distribution analysis is 31 (18 pLLSs and 13 LLSs; two pLLSs having similar metallicity and are only separated by $\sim$50 \km\ in the redshift rest-frame of the absorbers). We emphasize that our study contrasts from the recent HD-LLS survey at $z>2$ (\citealt{prochaska15}; \citealt{fumagalli16}, hereafter \citetalias{fumagalli16}) or from the survey of low-metallicity LLSs at $3.2\la z \la 4.4$ \citep{cooper15,glidden16}. The HD-LLS survey targets \hi-selected LLSs and SLLSs with $\mlnhi > 17.2$ at $z\sim 2.5$--3.0, but only 9 LLSs have $\mlnhi \sim 17.5$, while all the others have $\mlnhi \ga 18$. Similarly the \citeauthor{cooper15} study also  targeted a sample of 17 high \nhi\ LLS (typically $\mlnhi \sim 17.5$), but selected them on the absence of metal absorption in Sloan Digital Sky Survey (SDSS) spectra, i.e., they targeted a priori low-metallicity LLSs. These programs are therefore complementary to ours and we will use their results for comparison with our samples. 

Our paper is organized as follows. In \S\ref{s-data} we describe the new and archival pLLS and LLS samples. In \S\ref{s-metallicity}, we describe the different steps to estimate the metallicities of the absorbers with additional technical details (including the description of each absorber) provided in the Appendix for interested readers. Our main results are presented in \S\S\ref{s-results} and \ref{s-prop} where we discuss the metallicity distribution of the pLLSs and LLSs at $2.3<z<3.3$ and the evolution of their properties. In \S\ref{s-disc} we discuss some of the implications of our new observational results. Finally, in \S\ref{s-sum} we summarize our main results.

\begin{figure*}
\epsscale{1} 
\plotone{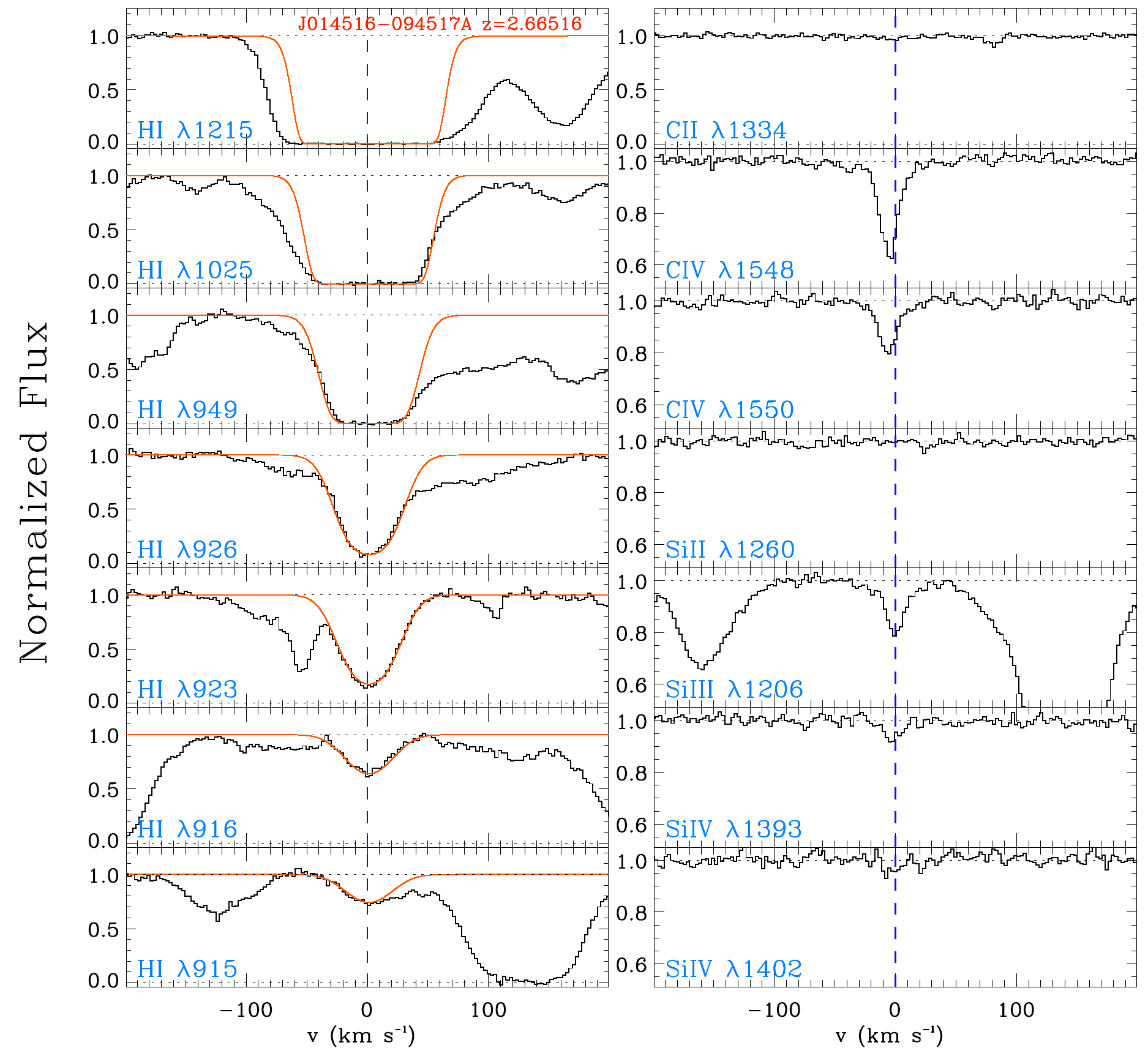}
 \caption{Example of normalized \hit\ (left) and metal-line (right) profiles of a pLLS with $\mlnhi \simeq 16.17$. The red lines are the profile fits to the \hit\ lines; in this case the most constraining transitions are $\lambda$$\lambda$926, 923, 916, 915. For this pLLS, the metal-line absorption is simple with a single component observed between $-25 \le v\le +20$ \km, which aligns well with the \hit\ transitions (we note that \civt\ is slightly shifted in this case by 4 \km). The absorption features observed  outside the velocity range $-25 \le v\le +20$ \km\ are unrelated to this pLLSs. }
 \label{f-example1}
\end{figure*}

\section{Data, sample selection and definition}\label{s-data}
With this pilot study, we assemble a sample of pLLSs and LLSs at $2<z<3.5$ similar in size and \nhi\ coverage to the original sample of pLLSs and LLSs in \citetalias{lehner13}. Our final sample for this study consists of 25 new \hi-selected absorbers with $16.1 \le \mlnhi \le 18.4$ and 7 from the literature with $16.4 \le \mlnhi \le 18.6$. We note that some of the high \nhi\ absorbers in the new sample were part of the LLS survey by \citet{steidel90}, but, in the present work, all the \hi\ and metal column densities were estimated using high resolution Keck spectra; the Steidel's study used much lower (35--80 \km) resolution observations, which led to metallicities being typically crudely estimated. 

 For the literature sample, we searched for \hi-selected absorbers with $\mlnhi \ga 16.1$, where we carefully excluded any absorbers that were selected for D/H or using metal diagnostics to preselect them. Two pLLSs are drawn from \citet{crighton13,crighton15}. The rest of the sample comes from our KODIAQ survey used to search for \ovi\ absorption in \hi-selected LLSs with five LLSs ($17.75 \la \mlnhi \la 18.60$)  \citep{lehner14}. 

Many of the other pLLSs/LLSs found in the KODIAQ database could not be used to study \ovi\ owing to the contamination of the \lyaf\ near the \ovi\ doublet transitions, but are useful for studying the metallicity distribution of these absorbers. In this sample, we selected pLLSs and LLSs for which we could derive \nhi\ reasonably well (specifically with a 1$\sigma$ error less than 0.3 dex, see \S\ref{s-nhi}) and estimate column densities (or column density limits) for \siii, \siiii, and \siiv\ (at least two of these ions are required to be uncontaminated), which are key ions to derive the metallicity of the pLLSs and LLSs at $z\sim 2$--3 (see \S\ref{s-nmetal}). 

All the new data presented here are from our KODIAQ database as part of our new KODIAQ Z survey \citep{lehner14,omeara15}. In short, these data were acquired with the HIgh Resolution Echelle Spectrometer (HIRES) \citep{vogt94} on the Keck\,I telescope on MaunaKea. These data were obtained by different PIs from different institutions with Keck access, and hundreds of spectra of QSOs at $0<z<6$ (most being at $z\sim 2$--$4$) were collected. As part of our previous NASA KODIAQ program, we have uniformly reduced, coadded, and normalized the Keck HIRES QSO spectra (for a full information regarding the data processing, see \citealt{omeara15}).  A significant fraction of the reduced KODIAQ data is now publicly available from the KOA \citep{omeara15}.\footnote{Available online at http://koa.ipac.caltech.edu/Datasets/.}

Before proceeding to our main analysis, we emphasize two aspects of our sample of the pLLSs and LLSs. First, there is no proximate pLLS or LLS in our sample, i.e., all the absorbers in our sample have velocity separations from the redshift QSOs well above 3000 \km. Second, as we emphasize further below,  we derive the column densities of  \hi\ and the metal lines in the main absorption associated with the pLLSs or LLSs, so the integration of the velocity profiles are over about 40 to 130 \km. This contrasts from the HD-LLS survey \citep{prochaska15}, where they consider that a LLS is all of the optically thick gas within a velocity interval of 500 \km\ from the redshift of the LLS. Owing to that we use higher resolution spectra in our survey and that the \nhi\ values are typically below $10^{18}$ cm$^{-2}$, we can consider reliably smaller velocity intervals. However, we note there is one case in our sample where a pLLS has evidence for two pLLSs ($z_{\rm abs}=2.46714$ toward J144453+291905), but the signal-to-noise (S/N) level is not good enough to accurately model them separately. There is also one case where two pLLSs are separated only by 50 \km\ ($z_{\rm abs}=2.43307$ and 2.43359 toward J170100+641209) and where we find a similar metallicity for each absorber; in that case we only kept one for our analysis of the metallicity distribution (there is also one similar case in the  \citealt{crighton15}, but in this case we adopted their results based on the total column density since there was little variation in the metallicity across the velocity profile). Finally, for two cases, a pLLS is associated with a SLLS, i.e., there is a velocity separation less than 300 \km\ between the pLLS and SLLS (one in our new sample -- $z_{\rm abs} = 2.66586$ toward J012156+144823, see Appendix, and one in \citealt{crighton13}). It is unclear at this stage if this could bias in any ways the sample, but since there are only two such cases presently, any effect would be marginal (in the case of the \citealt{crighton13} sample, the metallicity of pLLS is factor 50 than the SLLS, and hence the two absorbers do not have the same origin). In the future, with larger samples, we will be able to investigate more systematically pLLSs in the redshift vicinity of SLLSs or DLAs. 

\begin{figure*}
\epsscale{1} 
\plotone{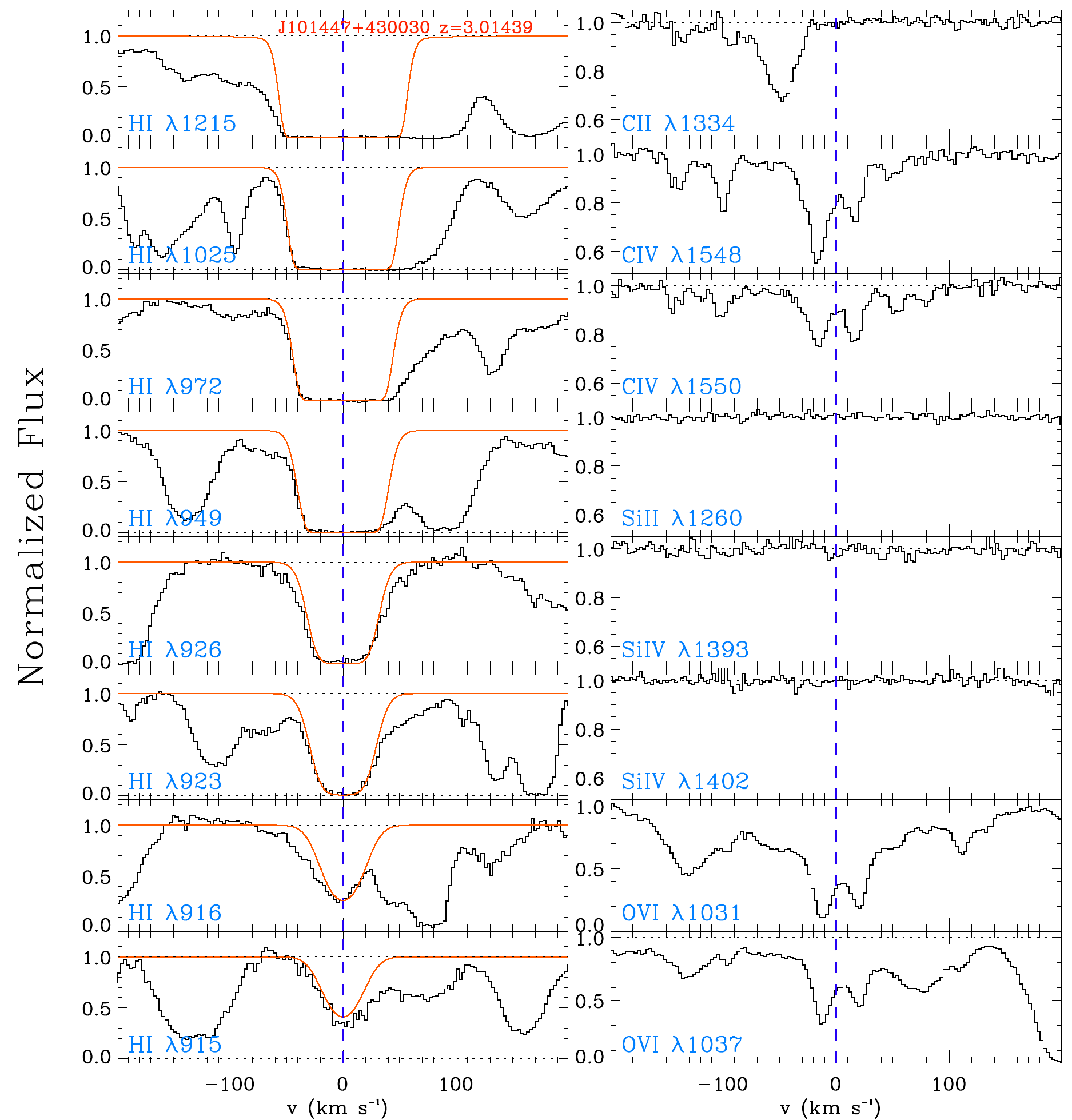}
 \caption{Same as Fig.~\ref{f-example2} but for stronger pLLS with $\mlnhi \simeq 16.63$. Despite that the \hit\ transitions are all contaminated to some level, the use of many transitions allows us to determine accurately \nhi. For this pLLS, the metal-line absorption consists of two main components observed between $-45 \le v\le +35$ \km. Note that in this case, there is evidence for weaker \hi\ absorption and metal-line features below $-45$ \km\ and above $+35$ \km\ (in particular  \civt\ and \ovit\ have strong absorption from about $-160$ to $+100$ \km). For our analysis of the metal lines, we only consider the absorption at  $-45 \le v\le +35$ \km, which is associated with the main component of the pLLS. }
 \label{f-example2}
\end{figure*}

\section{Estimation of the metallicity}\label{s-metallicity}
The most robust approach to measure the metallicity of the pLLSs and LLSs would be to use the \oi/\hi\ ratio given that charge exchange reactions with hydrogen ensure that the ionizations of \hi\ and \oi\ are strongly coupled. However, for absorbers with $\mlnhi \la 17.5$, \oi\ is rarely detected, and the limit that can be placed on $N_{\rm O\,I}$ is generally not sensitive enough. Hence to determine the metallicity of the pLLSs and LLSs, we have to compare the column densities of metal ions with \hi. Since the pLLSs and LLSs are not pre-dominantly neutral like DLAs, but nearly completely ionized, we need to constrain the ionization of this gas to be able to derive its metallicity  (e.g., \citealt{prochaska99,lehner09,lehner13,prochaska15}; \citetalias{fumagalli16}; and see below for more details). LLSs and pLLSs are often multiphase, with absorption seen in different ionization stages, and the low to intermediate ions (e.g., \siii, \siiii, \siiv, \cii, \ciii, and sometimes \civ) and high ions (\ovi) often show distinct kinematics (e.g., \citealt{lehner09,lehner13,fox13,crighton13}; \citetalias{fumagalli16}). This is illustrated in Figs.~\ref{f-example1} and \ref{f-example2}, where we show two examples of pLLSs at $z\sim 3$ from our new sample with $\mlnhi \simeq 16.17$ and 16.63, respectively. In the left panel of these figures, the \hi\ transitions used to determine the \hi\ column density are shown; the right panel shows some of the metal ions used to determine the metallicity. Other examples of high-$z$ LLS absorption profiles can be found, for example, in \citet{lehner14}, \citet{prochaska15}, and \citet{crighton13,crighton15} as well as in the Appendix for the metal lines. For the ionizing radiation field and for pLLSs with typical metallicities at $z \sim 2$--3 (about 0.1\%  solar or $\xh = -2$, see below and \citetalias{fumagalli16}), even strong transitions like \cii\ $\lambda$1334 and \siii\ $\lambda$1260 are often not detected, so we have to use  \siiii\ and \siiv\ to determine the metallicity. However, as in our study at low redshift \citep{lehner13}, we typically do not use high ions (specifically \ovi\ at $z\sim 2$--3) because the distinct kinematics of these ions (see  Fig.~\ref{f-example2} and  \citealt{lehner14}) imply that the bulk of the highest ions (i.e., \ovi) are not produced by the same mechanism that ionizes the lower ions in the pLLSs/LLSs or at the same density. 

In order to estimate the metallicity, we therefore need accurate column densities of \hi\ and metal ions. We describe in \S\ref{s-nmetal} and \S\ref{s-nhi} how we estimate the column densities of the metal ions and \hi. To correct for the large ionization when comparing \hi\ to metal ions  (e.g., \siii, \siiii, \siiv, \cii, \ciii, \civ) to determine the metallicity, we use Cloudy \citep{ferland13} models; a full description of this method and its limitations are presented in \S\ref{s-cloudy}.

\subsection{Metals and their column densities}\label{s-nmetal}
The main ions and transitions used in our study are  \siii\ $\lambda$$\lambda$1190, 1193, 1260, 1304, 1526, \siiii\ $\lambda$1206, \siiv\ $\lambda$$\lambda$1393, 1402, \cii\ $\lambda$$\lambda$1036, 1334, \ciii\ $\lambda$977, and \civ\ $\lambda$$\lambda$1548, 1550. In some cases, we can also use \oi\ $\lambda$$\lambda$1039, 1302, \alii\ $\lambda$1670, \feii\ $\lambda$1608, \feiii\ $\lambda$1122. We also consider \ovi\ $\lambda$1031, 1037 and \nv\ $\lambda$$\lambda$1238, 1242 in order to assess whether \civ\ is likely to arise in the same gas-phase as the low ions. In the Appendix, we show for each pLLS or LLS the normalized profiles of the metal ions or atoms and discuss the specific ions used to determine the metallicity. We emphasize that understanding the physical conditions of all the gas-phases is beyond the scope of this paper. However, to determine the metallicity requires one to determine the column densities of the metal ions that are tracing the ionized gas associated with the \hi\ of the pLLS or LLS. Following \citetalias{lehner13}, the preferred species to constrain the ionization parameter (see below) are those for which the velocity structures of their profiles best follow the \hi\ velocity profiles and that are produced mostly by a single phase ionization model. 

To estimate the column density of the metal ions, we use the apparent optical depth (AOD) method described by \citet{savage91}. The absorption profiles are converted into apparent column densities per unit velocity, $N_a(v) = 3.768\times 10^{14} \ln[F_c(v)/F_{\rm obs}(v)]/(f\lambda)$ cm$^{-2}$\,(\km)$^{-1}$, where $F_c(v)$ and $F_{\rm obs}(v)$ are the modeled continuum and observed fluxes as a function of velocity, respectively, $f$ is the oscillator strength of the transition and $\lambda$ is the wavelength in \AA\ (the atomic parameters are from \citealt{morton03}). Although the KODIAQ spectra are normalized \citep{omeara15}, we still model the continuum with a Legendre polynomial within $\pm 500$--2000 \km\ of the absorption feature of interest since the original continuum model may have sometimes over/under fitted some regions of the spectrum.\footnote{In this paper, we use high S/N data, so the continuum errors are typically at the 5\% level or less depending on the redshift and if the feature of interest is deep in the LYAF or not. \label{foot-cont}} The velocity ranges used to model the continuum depend on the number of absorbing features and the overall complexity of the continuum in this region of the spectrum. To determine the total column densities, we integrate the profiles over the velocities that correspond to the main absorption of the \hi\ of the pLLS or LLS. In the Appendix, we discuss for each pLLS/LLS the velocity structure of the metals and \hi\ and show the integration range used to estimate $N_a$ (see the listed values in Table~\ref{t-metal}, which can vary somewhat between different ions); typically the integration range is over $\la \pm 50$ \km\ in the rest-frame of the absorber. There can be several velocity components within that velocity range, but we do not consider higher-velocity components that correspond to typically weaker \hi\ absorbers clustered around the pLLSs or LLSs since the metallicity can be substantially different in these higher velocity components relative to the pLLSs or LLSs \citep[e.g.,][]{prochter10,crighton13}. 

For doublets or ions with several available atomic transitions (e.g., \civ, \siiv, \siii), the levels of contamination or saturation can be assessed directly by comparing the $N_a$ values. In that case if there is no evidence of contamination, the absorption is typically resolved,  i.e., there is no hidden saturation in the absorption profiles. For ions or atoms with only a single transition available, we require  similar velocity structures between different species in the velocity intervals used for integrating $N_a(v)$ to rule out contamination from unrelated absorbers. If the absorption reaches zero flux, the absorption is saturated, and we can only estimate a lower limit on the column density using the AOD method. If the peak optical depth is $\tau_\lambda \la 2$ or similar to that of absorption lines observed with two or more transitions where there is no evidence of saturation, we infer that the absorption is not saturated. For strong absorption ($\tau_\lambda \ga 1$--2), however, we allow in the photoionization modeling for the possibility that the line is saturated if needed by the models (i.e., we treat the column densities as possible lower limits). 

 In many cases, absorption from an ion or atom is not detected. If there is no contamination, we can estimate 2$\sigma$ upper limits on the equivalent widths, simply defined as the 1$\sigma$ error times 2. The 1$\sigma$ error is determined by integrating the spectrum over a similar velocity interval to that of a detected ion or over $\pm 20$ \km\ when no metals are detected in the absorber based on the typical smallest velocity intervals in other pLLSs/LLSs with detection of metals. The 2$\sigma$ upper limit on the column density is then derived assuming the absorption line lies on the linear part of the curve of growth. In Table~\ref{t-metal}, we summarize our apparent column density estimates of the metals as well the velocity interval used to integrate the profiles. For species with more than one transition, we list the results for each transition and in the row with no wavelength information the adopted weighted average column densities and velocities (see notes in this table for more information). Note that the errors are $1\sigma$ errors and include statistical and continuum placement errors following the methodology described in \citet{sembach92}. These errors do not, however, include errors arising from the original continuum fits to coadd the data (see \citealt{omeara15} and footnote~\ref{foot-cont}).

\subsection{\hi\ column density}\label{s-nhi}
\begin{figure}
\epsscale{1.2} 
\plotone{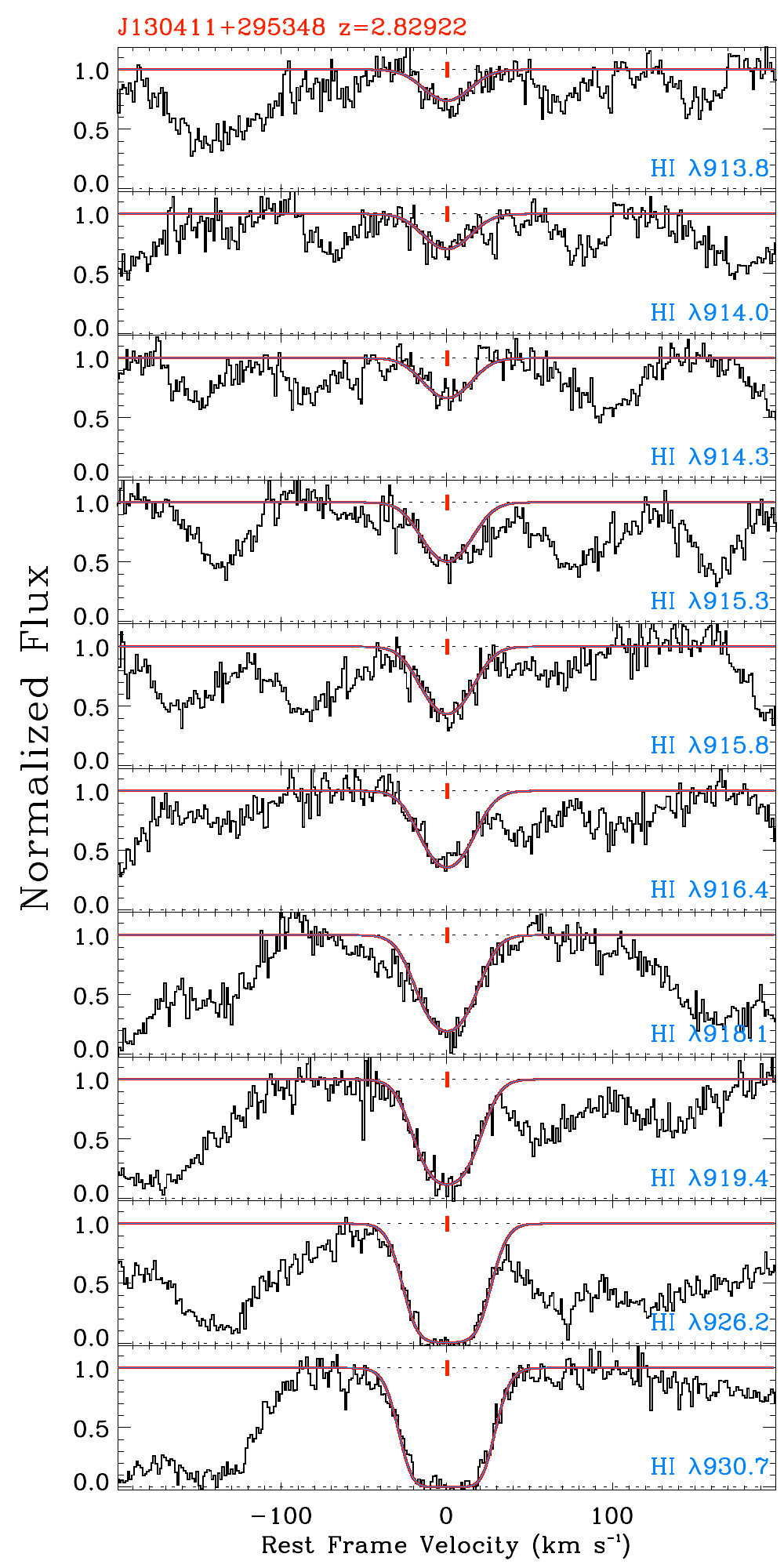}
 \caption{Example of an unusual pLLS with $\mlnhi \simeq 16.39$ where a large number of transitions shows little contamination (note that at $z<1$, it is typically not possible to model \hit\ transitions below 916 \AA\ as a consequence of the lower resolution of the data that blends these transitions). }
 \label{f-example3}
\end{figure}

The estimation of \nhi\ for each LLS ($\mlnhi \ge 17.2$) was made using a procedure similar to that described in \citet{lehner14}.  We use the graphical package {\sc x\_fitlls}\footnote{As part of the {\sc xidl} distribution package available at http://www.ucolick.org/$\sim$xavier/IDL/xidl\_doc.html} that allows us to create Voigt profiles to model the data. We iteratively varied the redshift, $b$-value, and \nhi\ of each system until a good fit was obtained. In many cases, the absorption in a LLS is complicated, requiring multiple absorption lines to produce a good fit.  For the LLSs presented here, we consider all absorption that produces significant absorption (normalized flux at line center $> 0.5$) through at least Lyman-5 (i.e., all components with $\mlnhi>15.0$) that might affect our total \nhi\ estimate. In most cases, such absorption impacts the total \nhi\ estimate at a level well below our $1\sigma$ error estimate on the \nhi, but in some cases multiple components of similar strength in \nhi\ are seen and cannot be ignored in the final \nhi\ estimate.  Since we are fitting the absorption of the LLSs by eye (as opposed to using a reduced-$\chi^2$ approach, see below), we adopt very conservative errors, with a minimum error on the \nhi\ for any LLS of $\sigma=0.15$ fitted using this methodology.  We finally note that we must appeal to further constraints to accurately determine \nhi\ for the strong LLSs, as the higher order Lyman series lines remain saturated for many more transitions than the pLLS or weak LLSs (see below).  We have, however, two important constraints. First, the onset of weak damping features in the \lya\ line  can be used to constrain the \nhi\ from above, as if the \nhi\ is too large, excess absorption appears on either side of the line-center. Second, the break in flux level below the Lyman limit can be used to determine \nhi\ if there is enough S/N in the data and no nearby absorption from other strong \nhi\ systems.

For the pLLSs ($16.2 \le \mlnhi < 17.2$) and one LLS, the primary tool used to constrain \nhi\ are the higher order Lyman series transitions (see Figs.~\ref{f-example1}, \ref{f-example2}, \ref{f-example3}). Two authors (O'Meara, Lehner)  undertook the analysis of the pLLSs where  the continuum placement near each \hi\ transition and profile fits to the pLLSs were independently assessed.\footnote{The only exception is the pLLS at $z=2.90711$ toward J212912-153841 where the S/N is too low to use the high order Lyman series transitions. In that case, we use the combined information of the Lyman series transitions and the flux decrement at the Lyman limit.} O'Meara used the same method described above for the LLSs, but instead fitted high order Lyman series transitions. For example, at the resolution of our HIRES data, a pLLS absorber with $\mnhi=16.35$ and $b =20$ \km\ becomes unsaturated (the normalized flux at the line-center being $>0.1$) at Lyman-9. This and higher order Lyman series transitions can then be used to accurately determine the combination of \nhi, $b$, and $z$ (or $v$ in the redshift rest-frame of the absorber) that best fits the observed absorption (see Fig.~\ref{f-example3}). Lehner fitted the \hi\ profiles with a minimum reduced-$\chi^2$ method using a modified version of the code described by \citet{fitzpatrick97}. The best-fit values describing the pLLSs were determined by comparing the model profiles convolved with an instrumental Gaussian LSF with the data. The three parameters $N_i$, $b_i$, and $v_i$ for each component, i (typically $i=1,2$), are input as initial guesses and were subsequently varied to minimize $\chi^2$. Since the Lyman series transitions are often blended with the \lya\ and \lyb\ forest absorbers, the fitting was an iterative process to select transitions that were not blended or with minimum blending. In the case of small blends, we iteratively masked the blended regions. Figs.~\ref{f-example1} and \ref{f-example2} show 2 pLLSs with various levels of contamination, while Fig.~\ref{f-example3} shows a rare pLLS where 10 Lyman series transitions have little contamination. Despite some contamination, the use of different \hi\ transitions with small oscillator strengths allows us to determine accurately \nhi. For each pLLS, the independently derived \nhi\ values were in excellent agreement. We adopted \nhi\ and errors from the Voigt profile fitting with the minimum reduced $\chi^2$. 

\begin{figure}
\epsscale{1.2} 
\plotone{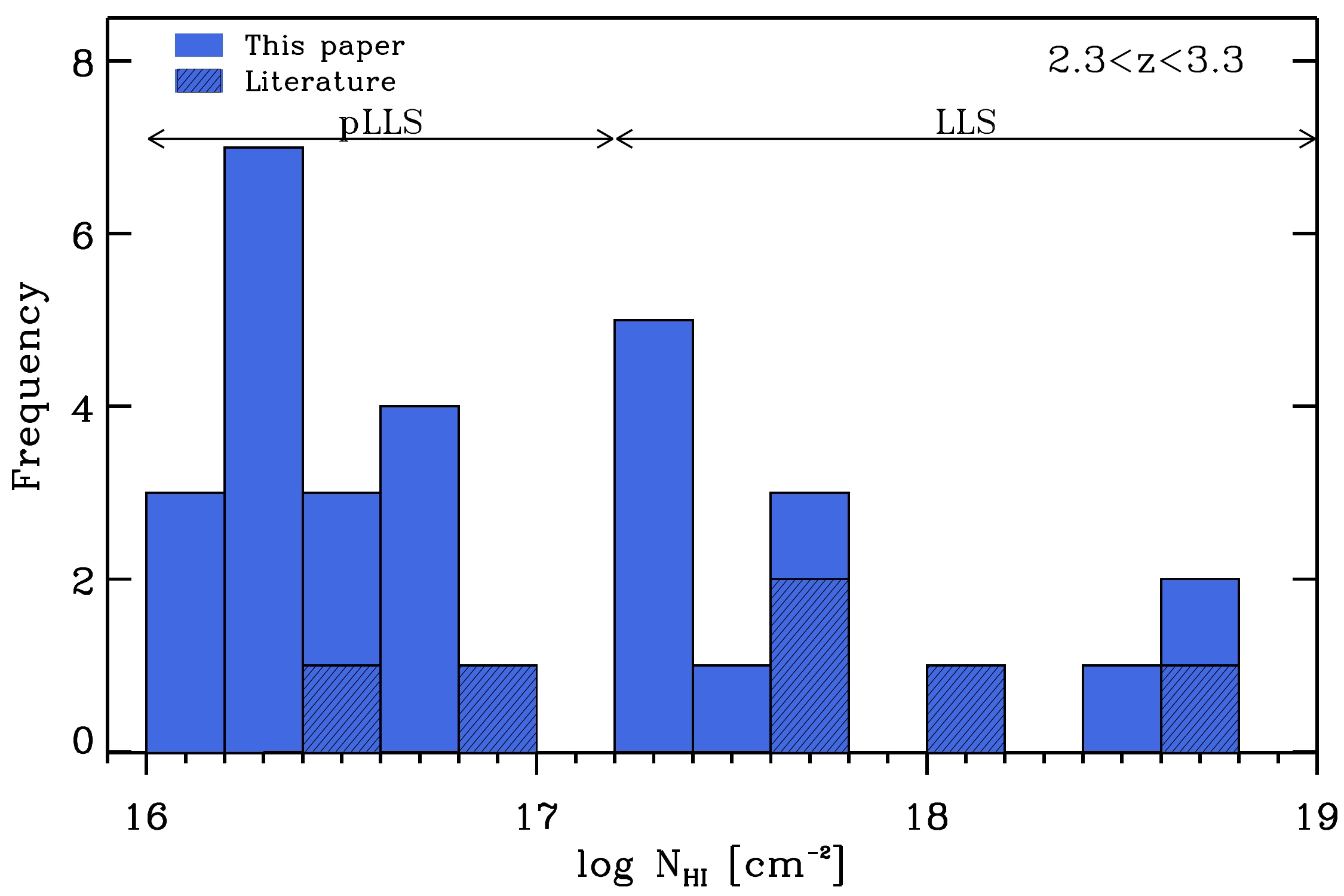}
 \caption{Distribution of the \hit\ column density in our sample at $2.3<z<3.3$. For comparison, in the same redshift interval, the HD-LLS survey has 9/38 (24\%) LLSs around $\mlnhi \sim 17.5$ and 29/38 (76\%) with $18\la \mlnhi \la 18.5$.
 \label{f-nhdist}}
\end{figure}

Our results are summarized in Table~\ref{t-nhi} and in Fig.~\ref{f-nhdist} where we show the \hi\ column density distribution for the entire sample of pLLSs and LLSs. There are 32 \hi-selected absorbers listed in Table~\ref{t-nhi}, 19 pLLSs ($16.2 \le \mlnhi < 17.2$) and 13 LLSs ($\mlnhi \ge 17.2$). However, two pLLSs are at essentially the same redshift (separated by about 50 \km) and have similar metallicities; we therefore treat these pLLSs as one, so that our total sample for the rest of the paper is 31. This is similar in size to the \citetalias{lehner13} sample of pLLSs and LLSs at $z<1$ (28 absorbers in total, 24 pLLSs and 4 LLSs). Our newer sample at $z<1$ has now doubled with 44 pLLSs and 11 LLSs \citepalias{wotta16}. Our sample is also complementary to the HD-LLS survey, which,  by definition of their sample, targets only LLSs with all but 9 LLSs at $z\sim 2.5$--3.3 having $\mlnhi \ga 18$ (\citealt{prochaska15}; \citetalias{fumagalli16}). 

\subsection{Photoionization modeling and metallicity determination}\label{s-cloudy}
With the column densities of \hi\ and metals determined, we can  estimate the metallicity of each pLLS or LLS. This requires large ionization corrections since  the fraction of H that is ionized always exceeds 90\% and is often close to 100\% (i.e., $\mnhii \gg \mnhi$). To determine the metallicity we follow closely \citetalias{lehner13},  modeling  the ionization using Cloudy \citep[version c13.02;][]{ferland13} and assuming the gas is a uniform slab geometry photoionized   by the  Haardt-Madau  background radiation field from quasars and galaxies (HM05, as implemented within Cloudy -- see also \citealt{haardt96,haardt12}; by adopting HM05 we also reduce any systematics in the comparison with the low redshift pLLSs/LLSs studied by \citetalias{lehner13} and \citetalias{wotta16}). For each absorber, we vary the ionization parameter, which is by definition the ratio of H ionizing photon density to total hydrogen number density ($U =n_\gamma/n_{\rm H}$), and the metallicity (we use the usual notation  for the metallicity $\xh \equiv \log N_{\rm X}/N_{\rm H} - \log ({\rm X/H})_{{\sun}} $, where X is a given element) to search for models that are consistent with the constraints set by the column densities determined from the observations.  

We assume solar relative heavy element abundances from \citet{asplund09}, i.e., we do not include  a priori the effects of dust or nucleosynthesis on the relative abundances. We note that for the main elements (C, Si, see below) that we use to model the photoionization and for the densities that the pLLSs and LLSs typically probe, the dust depletion levels of C and Si are expected to be small. In the Milky Way,  the depletions observed in the so-called ``warm-disk" and ``cool-halo" clouds for Si and C are $\la 0.3$ dex \citep[e.g.,][]{savage96,welty99,jenkins09}. At the studied redshift intervals in our survey, even smaller depletion levels of Si are typically observed in the denser environments probed by DLAs and SLLSs \citep[e.g.,][]{ledoux02,prochaska03a,rafelski12,quiret16}; e.g., \citet{rafelski12} found on average ${\rm [Si/S]}\simeq 0.0 \pm 0.2$ for gas metallicities $-2.3 \la {\rm [S/H]}\la -0.3$. Furthermore, \citetalias{fumagalli16} has shown that the strong LLSs reside typically in dust-poor enviromnents.  We nevertheless consider these possibilities a posteriori (especially for carbon that can have a different nucleosynthesis history than $\alpha$ elements as silicon or oxygen for example). This can be done a posteriori because the dust depletion or nucleosynthesis effects should affect all the ionization levels of a given element by the same factor. A posteriori, we find that typically dust depletion does not need to be invoked to explain the relative abundances of the pLLSs and LLSs in our sample, a finding consistent with the results from \citetalias{fumagalli16}.

The metallicity for each pLLS or LLS is determined using  $\alpha$ elements (usually Si), but the ionization model is constrained using the suite of Si and C ions (\siii, \siiii, \siiv, \cii, \ciii, \civ), and sometimes other atoms or ions (e.g., \oi, \alii, etc.). In the Appendix, we provide  the set of ions that determines $U$ and  $\xh$ for each LLS or pLLS. In Table~\ref{t-nhi}, we list the derived metallicities while in Table~\ref{t-cloudy} of the Appendix, we provide  for each pLLS and LLS the Cloudy output parameters from our models (total column density of H  -- \nh, $\xh$, $[{\rm C}/\alpha]$, $U$, ionized fraction -- \nhii/\nh, temperature -- $T$, and the linear scale of the absorber -- $l \equiv N_{\rm H}/n_{\rm H}$).

The errors on the metallicity and $U$ (listed in Table~\ref{t-nhi} and Appendix) reflect the range of values allowed by the $1\sigma$ uncertainties on the observed column densities. They do not include errors from the limitation of the models used to estimate the ionization corrections, which are about 0.3--0.5 dex on the metallicity (see \citetalias{lehner13}; \citetalias{wotta16}). As discussed in \citetalias{lehner13}, uncertainties in the assumed radiation field largely do not affect the {\it shape}| of the metallicity distribution. \citetalias{wotta16} explore the effect of changing the ionizing background from HM05 to HM12 \citep{haardt12} for the pLLSs and LLSs at $z\la 1$ and found that on average it would increase the metallicity of the pLLSs and LLSs by about $+0.3$ dex, well within the 0.3--0.5 dex uncertainty quoted above. This is, however, a systematic effect, i.e., both low and high metallicity absorbers are affected the same way, and hence the overall shape of the metallicity distribution would be very similar. \citetalias{fumagalli16} also provide a thorough analysis of a large sample of LLSs where they use several ionization models and Bayesian techniques to derive the physical properties and metallicities of the LLSs. They find as well that the  metallicity estimates are typically not very sensitive to the assumptions behind the ionization corrections. 

\section{Metallicity of the \lowercase{p}LLS\lowercase{s} and LLS\lowercase{s} at $2.3<\lowercase{z}<3.3$}\label{s-results}
\subsection{Metallicity distribution of the pLLSs and LLSs}\label{s-mdf}
\begin{figure}
\epsscale{1.2} 
\plotone{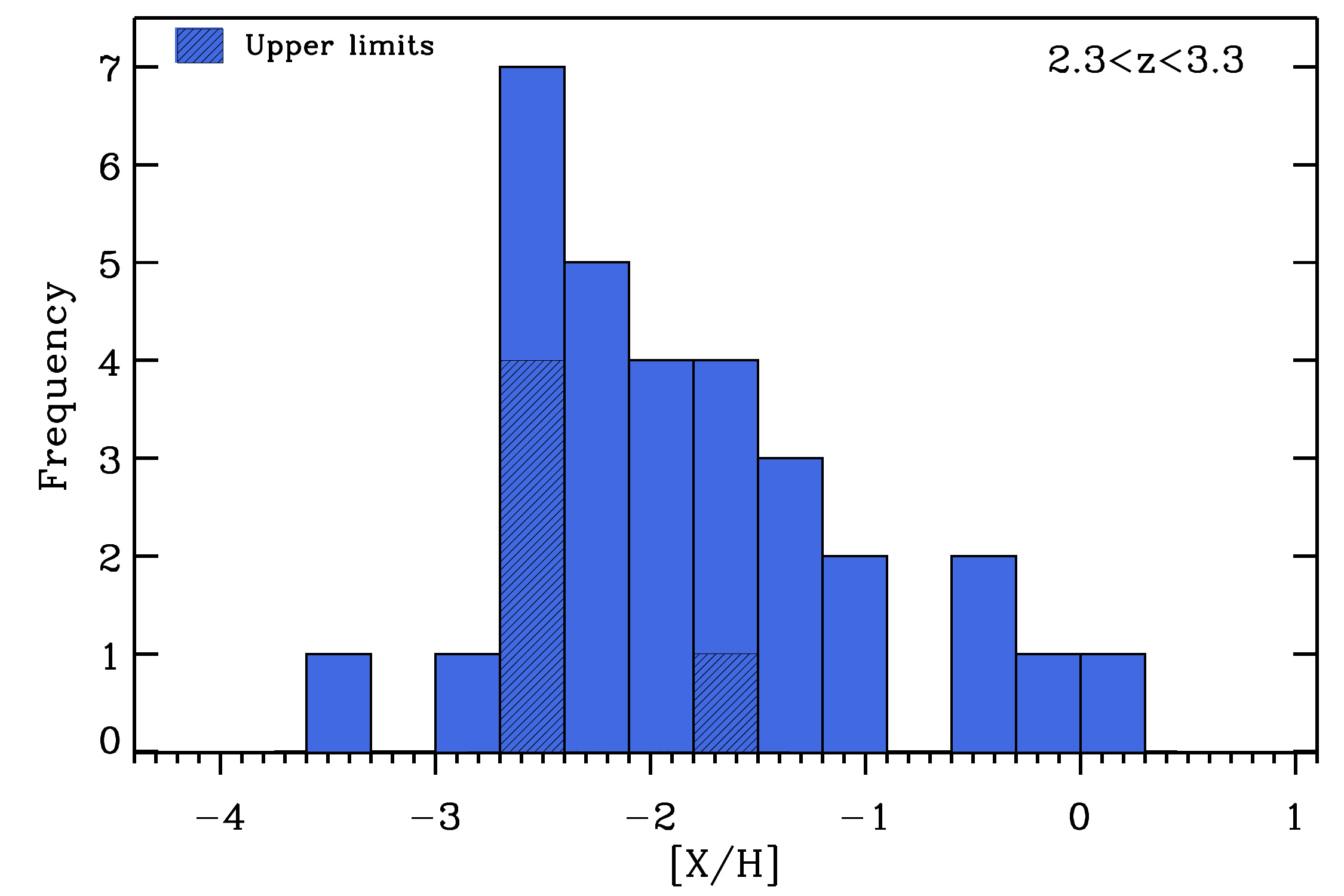}
 \caption{Distribution of the metallicity of the \hit-selected pLLSs and LLSs at $2.3 < z < 3.3$.}
 \label{f-zdist}
\end{figure}

Figure~\ref{f-zdist} shows the metallicity distribution function (MDF) for the 31 \hi-selected pLLSs and LLSs in our sample at $2.3 < z < 3.3$ summarized in Table~\ref{t-nhi}. Visually, the MDF is unimodal (see below). The MDF extends from $-3.5$ dex ($0.03\% Z_{\sun}$) to $+0.2$ dex ($1.6 Z_{\sun}$), but most of the values are dispersed around $-2$ dex. Using the Kaplain-Meier (KM) product limit estimator from the survival analysis (\citealt{feigelson85,isobe86})   to account for the upper limits in the sample, we estimate for the pLLSs {\it and}\ LLSs that  $\langle \xh\rangle = -2.00 \pm 0.17$ (where the quoted error is the KM error on the mean value). Treating the 5 upper limits as values, the median and standard deviation are $-2.05$ and 0.83 dex, respectively (under that assumption the mean of the MDF would be $-1.89$ dex). 

There is no evidence of a strong dip in the distribution as observed at low redshift (\citetalias{lehner13,wotta16}), and there is a prominent peak near the mean. A Dip test  \citep{hartigan85} shows that the significance level with which a unimodal distribution can be rejected is only 26\%.\footnote{See \citealt{muratov10} for the description of the Dip test code.} Treating censored data as actual values, a Kolmogorov-Smirnov (KS) test  finds the metallicity distribution is not inconsistent with a normal distribution with $p$-value $p= 0.39$ where the normal distribution has a mean\,$= -1.8$ and $\sigma = 0.9$ With future larger KODIAQ Z samples, we will be able to determine more robustly the shape of the MDF of both the pLLSs and LLSs. With the current sample, the MDF of the pLLSs+LLSs at $2.3 < z < 3.3$ can therefore be described by a unimodal distribution (possibly as a Gaussian distribution) with a large spread in both high and low metallicities.

\begin{figure}
\epsscale{1.2} 
\plotone{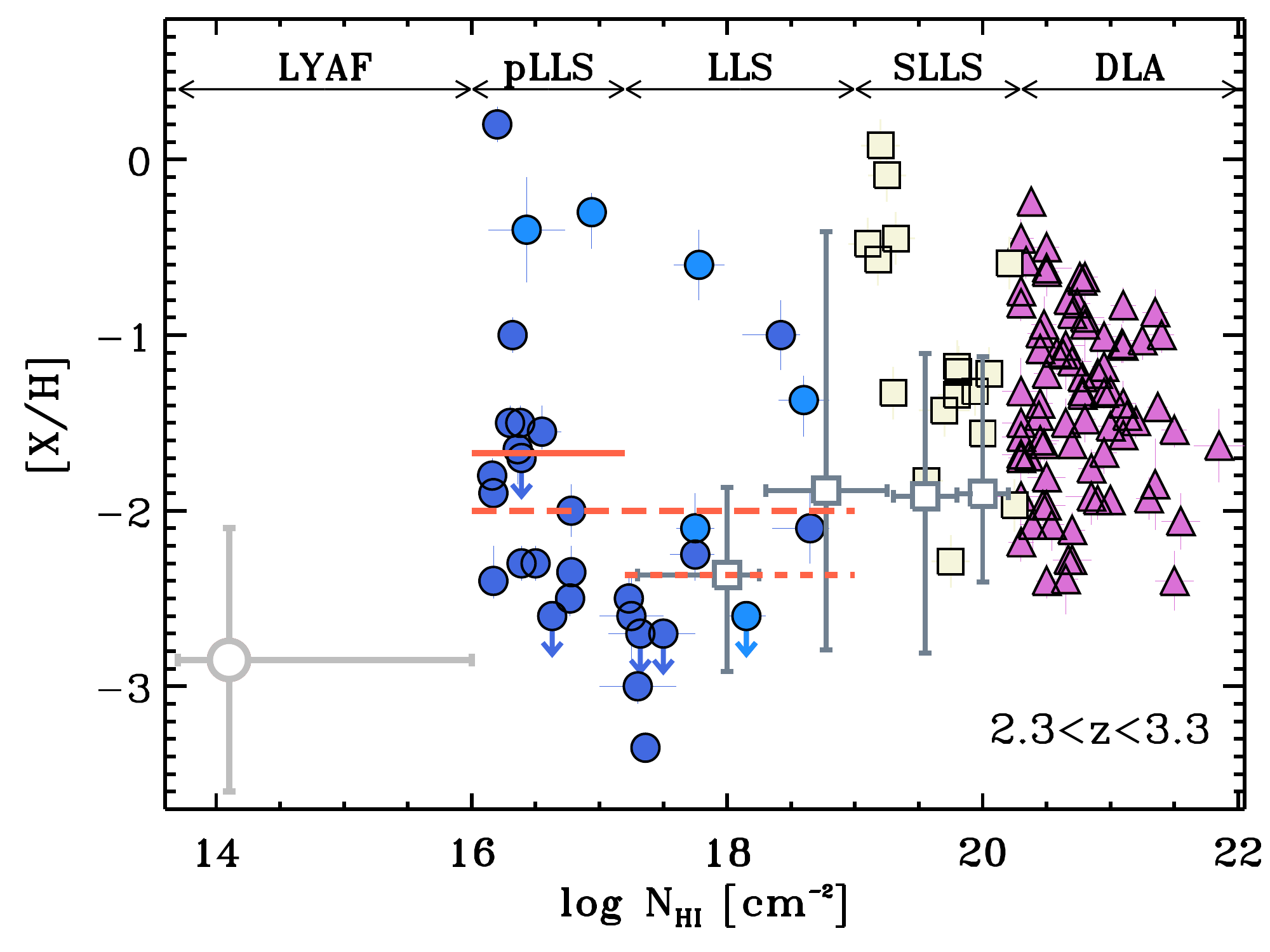}
 \caption{Metallicity as a function of the \hit\ column density for absorbers at $2.3 \la z \la 3.3$. The grey open circles are for the LYAF absorbers from \citet{simcoe04}. The light blue pLLS data are from \citet{crighton13,crighton15} and LLS data from \citet{lehner14}. The dark blue data are from this work. The grey squares are adapted from \citetalias{fumagalli16} (see text for more details). The light-yellow squares are from the survey and compilation from \citet{quiret16} (see text for more details). The orchid triangles are from \citet{rafelski12}. The grey squares and circle are centered near the most typical \nhi\ values within the range of values described by the horizontal bar of each data point. The red solid, long-dash, and short-dash lines are the mean of the pLLSs, pLLSs+LLSs, and LLSs, respectively. 
 \label{f-metvsnh1}}
\end{figure}

\subsection{Variation of the metallicity with \nhi}\label{s-metvsnhi}

In Fig.~\ref{f-metvsnh1}, we show the distribution of the metallicity against \nhi\  at $2.3 < z < 3.3$ , which allows us to separate the pLLSs and LLSs (and other absorbers) and to visualize the unbinned measurements. There is a large spread in the data for both the pLLS and LLS samples. In Table~\ref{t-metavg}, we list the mean, median, standard deviation, and fraction of very metal poor (VMP) absorbers with $\xh \le -2.4$ (value corresponding to $2\sigma$ below the mean metallicity of the DLAs). The LLSs and pLLSs have similar dispersions in their metallicity distributions, but from the KM method, we estimate that the mean metallicity of the LLSs is a factor 5 smaller (0.7 dex) than that of the pLLSs, $\langle \xh \rangle_{\rm LLS} = -2.37 \pm 0.24$ vs. $\langle \xh \rangle_{\rm pLLS} = -1.67 \pm 0.18$ (although they overlap within less than $2\sigma$ KM error).  There is also less evidence of VMP $\xh \le -2.4$ pLLSs than LLSs (6\% vs. 43\%). A Gehan's generalized Wilcoxon test and log-rank tests (which take into account that there are censored data -- upper limits -- in both the pLLS and LLS samples, see \citealt{feigelson85}) indicate a marginal statistical difference between the MDFs of the pLLSs (18 data points including 2 upper limits) and LLSs (13 data points including 4 upper limits) at significance levels $p=6.5$\% and $2.7$\%, respectively.  Yhe samples of LLSs and pLLSs are still small and there is a large overall dispersion in the metallicity distribution of both the pLLSs and LLSs; hence we consider any difference between the pLLS and LLS samples as tentative and marginal.

In Fig.~\ref{f-metvsnh1}, we also show the metallicity for lower and higher \nhi\ absorbers. For the LYAF, we show the mean and standard deviation from  \citet{simcoe04} who determined in the spectra of 7 QSOs the metallicity using \ovi\ and \civ\ for absorbers with $13.6 \la \mlnhi \la 16$ (most between  $13.6 \la \mlnhi \la 14.4$, which is highlighted by the asymmetric error on the horizontal axis) at $z\sim 2.5$. We also note the pixel optical depth method leads to similar results at $z\sim 3$ \citep{ellison00,schaye03,aguirre04}. In the LYAF sample, about  60--70\% of the LYAF absorbers are enriched to (observable) levels of ${\rm [O/H]} \ga -3.5$, while the remaining have even lower abundances. The LLSs and SLLSs shown with grey squares and associated vertical error bars are from the HD-LLS survey and represent the medians and the 25th/75th percentiles of the composite posterior metallicity PDFs (\citetalias{fumagalli16}; the horizontal error bars show the range of \nhi\ and are centered on the average  \nhi\ values). For completeness and reference, we also show in this figure (in light-yellow squares) the SLLS metallicities recently compiled from the literature as well as a few new metallicity estimates by \citet{quiret16}. For that sample, we only consider metallicities that were derived using an $\alpha$-element (i.e., \oi, \siii, \mgii) and within the redshift interval $2.3<z<3.3$. We have also attempted to remove from that sample any proximate SLLSs or absorbers that may be possibly biased (e.g., a D/H target). In that sample, the 5 estimated metallicities with \oi\ are all for SLLSs  with $19.75 \le \mlnhi \le 20.05$ and resulted in metallicities within the range $-2.3 \la \xh \la -1.2$. Note that for several of these metallicites (including those derived with singly ionized species) no ionization correction was realized, which may play in part a role in some of the observed elevated values ($-0.5 \la \xh \la +0.1$), especially since 5 of these have comparatively  low \nhi\ values with  $19.1\le \mlnhi \le 19.3$. Owing to the clean selections of the LLSs and SLLSs and the uniform analysis of the HD-LLS survey (both similar to the KODIAQ Z survey), we favor HD-LLS survey for comparison with our sample.  For the DLAs, we use the measurements and compilation from \citet{rafelski12}.\footnote{We note that \citet{quiret16} also compile all the existing DLA metallicities from the literature. Unfortunately, for our purposes, this compilation lacks key information regarding any selection biases (e.g., D/H targets, DLAs pre-selected owing to the absence of metal absorption in SDSS spectra, etc.).} In Table~\ref{t-metavg}, we summarize the mean, median, and dispersion for each of these classes of absorbers. We also estimated the fraction of VMP DLAs with $\xh \le -2.4$ (see Table~\ref{t-metavg}), which by definition of this threshold value ($2\sigma$ below the mean metallicity of the DLAs) is small. For the HD-LLS survey, owing to the method used to determine the metallicity, we list in Table~\ref{t-metavg} the probability of finding absorbers lower than $\xh \le -2.4$.

Considering the entire range of \nhi\ plotted in Fig.~\ref{f-metvsnh1} ($14 \la \mlnhi \la 22$) at $2.3 < z < 3.3$, several immediate conclusions can be drawn: 1) there is a gradual decrease in the mean (or median) metallicity from the DLAs to the LYAF (with possibly the exception of the pLLSs, but see above); 2) the dispersion around the mean for the LYAF, pLLSs, LLSs, and SLLSs is large (about 0.8 dex on average), but for the DLAs the dispersion is a factor 2 smaller ($\sim$0.5 dex); 3) there is a substantial fraction of LYAF, pLLSs, LLSs, and SLLSs that has metallicities below $\xh \le -2.4$ while $<3\%$ of the DLAs have such low metallicities; 4) only for the LYAF, pLLSs, and LLSs, there is evidence of metallicity below $\xh \simeq -3$ (see Fig.~\ref{f-metvsnh1}): for the pLLSs and LLSs, the fraction with $\xh \le -3$ is in 2.5--17.7\% (68\% confidence interval), while $\sim 30\%$ of the LYAF absorbers have $\xh  \la -3.5$ \citep{simcoe04,simcoe11b}.

%\clearpage
\section{Redshift evolution of the  \lowercase{p}LLS\lowercase{s} and LLS\lowercase{s}}\label{s-prop}
Our selection of the pLLSs and LLSs at $z<1$ and $2.3<z<3.3$ follows the same criteria: first, they are \hi-selected to have \hi\ column densities between $16 \la \mlnhi <19$; second, the \hi\ column density can be estimated reasonably accurately (within $\sim$0.3 dex and often better than 0.1 dex); and third, there is enough information from the metal lines to derive sensitively the metallicities. Therefore we can directly compare the high and low redshift samples to study the evolution of the metallicity for these systems.  However, the overdensities of the structures change as function of $z$. At $z\sim 0.7$ the critical density of the universe is about a factor 8 lower than at $z\sim 2.8$. Using, e.g., the empirical relationship for the overdensity derived by  \citet{dave99} for absorbers with $12.5 \la \mlnhi \la 17.5$, $\delta_{\rm H} \equiv (n_{\rm H} - \bar{n}_{\rm H})/\bar{n}_{\rm H} \sim 20 \mnhi/(10^{14}\,{\rm cm}^{-2}) 10^{-0.4 z}$, the change in $\delta_{\rm H}$ is similarly a factor $\sim$8  between the mean redshifts of the \citetalias{wotta16} ($\langle z \rangle = 0.7$) and this study ($\langle z \rangle = 2.8 $). This implies that absorbers at some given \nhi\ at high and low redshifts are not necessarily physically analogous \citep[see also][]{dave99}. For the LYAF absorbers, SLLSs, and DLAs, the redshift evolution of the density does not change the fact that  LYAF absorbers trace very diffuse gas ($\delta_{\rm H}\ll 100$) and SLLSs/DLAs trace virialized structures ($\delta_{\rm H}\gg 100$) at both high and low $z$. On the other hand, for the LLSs and especially the pLLSs, while at $z<1$ they probe gas well within the CGM of galaxies, at $z\sim 2.8$, $\delta_{\rm H}$ can be $\la 100$, and hence pLLSs could probe more diffuse ionized gas at $z>2$. KBSS shows that only half of the absorbers with $\mlnhi >15.5$ are found in the CGM of ({\it massive}) galaxies at $z\sim 2$; the other half may probe more diffuse gas or the CGM of dwarf galaxies \citep{rudie12}. Hence while high $z$ LLSs and pLLSs are by definition at the interface between the denser and more diffuse gas, they may not trace necessarily the same dense CGM of galaxies as their counterparts at $z<1$. We keep this caveat in mind as we now review the evolution of the properties of the pLLSs and LLSs with $z$.

\subsection{Evolution of the physical properties with $z$}\label{s-difference}

While the main goals of our study are to determine the shape of the metallicity distribution of the pLLSs/LLSs at high $z$ and how it evolves with $z$, we can also highlight similarities and differences in other properties (densities, $U$, etc.) of the pLLSs and LLSs at low and high $z$. In Table~\ref{t-cloudyavg}, we summarize the mean, median, standard deviation, and minimum, maximum values of \nhi\ and several physical parameters derived from the Cloudy models for the pLLS/LLS samples at $z<1$ (from \citetalias{lehner13}) and $2.3<z<3.3$ (this paper as well as the results from \citealt{crighton13,crighton15,lehner14}). Note that here we have treated upper or lower limits as actual values, but this has limited effect on the statistics and comparison.\footnote{We have removed for this analysis the two absorbers where we set by hand $\log U \ge -4$ owing to too little constraints from the observations; including these would, however, not have changed the results.} For example, we find for the sample of pLLSs and LLSs at $2.3 < z < 3.3$ $\langle \log U \rangle = -2.35 \pm 0.12$ using the KM estimator instead of $-2.4$ assuming that the lower limits are actual values.  As demonstrated by \citetalias{fumagalli16}, we emphasize that while the metallicities derived from the Cloudy simulations are quite reliable, there is a degeneracy between ionization parameter and intensity of the radiation field, which hinders robust estimates of the densities and sizes of the absorbers. Hence the hydrogen 
density ($n_{\rm H}$) and linear scale ($l \equiv N_{\rm H}/n_{\rm H}$)  are not as robustly derived as the metallicities or the total H column density ($N_{\rm H}$).

\begin{figure}
\epsscale{1.2} 
\plotone{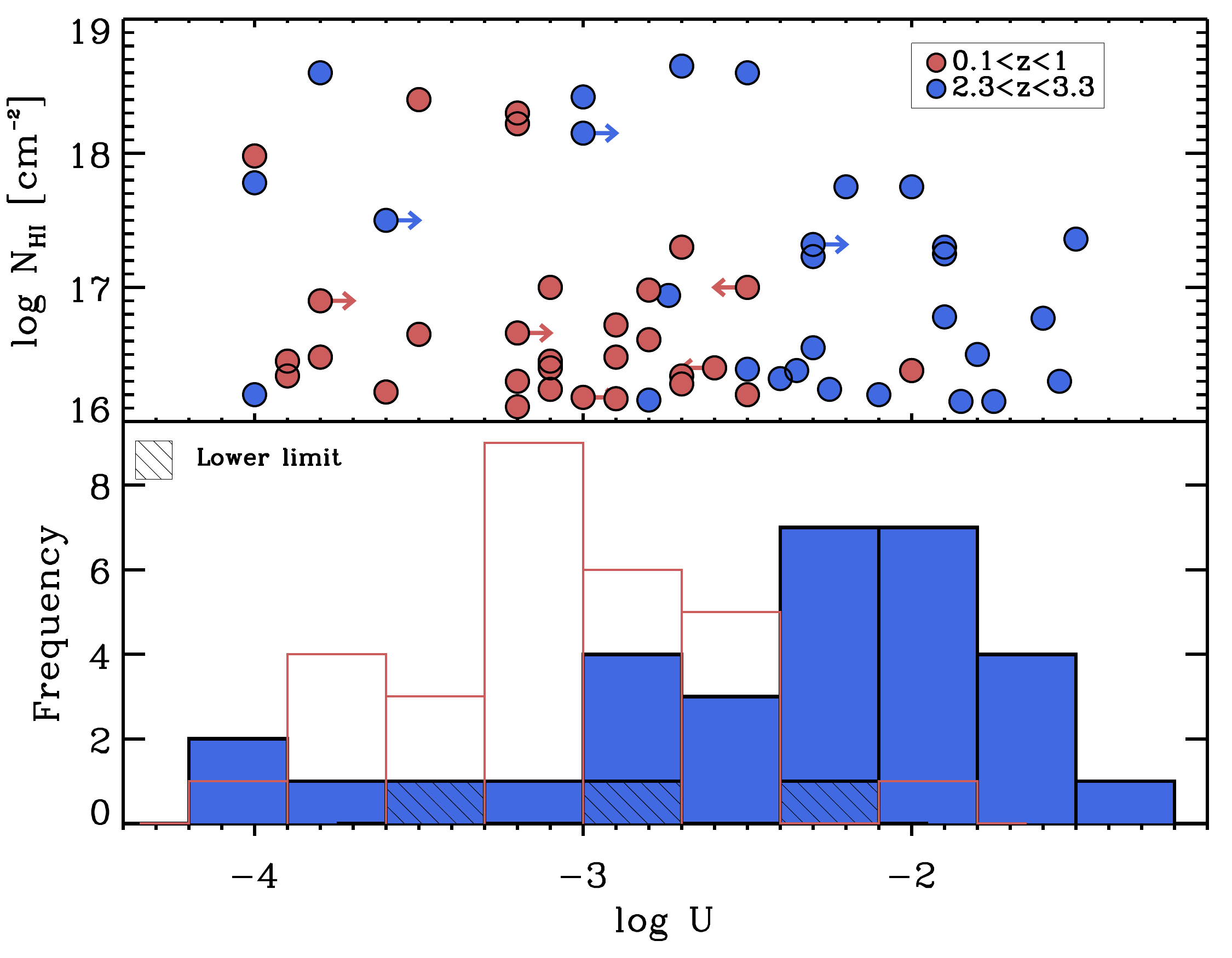}
 \caption{The \hit\ column density as a function of $\log U$ ({\it top}) and distribution of $\log U$ for the pLLSs and LLSs ({\it bottom}) at $2.3 < z < 3.3$ from our sample and at $z<1$ from \citetalias{lehner13}. Note that lower/upper limits are not shown in the bottom panel for the $z<1$ sample for clarity, but can be identified from the top panel.   }
 \label{f-udist}
\end{figure}

Unsurprisingly, the statistics for \nhi\ at low and high $z$ are not too dissimilar owing to a similar initial selection of the pLLSs and LLSs (see Table~\ref{t-cloudyavg}). A two-sided KS test on the \nhi\ low and high $z$ samples gives a maximum deviation between the cumulative  distributions $D= 0.28$ and a $p$-value $p=0.16$, implying no significant difference between the \nhi\ samples at low and high $z$. On the other hand, the ionization parameter derived from the Cloudy simulations evolves significantly with $z$.  In Fig.~\ref{f-udist}, we show the histogram distribution of $U$  and distribution of $U$ against \nhi\ for the pLLSs and LLSs in our sample at $2.3<z<3.3$ (see Appendix) and the \citetalias{lehner13} sample at $z\la 1$. There is some evidence that strong LLSs with $\mlnhi \ga 18$ have smaller $U$-values at any studied $z$, but the sample of these strong LLSs is still small. For absorbers with $\mlnhi \la 18$, there is no obvious trend between $U$ and \nhi\ at any $z$.  Most of the pLLSs/LLSs at $2.3<z<3.3$ have $-3\la \log U \le -1.5$ (consistent with the early compilation made for the LLSs by \citealt{fumagalli11b} and from the HD-LLS analysis, see \citetalias{fumagalli16}) while at $z<1$, most have $-3.8 \le \log U \le -2.5$. A two-sided KS test on the $U$ samples at low and high $z$ gives $D =0.58$ and $p=4.0\times 10^{-5}$, implying a significant difference in the $U$ distributions at low and high $z$. The mean and median of $\log U$ are a factor 10 times larger at $2.3<z<3.3$ than at $z<1$.  The higher $U$-values at high redshift explain why highly ionized species (\siiv, \civ) can be modeled by photoionization, while a single-phase photoionization model typically fails to produce the same highly-ionized species (especially \civ) at $z<1$ for the pLLSs and LLSs (\citetalias{lehner13} and see also \citealt{fox13}).  

\begin{figure}
\epsscale{1.2} 
\plotone{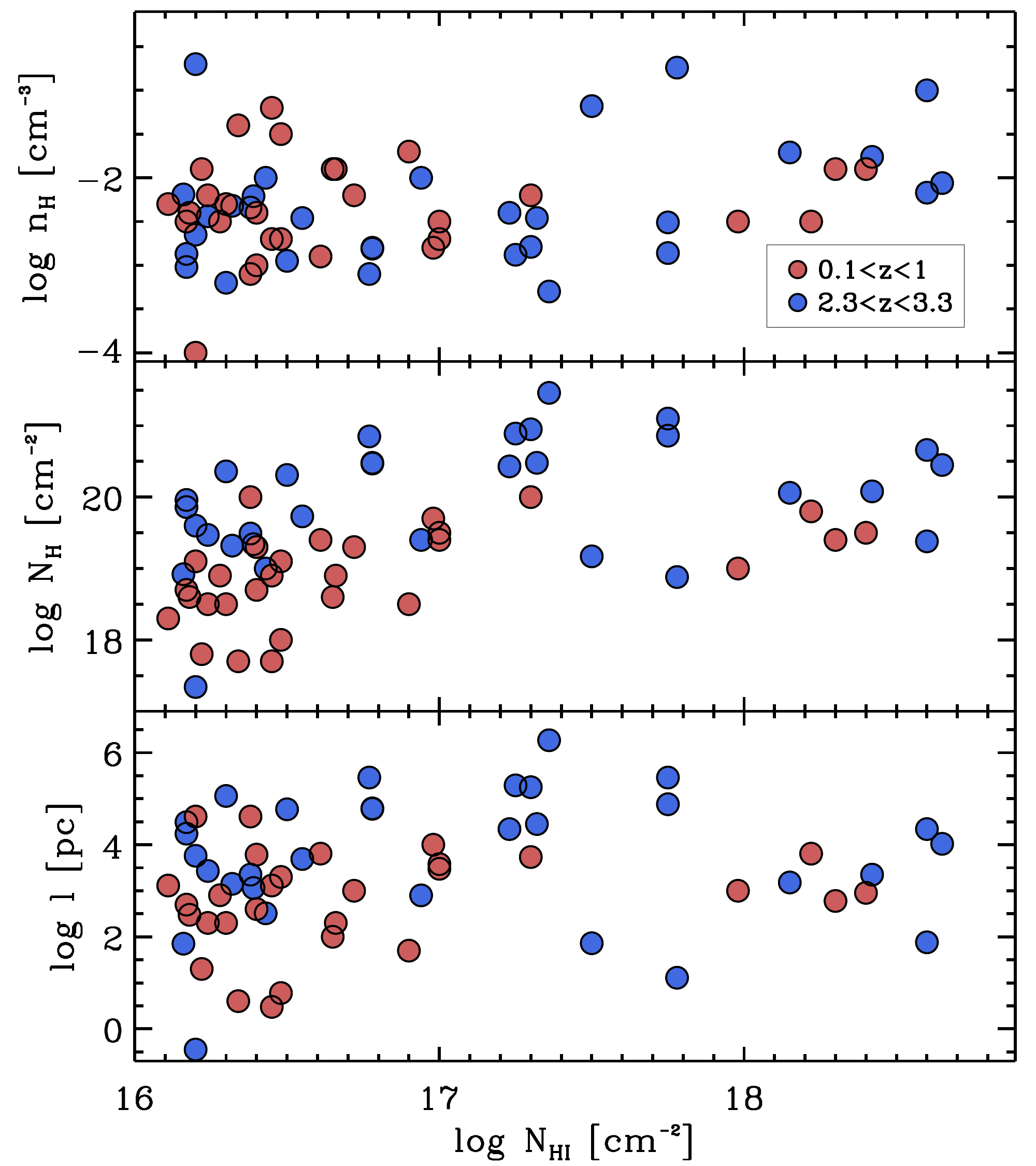}
 \caption{The hydrogen density ({\it top}), hydrogen column density ({\it middle}), and physical scale ({\it bottom}) as a function of the \hit\ column density for the pLLSs and LLSs at $2.3 < z < 3.3$ from our sample and at $z<1$ from \citetalias{lehner13}.}
 \label{f-ldist}
\end{figure}

 In Fig.~\ref{f-ldist}, we show the hydrogen density, hydrogen column density, and physical scale as a function of the \hit\ column density for the pLLSs and LLSs at $2.3 < z < 3.3$ from our sample and at $z<1$ from \citetalias{lehner13} (note that we ignore the very few lower/upper limits in this figure).  For the densities, while there are few more high $n_{\rm H}$ values at $z<1$ for weak pLLSs, overall $n_{\rm H}$ at high and low redshifts overlaps and have the same mean $\langle \log n_{\rm H} \rangle \simeq -2.3$ with a dispersion of about $0.6$ dex. These densities are very similar to the densities estimated by \citetalias{fumagalli16} for stronger LLSs. A two-sided KS test on the $n_{\rm H}$ samples at low and high $z$ gives $D =0.18$ and $p=0.65$, implying indeed no significant difference in the $n_{\rm H}$ distributions at low and high $z$. 

 For the total H column densities, their typically values are higher at high redshift than at low redshift over the entire \nhi\ range probed by the pLLSs and LLSs. On average, $N_{\rm H}$  is a factor $\sim$10 times larger at high $z$ than low $z$.  A similar trend is also observed for $l$ where large-scale structures ($l>10$--$100$ kpc) for the pLLSs and LLSs are not rare at $z\ga 2.4$ (a result also found by \citetalias{fumagalli16} and \citealt{cooper15} at higher $z$ and for the LLSs at the boundary with the SLLSs). In the pLLS regime, while there is a large fraction of low-$z$ pLLSs with $l \la 1$ kpc, there is also an overlap between high- and low-$z$ pLLSs with $1 \la l < 100$ kpc. A two-sided KS test on the $N_{\rm H}$ and $l$ samples at low and high $z$ gives $p=0.0002$ and $p=0.003$, respectively, implying in both cases significant differences in the distributions of these quantities at low and high $z$.

Finally, the last entry of Table~\ref{t-cloudyavg} shows that the temperature of the gas probed by the pLLSs and LLSs is higher at high $z$, but with a similar large dispersion at both low and high $z$.  \citetalias{fumagalli16} found that the  probability distribution function of the gas temperature peaks strongly at a similar value for the photoionized gas than the mean of our high redshift sample.  A two-sided KS test on the temperatures samples at low and high $z$ gives $D =0.61$ and $p=1.1\times 10^{-5}$, implying a significant difference in the $T$ distributions at low and high $z$.

Hence this strongly suggests based on simple overdensity arguments and the Cloudy results that the pLLSs and LLSs have different physical parameters at high and low $z$ (except for the densities), implying that the pLLSs and LLSs  at $z>2$ do not evolve directly into their low $z$ analogs. Using the empirical relationship from \citet{dave99}, the pLLSs and LLSs at $z\sim 2.8$ should evolve into strong LYAF absorbers ($\mlnhi \ga15$) and pLLSs at $z \sim 0.7$, respectively.

\begin{figure}
\epsscale{1.2} 
\plotone{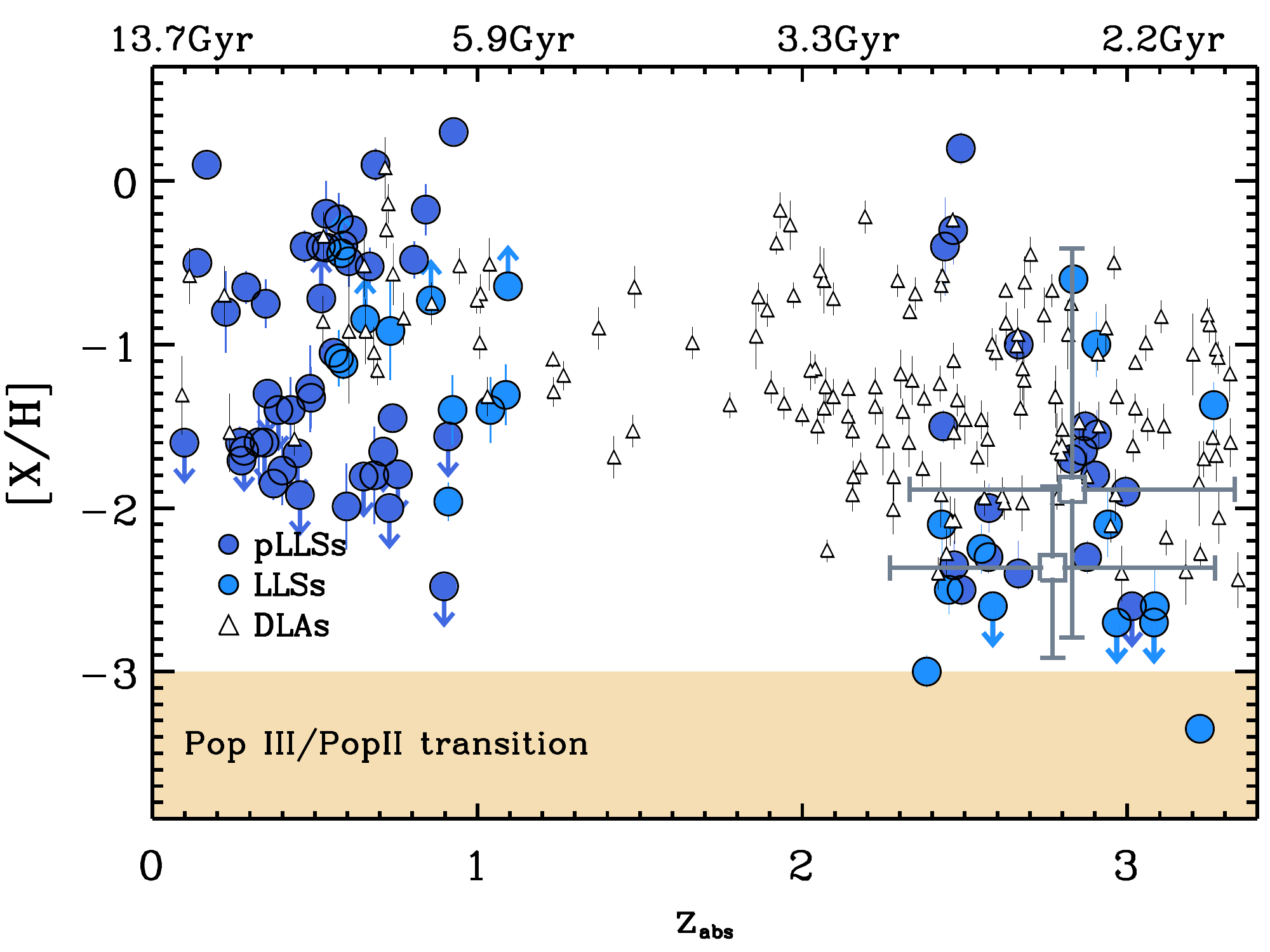}
 \caption{Metallicity as a function of the redshift (time since Big Bang is indicated on the top axis). The pLLS+LLS data at $2.3 < z < 3.3$ are from  this work and at $z<1$ are from \citetalias{wotta16} and \citetalias{lehner13}.The grey squares are for the LLSs at $2.3 < z < 3.3$ with $17.30 \le \mlnhi < 18.3$ ({\it bottom}) and  $18.30 \le \mlnhi < 19.3$ ({\it top}) from the HD-LLS survey (\citetalias{fumagalli16}; the slight redshift difference between the two data points is only artificial to be able to more easily separate them). The DLA data (open black triangles) are from \citet{rafelski12}.
 \label{f-metvsz}}
\end{figure}

\subsection{Evolution of the metallicity with $z$}\label{s-metvsz}
The cosmic evolution of the DLAs  \citep[e.g.,][]{prochaska03,rafelski12,battisti12,jorgenson13}  and SLLSs  (e.g., \citealt{som13,som15}; \citetalias{fumagalli16}; \citealt{quiret16})  have been well studied for several years.  In Fig.~\ref{f-metvsz}, we show the metallicity evolution of the pLLSs and LLSs as a function of redshift (and look-back time) where the low and high $z$ absorbers were selected and analyzed using the same methodology. At all $z$ the peak-to-peak scatter in the metallicities of the pLLSs and LLSs is large (over 2 dex spread in $\xh $). Owing to this large scatter, there is an overlap in the MDFs of the pLLSs and LLSs at low and high $z$, but the MDF is also changing drastically with $z$: at $2.3 < z < 3.3$, the MDF is unimodal, peaking at  $\xh \la -2$ with a long tail to higher metallicities, while at low $z$, the MDF is bimodal, peaking at $\xh \simeq -1.8$ and $-0.3$ with about the same number of absorbers in each branch of the distribution (see also \citetalias{lehner13,wotta16}). At low $z$, only one system has a metallicity well below $\xh \simeq -2$, although there are several upper limits near this lower bound metallicity.  The quasi-absence of very low metallicity gas at $z<1$ can be attributable in part to the lower sensitivity of the UV data (typically, S/N\,$\la 20$--$30$ for \hst/COS observations compared to $\ga 30$--$100$ for data obtained with Keck HIRES, see \citetalias{lehner13} and \citealt{omeara15}), but it is also possible that low metallicity gas with $\xh \la -2$ is rare at low $z$.

As noted above, pLLSs and LLSs at low $z$ are probably not always their direct high redshift analogs. Based on the overdensity argument, LLSs at $2.3<z<3.3$ could evolve into the low $z$ pLLSs. Using the results from this work (see Fig.~\ref{f-metvsnh1} and \S\ref{s-metvsnhi}) and \citetalias{fumagalli16}, the MDF of the LLSs at $2.3<z<3.3$ is consistent with a unimodal distribution, significantly different from the bimodal MDF of the pLLSs at $z\la 1$  \citepalias{wotta16}. Therefore, even considering the redshift evolution of the cosmic structures, there is a significant evolution of the MDF of the LLSs with $z$. 

The change in the MDF of the pLLSs and LLSs between $2.3 < z < 3.3$ and $z<1$ is also quite significant and distinct from DLA and SLLS evolution. The MDF of the pLLSs and LLSs is not simply shifting to higher metallicity as observed for the SLLSs and DLAs, but the shape of the MDF is evolving significantly to lower $z$.  In Fig.~\ref{f-metvsz}, we also show the redshift evolution of DLA metallicities from the \citet{rafelski12} survey for comparison. As noted by \citet{rafelski12} and others, there is an overall increase of the metallicity with decreasing $z$, but the shape of the MDF for the DLAs does not evolve with $z$; it is  unimodal with similar scatter about the mean at all redshifts. This scatter in metallicities is also smaller than that observed for the pLLSs and LLSs. The ``lower envelope" of the metallicity  of the DLAs (mean metallicity of the DLAs minus $2\sigma$) changes from $\xh \simeq -2.4$ at $2.3<z<3.3$ to $-1.4$ dex at $z\la 1$. Below these metallicities at the respective redshifts, there is a large number of pLLSs or LLSs, implying that a large fraction of the pLLSs and LLSs follows a different metal enrichment than the DLAs. At all $z$, however, there is also a large overlap in the metallicities of the DLAs and the more metal-enriched pLLSs and LLSs; these higher-metallicity pLLSs and LLSs may follow a similar metal enrichment evolution similar to that of the DLAs.

\subsection{Relative Abundances of C/$\alpha$}\label{s-calpha}

\begin{figure}
\epsscale{1.2} 
\plotone{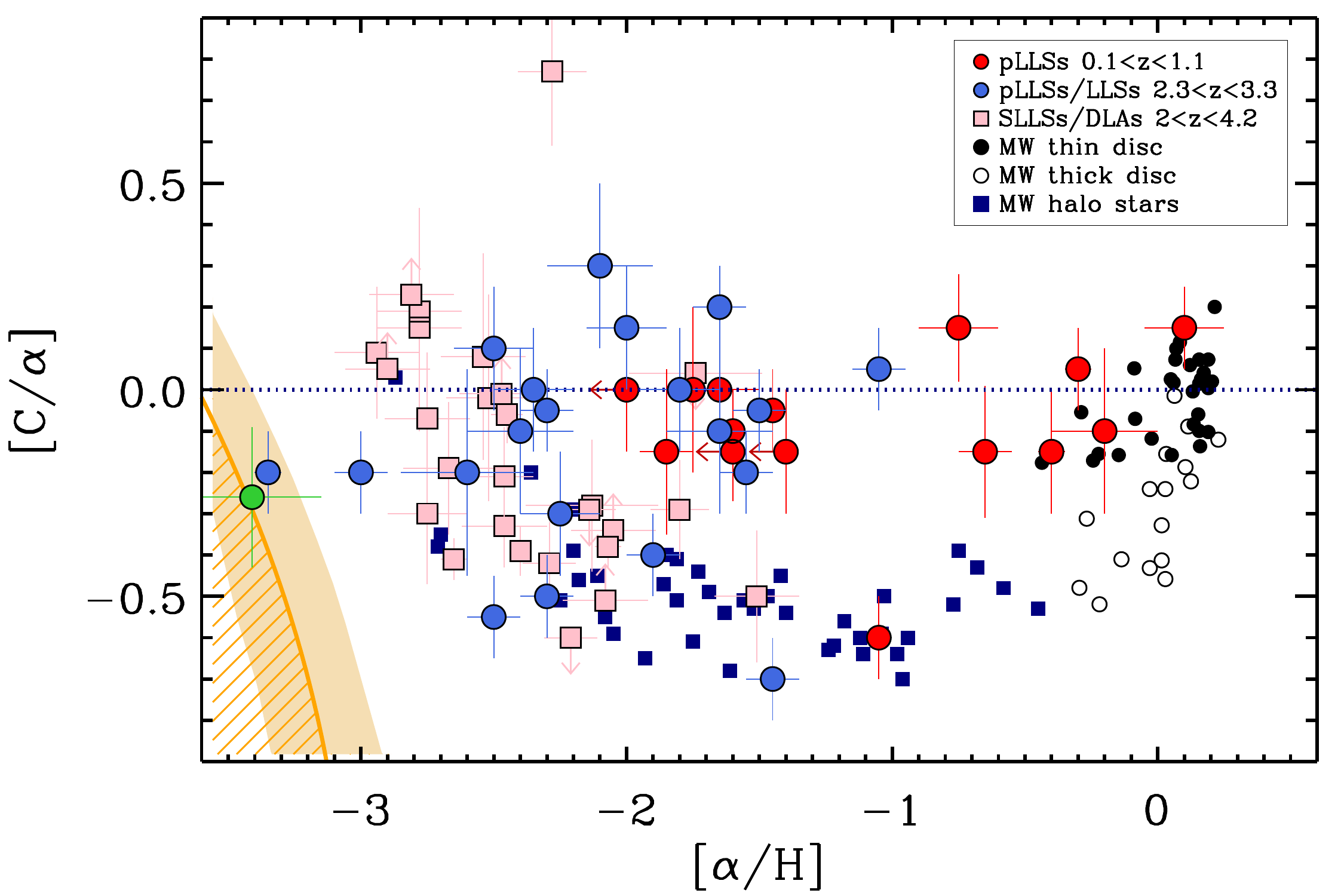}
 \caption{Evolution of  [C/$\alpha$] as a function of the metallicity $[{\rm \alpha/H}]$ for various types of absorbers and stars indicated in the legend (see text for more details and references; the green data point is a LLS at $z\simeq 3.5$ from \citealt{crighton15}). The hatched orange region is the “transition discriminant” criterion \citep{frebel07}; any gas in this region may have been polluted by Pop III stars (see text). 
 \label{f-calpha}}
\end{figure}

So far we have only presented the results for the absolute abundances of the gas. Although we have limited information on the relative abundances, at both high and low redshifts (see \citetalias{lehner13}), we have some constraints on the C/$\alpha$ ratio. This ratio is a good indicator of the nucleosynthesis history since in low density, diffuse gas, carbon and the $\alpha$ elements used in these works are not expected to be strongly depleted into dust grains (see \S\ref{s-cloudy}), and hence this ratio provides additional information regarding the origin of the gas. For the pLLSs and LLSs, this ratio was principally derived from the photoionization models (see \S\ref{s-cloudy}). In these  models,  C/Si was set a priori to a solar value, but was allowed to vary in order to determine the best $U$, $\xh$-values that fit the data.   Although, this ratio is derived using photoionization models and subtle changes in the radiation field could change its value, we feel it is robustly derived for the following reasons. Firstly, \citetalias{wotta16} show that while modifying the radiation field from HM05 to HM12 can change ${\rm [\alpha/H]}$ in a systematic manner by about $+0.3$ dex, it does not affect as much the  C/$\alpha$ ratio. Secondly and independently from any ionization assumption, we can directly estimate C/$\alpha$ from the observations using the column density ratios $(N_{\rm CII} + N_{\rm CIII} +N_{\rm CIV})/(N_{\rm SiII} + N_{\rm SiIII} +N_{\rm SiIV})$ at $z>2$ and $(N_{\rm CII} + N_{\rm CIII})/(N_{\rm SiII} + N_{\rm SiIII} )$ at $z<1$ (\civ\ and \siiv\ are not considered at lower redshift because these are typically produced in a different gas-phase, see \citetalias{lehner13}).  We summarize these results in Table~\ref{t-calpha}. There is only a small fraction of the sample where we have simultaneously column densities for all these ions, but it is striking that for all but one, the direct and modeling methods provide consistent results (the only discrepancy toward J131215$+$423900  could be possibly arising owing to some contamination in the \ciii\ $\lambda$977 absorption). As a reminder for the pLLSs and LLSs, at high redshift, the $\alpha$-element is mostly Si, but at low redshift it can also be O, Mg, and/or S depending on the system (see \citetalias{lehner13}).

In Fig.~\ref{f-calpha}, we show [C/$\alpha$] vs. [$\alpha$/H] for the pLLSs and LLSs from both the high- and low-redshift samples from this and \citetalias{lehner13} surveys (note that the most metal poor LLS in this  figure is from \citealt{crighton15}). We note that in the regions of overlapping metallicities, there is no obvious difference between the low and high redshift samples, and we therefore treat them together in the remainder of this section. For comparison, we also show the results for high redshift DLAs and SLLSs and Milky Way (MW) stars. For the DLAs and SLLSs, we use the results from \citet{pettini08}, \citet{penprase10}, and \citet{cooke11a} (and references therein and see also \citealt{becker12} for $z\ga 5$ measurements). For the MW thin and thick stars, we use the results from \citet{bensby06}, and for the MW halo stars, \citet{fabbian09} and \citet{akerman04}. For the stars, $\alpha$ is O, while for the DLAs and SLLSs, $\alpha$ is O or Si (changing O to Si or vice-versa for the DLAs would have little effect on the distribution of these data). As noted by \citet{pettini08}, \citet{penprase10}, and  \citet{cooke11a}, the  metal-poor SLLSs/DLAs follow well the overall behavior of  [C/$\alpha$] with [$\alpha$/H] having a similar dispersion as observed in the MW  metal-poor stars and confirm the overall increase of  [C/$\alpha$] seen in metal-poor stars \citep{akerman04,spite05}. Where DLAs and stars overlap (${\rm [O/H]}\la -1.5$), the overall agreement in the distribution of C/$\alpha$ suggests a universal origin for the production of C relative to $\alpha$-elements \citep{cooke11a}. 

The overall trend observed in Fig.~\ref{f-calpha} in the stellar and SLLS/DLA samples can be separated in roughly two regions. {\it Region 1}: At $-3\la [{\rm \alpha/H}] \la -1$, [C/$\alpha$] decreases with increasing metallicity from super-solar values to about $-0.7$ dex. {\it Region 2}: at $-0.7\la [{\rm \alpha/H}] \la +0.2$, [C/$\alpha$] increases with increasing metallicity from about $-0.6$ dex to super-solar values. The behavior in region 2 has been well known for some time and is thought to occur as a result of the delayed release of carbon from low- and intermediate-mass stars combined with a strong metallicity dependence of the yields of carbon by massive stars with mass-loss \citep[e.g.,][]{akerman04,fabbian09}. The increase of [C/$\alpha$] to lower metallicity at $ [{\rm \alpha/H} ] \la -1$ was somewhat surprising at first, but has now been confirmed independently in both stellar atmospheres and SLLSs/DLAs. One possible interpretation for the high values of C/$\alpha$ at low metallicity could  be the leftovers from the enhanced production of C (relative to $\alpha$-elements, and in particular O) in Population III (Pop III) stars. As shown by \citet{frebel07} and \citet{bromm03}, the gas progenitor of Pop III stars must have had high C abundance to efficiently cool the gas in order to actually form stars and to drive the transition from Pop III to Pop II stars (see also \citealt{cooke11a} for more discussion).  We show in Fig.~\ref{f-calpha} that condition (hatched orange region) defined as the ``transition discriminant" criterion. No Pop II stars should be found in that zone, but any gas in this region will likely have been polluted by Pop III stars (twoo LLSs are found in that``forbidden'' zone, see Fig.~\ref{f-calpha} and below).

Considering now the pLLSs/LLSs, about half the sample of the pLLSs and LLSs follows a similar distribution to that observed for the DLAs and stars over the entire range of metallicity, i.e.,  $-2.8\la [{\rm \alpha/H}]\la 0$. For these, their chemical enrichment (at least of C and $\alpha$-elements) appears to be similar to that of the MW stars and the bulk of the SLLSs/DLAs. However, the other half --- mostly clustered at $-2.2\la [{\rm \alpha/H}] \la -0.5$ and $-0.2\la [{\rm C/\alpha}] \la +0.2$ --- does not follow the trend observed in MW stars or DLAs as first pointed out by L13.  These gas clouds are carbon-enhanced by a factor $\ga 2$--5 ($\ga 0.3$--$0.7$ dex) compared to stars or most DLAs with similar  $[{\rm \alpha/H}]$. This effect is not artificially caused by the ionization modeling since near solar [C/$\alpha$] over $-2\la [{\rm \alpha/H}] \la -1$ are confirmed directly by the observations (see Table~\ref{t-calpha}), and hence the carbon-enhancement observed at $-2.2\la [{\rm \alpha/H}] \la -1$ is real. 

Finally, we highlight the lowest metallicity LLS in our sample with ${\rm [\alpha/H]} = -3.35 \pm 0.05$ and ${\rm [C/\alpha]} = -0.20 \pm 0.10$ at $z_{\rm abs} = 3.22319$ observed toward J095852+120245 that lies in the Pop III/Pop II transition (orange-zone in Fig.~\ref{f-calpha}). The properties of this LLS are reminiscent of another one at $z_{\rm abs}=3.53$ with  ${\rm [\alpha/H]} = -3.41 \pm 0.26$ and $ {\rm [C/\alpha]} = -0.26 \pm 0.17$ described by \citet{crighton16} (shown with green data point in Fig.~\ref{f-calpha}). This implies that there are now two LLSs at $z\sim 3.4$ with expected [C/$\alpha$] and ${\rm [\alpha/H]}$ that are consistent with gas polluted from Pop III stars.

\section{Discussion}\label{s-disc}
Our present study explores the properties (in particular the metallicity) of the pLLSs and LLSs at $2.3<z<3.3$, a redshift epoch corresponding to the ascending part of the cosmic  star formation rate (SFR) density, near its peak  \citep[e.g.,][]{madau14}. Our previous studies  \citepalias{lehner13,wotta16}  have explored the metallicity of the pLLSs and LLSs with similar \nhi\ at $z<1$ where the cosmic SFR density has significantly decreased. According to cosmological simulations, the exchanges of matter in and out through the CGM play critical roles in the evolution of galaxies and in the evolution of the cosmic star-formation \citep[e.g.,][]{keres05,dekel06,faucher-giguere11}. We therefore expect that some of the properties of the pLLSs and LLSs should be intimately coupled to those of star formation in galaxies. This should also be reflected in changes of the properties of the IGM/galaxy interface region as a function of $z$. As we lay out below, there are clear differences but also similarities between the low and high $z$ CGM probed by pLLSs and LLSs.

Before going further we emphasize that at both high and low redshift studies the samples were \hi-selected absorbers with $16.2 \la \mlnhi \la 18.5$ in order to avoid introducing any bias in the metallicity of the gas probed by these absorbers. We  also use the same technique to derive the metallicity of the absorbers, so any changes in the MDF of the pLLSs and LLSs as a function of $z$ should be genuine, not some effect from comparing different samples or metallicities derived using different techniques. However, owing to the redshift evolution of the universe, pLLSs and LLSs at high $z$ are not the direct analogs of the low redshift pLLSs and LLSs (see \S\ref{s-difference}). 

We also note that at low $z$ we make a direct association between the CGM and absorbers with $16.2 \la \mlnhi \la 18.5$ since all the $z<1$ pLLSs and LLSs with galaxy information have been found so far well within the virial radius of relatively bright galaxies ($0.2L^*$ to $>L^*$, see, e.g., \citetalias{lehner13}; \citealt{lehner09,cooksey08}). At high $z$, galaxy information is still scant.

Observations with the Multi Unit Spectroscopic Explorer (MUSE) found no bright, star forming galaxy in the vicinity of the most metal-poor LLS in our sample \citep{fumagalli16a}. This LLS could probe an IGM structure\footnote{The path length of $\sim$2 Mpc and density $n_{\rm H}\sim 5\times 10^{-4}$ cm$^{-3}$ derived using our Cloudy model for this absorbe rare consistent with an IGM origin. However, we note this absorber is unique among our sample.} or the CGM of a faint galaxy with a SFR $<0.2$ M$_{\sun}$\,yr$^{-1}$. Furthermore, we note that the KODIAQ \ovi\ survey of \hi-selected absorbers with $\mlnhi \ga 16$ shows a large fraction of the pLLSs and LLSs at high $z$ has strong and broad \ovi\ absorption associated with these absorbers, which contrasts remarkably from the \ovi\ properties in the IGM (typically much narrower and weaker). The strength and breadth of the \ovi\ make these absorbers likely probes of the CGM of some very actively star-forming galaxies (\citealt{lehner14} and see \S\ref{s-ovi}).  In any case and at all $z$, the pLLSs and LLSs are at the interface between the very diffuse IGM probed by LYAF absorbers and virialized structures probed by SLLSs and DLAs, and it is in this context that we discuss our results below.

\subsection{Evolution of the MDF of pLLSs and LLSs with $z$}

In the ascending part of the cosmic SFR density at $2.3<z<3.3$, we find that the MDF of the pLLSs/LLSs is heavily weighted to low metallicities, unimodally distributed around $\xh \simeq -2$. At $z\le 1$, well past the peak SFR density, the overall MDF has shifted to higher metallicity. For the pLLSs at $z<1$, the MDF is bimodal with about the same weight in each of the metallicity branches that peak at $\xh \simeq -1.8$ and $-0.3$, i.e., the low-metallicity branch has on average a metallicity 20 times lower than those in the high-metallicity branch \citepalias{wotta16,lehner13}. These results for the low-redshift universe show that there are clearly two main populations of gaseous flows through the CGM at $z<1$.  The metal-enriched CGM gas has properties consistent with those expected for matter being ejected by large-scale galaxy outflows, for matter being tidally-stripped from satellite galaxies, or for material tracing the remnants of earlier outflows that are recycled by galaxies. The other half has an extremely low metallicity for the $z<1$ universe. For all the cases so far, these metal-poor pLLSs and LLSs have been found well within the virial radius of some $>0.1 L*$ galaxy  and have column densities, temperatures, and metal-enrichment levels about consistent with cold accretion gas as observed in cosmological simulations at $z\sim 2$--3 and $z<1$ (see \citetalias{lehner13} and simulations by, e.g.,\citealt{fumagalli11a,vandevoort12b,shen13,hafen16}; and see also \S\ref{s-compsim}).

On average the metallicity of the gas also increases with increasing \nhi\ at  $z<1$ and $2.3<z<3.3$ (see Fig.~\ref{f-metvsnh1} and \citetalias{lehner13,wotta16}). As noted by \citetalias{wotta16}, the difference in the MDFs of the pLLSs/LLSs compared to the SLLSs and DLAs implies there is a fundamental change in the physical origins with \nhi. DLAs are likely probing gas that have been enriched recently at a given $z$, while the bulk of the LYAF probes typically the diffuse IGM with little metal content. The pLLSs and LLSs appear to probe both types of gas, recent metal-enrichment as well as very ancient metal-enrichment. The SLLSs  predominantly probe recent enrichment, but a non-negligible fraction may also be more pristine IGM-like metallicity (see Table~\ref{t-metavg}). 

Naively, if the interpretation that low-metallicity pLLSs and LLSs are mostly probing infalling gaseous streams or clouds, then the gas at the interface between galaxies and diffuse IGM at high $z$ would be infall-dominated at $2.3<z<3.3$. However, at these redshifts, the median metallicity of the pLLSs and LLSs is $\xh=-2.1$, and hence a large proportion of the pLLSs and LLSs has metallicity overlapping with those of the DLAs (Table~\ref{t-metavg} and see Figs.~\ref{f-metvsnh1} and \ref{f-metvsz}). At $z<1$, only the high metallicity branch overlaps with the DLA MDF \citepalias{wotta16}; the mean metallicity of the DLAs at $z<1$ is $\langle\xh\rangle \simeq -0.5$, very similar to that of the pLLSs/LLSs in the high metallicity branch. The mean metallicities of the DLAs and pLLSs/LLSs at $2.3<z<3.3$ are, however, much closer than at low redshift (a factor 4 compared to a factor 20). 

In view of the overlap of metallicities between pLLSs/LLSs and DLAs at high $z$, a better approach to separate at all $z$ potential metal-poor cold accretion candidates from other processes is to consider the fraction of VMP pLLSs/LLSs that we define as absorbers with metallicities $2\sigma$ below the mean metallicity of the DLAs in any given redshift interval. At $z<1$, that threshold is $\xh_{\rm VMP} \le -1.40$; at $2.3<z<3.3$, it is $\xh_{\rm VMP} \le -2.40$; and at $3.2 \le z\le 4.4$, it is $\xh_{\rm VMP} \le -2.65$.  At $2.3 <z<3.3$, the proportion of VMP pLLSs/LLSs is 25--41\% in our sample (see Table~\ref{t-metavg}). Similar numbers in the same redshift interval are found for the HD-LLS survey (31\% for the LLSs, 21\% for the SLLSs, see Table~\ref{t-metavg}). At $z<1$, \citetalias{wotta16} derive 28--44\% of the pLLSs are VMP.  Using the recent sample at $3.2 \le z\le 4.4$ of very strong LLSs from \citet{glidden16} ($\mlnhi\ge 17.8$, except for 2 systems), we calculate that the fraction of VMP strong LLSs is 18--34\% (sample size is 31 as we exclude the two SLLSs, which is similar to the present KODIAQ Z sample). Since many of these absorbers overlap with the SLLS regime, if we include only systems with $\mlnhi\le 19.2$ from the \citeauthor{glidden16} sample, then the fraction of VMP strong LLSs would be 30--51\% (sample size 20).\footnote{At $3.2 \le z\le 4.4$ with a smaller sample probing extremely strong LLSs ($17.8 \la \mlnhi \la 19.5$)  and an indirect method, \citet{cooper15} also found  28\%--40\% of the LLS population could trace VMP gas.} All these intervals are at the 68\% confidence level. 

While in the future we will improve the confidence intervals and refine these fractions over smaller redshift bins, it is striking that the proportion of VMP pLLSs and LLSs do not evolve much with redshift (although we emphasize the \nhi\ values sampled in the $3.2 \le z\le 4.4$  interval are quite higher than in our sample). The average metallicities of the VMP pLLSs/LLSs increase with increasing redshift, but their fractions remain about the same over 12 billion years.\footnote{We also note that the total hydrogen column densities or scale-lengths of the VMP pLLSs and LLSs evolve in the same way as for the more metal rich pLLSs and LLSs, i.e., on average $N_{\rm H}$ is 10 times larger  at $2.3<z<3.3$ than at $z<1$ and there is no obvious difference between the VMP pLLSs/LLSs and the rest of the sample.} These VMP pLLSs and LLSs have metallicities that are consistent with the IGM metallicities in each redshift interval (although at $z<1 $, the metallicity of the IGM is unknown as a result of the limited sensitivity of the space-based UV observations). Hence these VMP pLLSs/LLSs appear to be the reservoirs of metal-poor gas in the interface between galaxies and the IGM, which appear to remain constant over cosmic time and which may feed galaxies with metal-poor gas to continue to form stars over billions of years. These VMP pLLSs/LLSs are also good candidates for cold flow accretions as seen in cosmological simulations (see \S\ref{s-compsim}).

\subsection{The fraction of pristine gas at $2.3<z<3.3$}

We found two pLLSs and LLSs with no metals (see Appendix) that might be reminiscent of the pristine  LLSs that were discovered at $z=3.4$ and 3.1, down to a limit $\xh< -4.2$ and $< -3.8$ \citep{fumagalli11b}. Unfortunately, \siiii\ is contaminated for each of these cases, and hence we cannot place a stringent constraint on their metallicities. For example, the conservative limit on the LLSs at $z = 3.08204$ toward J025905+001121 is $\xh <-2.7$ (and $\log U\ge -3.6$); if instead we adopt the mean $\langle \log U\rangle  = -2.4$ derived in our sample, then $\xh < -4.1$ (see Appendix), a limit similar to those found by \citet{fumagalli11b}. 

To better understand the level of mixing of metals in the gas probed by pLLSs and LLSs in the early universe,  we will need a much larger sample to reliably determine the frequency of pristine gas at $2<z<4.5$ in the interface regions between galaxies and the LYAF. With our sample, we determine that the fraction of pLLSs/LLSs with $\xh \le -3$ is $3$--$18\%$ (2/31) at $2.3<z<3.3$ (68\% confidence interval), consistent with the \citetalias{fumagalli16} results for stronger LLSs. This fraction includes the lowest metallicity absorbers in our sample that have metals detected. If we push to metallicities down to $\xh \le -4$  to exclude any pLLS or LLS with some metals detected, that fraction becomes $\le 3\%$  (68\% confidence interval), implying that pristine pLLSs/LLSs at $2.3<z<3.3$ are rare. 

As noted by \citet{crighton16} \citep[see also][]{cooke11a}, the extremely metal-poor LLSs ( $\xh \sim -3.5$ at $z\sim 3$) with detected metal absorption may provide a new path to study the Pop III/Pop II metal-enrichment transition. The use of both the low metallicity and C$/\alpha$ ratio indeed provides a strong method to find metal-pollution at the transition from Pop III to Pop II star formation. In our sample of 31 pLLS/LLSs, we have found one such absorber (corresponding to a proportion of  $3$--$18\%$) with $\xh  \simeq -3.35$ and [C/$\alpha ]=-0.2$, both consistent with a Pop III origin.  

\subsection{Super metal-rich gas at $z\sim 2.5$}

On the other end of the metallicity spectrum, we have also discovered a supersolar pLLS ($\mlnhi \simeq 16.2$) at $ z = 2.48778$ toward J172409+531405. This absorber is extraordinary on several levels.  It has a metallicity of $\sim 1.6 Z_{\sun}$ at a redshift $z\sim 2.5$. This is the only pLLS with a detection of \oi, which is remarkable for such a low \nhi\ absorber. The physical-scale ($l\simeq 0.35$ pc), density ($n_{\rm H} \simeq 0.2$ cm$^{-3}$), and temperature ($T\simeq 6000$ K) are all extremely atypical for any pLLSs at any $z$. The non-detection of \feii\ implies a $\alpha$/Fe enhancement (or possibly some dust depletion of Fe relative to Si). This pLLS is detected in several ions and transitions, so its properties are well-constrained. It is a multiphase absorber since the \civ\ and singly-ionized species have very different velocity profiles (see Appendix)

This is clearly an outlier in our sample (1/31 or  $1$--$8\%$  at the 68\% confidence interval). Its properties (in particular its high metallicity and multiphase nature) suggests that it directly probes an active outflow from a proto-galaxy at $z\simeq 2.5$. As our KODIAQ Z survey will grow, we will more robustly determine the frequency and properties of both metal rich and pristine pLLSs and LLSs at $2 < z < 4$.

\subsection{C/$\alpha$ in pLLSs and LLSs over cosmic time}
The combined sample of pLLSs and LLSs at $2.3<z<3.3$ and $z<1$ shows that the  scatter in C/$\alpha$ with metallicity is very large at any $z$ and  C/$\alpha$ does not follow the trend observed in stars or DLAs (see Fig.~\ref{f-calpha} and Table~\ref{t-calpha}). Stated in another way, about half the sample of pLLSs and LLSs has an  enhanced  C/$\alpha$ ratio in the metallicity range $-2 \la \xh \la -0.5$ compared to Galactic halo stars and DLAs,  while the other half follows more closely C/$\alpha$ patterns seen in  Galactic metal-poor stars or DLAs. The enhanced  C/$\alpha$ ratio in the metallicity range $-2 \la \xh \la -0.5$ implies that this gas must have been polluted by preferential ejection of C from low metallicity galaxies. A recent study in fact shows that at least some local metal-poor dwarf galaxies have also enhanced C/$\alpha$ over similar metallicities \citep{berg16}. While their  C/$\alpha$ ratios are not as high as observed for the pLLSs and LLSs and their sample is small (12 galaxies), the absence of clear trend  between [C/$\alpha]$ and $[\alpha$/H]  is similar to that observed in pLLSs and LLSs. 

On the other hand, in the IGM (probed by the LYAF) at $z\sim 2.1$--$3.6$, using the pixel optical depth analysis of \civ, \ovi, and \siiv, low C/$\alpha$  ratios were derived:  $[{\rm C/Si}] = -0.77 \pm 0.05 $ and $[{\rm C/O}] = -0.66 \pm 0.06 $  \citep{schaye03,aguirre04,aguirre08}. As discussed in \citet{aguirre04}, they only use the \civ/\siiv\ and \ovi/\siiv\ ratio to determine C/Si and O/Si, respectively, which is dependent on the assumed ionizing background (and if collisional ionizing processes take place). While such low values are found for some of pLLSs and LLSs (see Fig.~\ref{f-calpha}), our results imply a very large scatter in  C/$\alpha$ that does not depend on the redshift or the metallicity. It would seem likely that this should also happen in the IGM.

\subsection{\ovi\ associated with pLLSs and LLSs}
\label{s-ovi}
Although we focus throughout on the metallicity of the cool gas of the pLLSs and LLSs, some of the surveys described above have also revealed that  \ovi\ absorption with overlapping velocities with \hi\ is found at any $z$ \citep{lehner13,lehner14,fox13}. When \ovi\ is detected, these pLLSs and LLSs have typically multiple gas-phases as evidenced by the presence of low ions (e.g., \cii, \siii, \siiii) and \ovi\ (or other high ions) that have often very different kinematics and cannot be explained by a single photoionization model (e.g., \citealt{lehner09,lehner13,crighton13,fox13}). At $z<1$, among the 23 pLLSs/LLSs with \ovi\ coverage, only 6 have no \ovi\ absorption, and hence the detection rate of \ovi\ absorption associated with pLLSs/LLSs is about 70\% and even higher (75--91\%) if only sensitive limits on \novi\ are considered
 \citep{fox13}.  At $2.3<z<3.6$, a similar number is found with the KODIAQ survey \citep{lehner14}. While there is a high frequency of \ovi\ absorption associated with pLLSs/LLSs at both high and low $z$, the similarities in the highly ionized gas properties between the high and low $z$  pLLS/LLS sample end there. 

The KODIAQ survey shows that for \hi-selected absorbers at  $z\sim 2$--3.5 with $\mlnhi \ga 16$, the \ovi\ absorption has typically total column densities $14.2 \la \mlnovi \la 15.5$ and full-widths $150 \la \Delta v_{\rm OVI} \la 500$ \km\ (\citealt{lehner14,burns14,lehner16a}; N. Lehner, J.C. Howk, J. O'Meara al. 2016, in prep., and see also Fig.~\ref{f-example2} and Appendix). More than half of the KODIAQ sample has $\mlnovi \ga 14.4$ and $\Delta v_{\rm OVI} \ga 300$ \km. The breadth and strength of the \ovi\ absorption in  strong \hi\ absorbers at $z\sim 2$--3.5 are quite similar to those observed in starburst galaxies at low redshift \citep[see, e.g.,][]{grimes09,tripp11,muzahid15} and remarkably  different from those of the \ovi\ absorption in the IGM at similar redshifts (typically $13.2 \la \mlnovi \la 14.4$ and $20 \la \Delta v_{\rm OVI} \la 100$ \km, see \citealt{simcoe02,muzahid12}). This strongly suggests that the bulk of the strong and broad \ovi\ associated with pLLSs and LLSs traces large-scale outflows from high-redshift star-forming galaxies. In contrast, at $z<1$, \ovi\ absorption in the pLLS sample has typically $ 50\la \Delta v_{\rm OVI} \la 150$ \km\ and  $13.8 \la \mlnovi \la 15$ \citep{fox13}. There is overlap between the low and high $z$ surveys, but broad and strong \ovi\ absorption associated with LLSs and pLLSs at $z<1$ is the exception, not the norm. Only two strong \hi\ absorbers with broad ($\Delta v \ga 300$ \km\ ) and strong \ovi\ absorption at $z<1$ have been reported so far, both associated with a massive, large-scale outflow from a massive star-forming galaxy \citep{tripp11,fox13,muzahid15}. Therefore randomly \hi-selected pLLSs and LLSs at $z<1$ and $2.3<z<3.3$ show a dramatic change not only in the MDF of their cool gas but also in the properties of the associated highly ionized gas. 

It is likely that the difference in frequency of strong and broad \ovi\ between the low and high $z$ pLLS/LLS surveys reflects the fact that low-$z$ galaxies are much more quiescent than their high-redshift counterparts. The weaker \ovi\ absorbers associated with pLLSs/LLSs at both low and high $z$ have, however, likely a wider range of origins; according to simulations these may include outflows, inflows, ambient CGM \citep[e.g.,][]{shen13,ford14}.

\subsection{pLLSs and LLSs in cosmological simulations}
\label{s-compsim}
With the first study that extends into the pLLS and low column density LLS regime  with $16.2\la \mlnhi \la 17.5$ at high $z$, we provide new stringent empirical results to test cosmological hydrodynamical simulations. In particular, we demonstrate there is a strong evolution of the metallicity of the pLLSs/LLSs with $z$, but also a remarkably constant fraction of VMP pLLSs/LLSs over cosmic time. For a large proportion of the pLLSs/LLSs at $z<1$ and $2.3<z<3.3$, C/$\alpha$  also does not follow the typical trend observed in metal-poor Galactic stars or high redshift DLAs (see Fig.~\ref{f-calpha} and Table~\ref{t-calpha}). As shown by \citet{bird14}, the simultaneous knowledge of the DLA MDF and column density function can provide strong constraints on the feedback model in cosmological simulations. The same applies for the pLLSs and LLSs for which the evolution of the MDF with $z$ starts to be constrained (and more refinement and improvement will come in the near future) and their column density function is also constrained over cosmic time \citep[e.g.,][]{lehner07,omeara07,prochaska10,ribaudo11,fumagalli13}. 

Simulations have already shown that pLLSs/LLSs may be used to trace cold flows \citep{faucher-giguere11,faucher-giguere15, fumagalli11a, fumagalli14,vandevoort12a, vandevoort12b, hafen16}. Simulated pLLSs and LLSs  at $z\sim 2$--3 and $z<1$ appear, however, to have too many metals (see also discussion in \citetalias{fumagalli16}). Only in simulations with very mild stellar feedback \citep{fumagalli11b}, there is some agreement between the observed and simulated metallicity distributions; in this simulation, cold streams are traced mostly by LLSs within 1 or 2 virial radii of galaxies where the gas has only been enriched to $\xh\simeq -1.8$ with similar scatter to that observed at high or low $z$. However, while mild feedback produces better agreement with the observed MDF at $z\sim 2$--3, the disagreement with the baryon fraction in stars worsens \citep{fumagalli11b}. The zoom-in Eris2 simulations by \citet{shen13} include much stronger galactic outflows (but possibly more realistic at these redshifts, see \citealt{lehner14}) and show that cold flows are metal-poor, but with a median value $-1.2$ dex, much larger than observed.  \citet{vandevoort12a} similarly show that cold mode accretion is generally metal-poor  with $\xh\sim -1.5$ for any halo mass at $0.8 R{\rm vir}$, and only for $R>R_{\rm vir}$ does the metallicity of the cold mode accretion go below $-2$ dex. The FIRE zoom-in simulations at $z<1$ have also recently studied the physical nature of the pLLSs and LLSs \citep{hafen16}. These simulations confirm the general interpretation of the bimodal metallicity distribution observed at $z<1$: very low metallicity LLSs are predominantly associated with inflows at $z<1$, but higher metallicity LLSs trace gas with roughly equal probability of having recycled outflows (inflows) or outflows. However, the simulated metallicity distribution is not bimodal and has a metallicity plateau between about $-1.3$ and $-0.5$ dex at $z<1$. Furthermore, while very low metallicity pLLSs and LLSs are prevalent in the observations, they are not in the FIRE simulations, showing again that the gas is typically too metal rich in simulations.  

Nevertheless despite some  disagreements between the simulations and the observations, there is a consensus in the simulations that a large fraction of the metal-poor LLSs and pLLSs should probe cold flow accretions onto galaxies. Future simulations with the goals of studying absorbers such as the pLLSs and LLSs (such as in \citealt{hafen16}) that include advanced radiative transfer techniques (crucial for correctly predicting the pLLS/LLS properties) and varying feedback prescriptions will help guiding the interpretation of these observational results, and in turn these observational results should help refining the sub-grid simulation physics and feedback prescriptions.

\section{Summary}\label{s-sum}
We have undertaken a study of the properties of the gas probed by pLLSs and LLSs at $2.3<z<3.3$ and the evolution of their properties over cosmic time. Here we present the first results from our KODIAQ Z survey with which we have assembled the first sizable sample of \hi-selected pLLSs and LLSs at $2.3<z<3.3$ with $16.2 \le \mlnhi \le 18.4$ (most with $16.2 \le \mlnhi \le 17.8$) for which we have determined the metallicity for each absorber. This sample of 31 absorbers therefore probes gas at the transition in \nhi\ between the LYAF  ($\mlnhi \la 16$) and stronger LLSs ($\mlnhi \ga 18.5$). It provides a direct comparison sample with the $z<1$ sample of \citetalias{lehner13} and \citetalias{wotta16} and complements other samples of typically stronger LLSs at similar and higher redshifts (\citetalias{fumagalli16}; \citealt{cooper15,glidden16}).

To derive the metallicity we have used Cloudy simulations assuming a single gas-phase model following the methodology of our early work at low redshift \citepalias{lehner13}. In particular we have used the same ionizing background (HM05) to avoid introducing additional systematics in our comparison between low and high redshift absorbers. As in \citetalias{lehner13}, we only model the absorption seen in the metals that is associated with the pLLS or LLS \hi\ absorption, i.e., the metallicity is determined by comparing estimated column densities of metal ions and \hi\ in the strongest \hi\ component (not over the entire velocity profile where metal-line absorption may be observed). Our main results  are as follows. 

\begin{enumerate}
\item Typically the following ions \siii\, \siiii, \siiv, \cii, \ciii, \civ\ associated with the pLLSs or LLSs at $2.3<z<3.3$ are satisfactorily modeled with ionization models with $\langle \log U \rangle \simeq -2.4$ (with a dispersion of 0.6 dex), which imply temperatures (1--$4) \times 10^4$ K. Based on these Cloudy models, about half of the sample has physical scale $l<10$ kpc and the other half $17<l<200$ kpc (see Table~\ref{t-cloudyavg}). 

\item We empirically establish that the metallicity distribution of the pLLSs and LLSs at $2.3<z<3.3$ is unimodal peaking at $\langle \xh \rangle  = -2.00 \pm 0.17$ (error on the mean from the survival analysis) with a standard deviation of $\pm 0.84$ dex. The mean and distribution are quite similar to those derived for the stronger LLSs ($17.5 \le \mlnhi \le 18.5$) from the HD-LLS survey over the same redshifts. On the other hand, the mean metallicities of the SLLSs  ($19 \le \mlnhi < 20.3$) and DLAs ($\mlnhi \ge 20.3$) at $2.3<z<3.3$ are higher, $-1.71$ and $-1.39$ dex, respectively (the dispersion of the metallicities for the DLAs is also also factor 2 smaller). For the LYAF ($\mlnhi \la 15.5$), the mean metallicity is significantly smaller at similar redshifts,   $\langle \xh \rangle  = -2.85$ (with a similar dispersion). The mean metallicity of the gas at $2.3<z<3.3$ therefore increases with increasing \nhi\ (with a possible exception for the pLLSs, although a larger sample will be needed to robustly determine this).

\item There is a substantial fraction ($25$--$41\%$) of VMP pLLSs and LLSs with metallicities $2\sigma$ below the mean metallicity of the DLAs  (i.e., $\xh \la -2.4$ at $2.3<z<3.3$).  These VMP pLLSs and LLSs are good candidates of metal-poor cold gas feeding galaxies as seen in cosmological simulations. 

\item At $2.3<z<3.3$, we determine that the fraction of pLLSs and LLSs with $\xh \le -3$, i.e., at the Pop III remnant level, is $3$--$18\%$ at $2.3<z<3.3$ (68\% confidence interval). The lowest metallicity LLS in our sample with a metallicity of $\xh \simeq -3.35$ has some metals detected with  ${\rm [C/\alpha]\simeq -0.2}$, consistent with a Pop III enrichment. There is no strong evidence ($\la 3\%$  at the 68\% confidence interval) in this sample of pristine pLLS or LLS (i.e., with no metal absorption) with $\xh \le -4$. 

\item About half the sample of the pLLSs and LLSs at $2.3<z<3.3$ and $z<1$ has C/$\alpha$ ratios similar to those derived for MW stars and SLLSs/DLAs with similar metallicities over the entire probed metallicity interval ($-3 \la \xh \la +0.5$). The other half has enhanced  C/$\alpha$ ratios (near solar values) in the metallicity range $-2 \la \xh  \la -0.5$, implying that this gas must have been polluted by preferential ejection of C  from low metallicity galaxies.

\item The comparison of the pLLSs and LLSs at $2.3<z<3.3$ and $z\la 1$ that were selected using the same selection criteria and analyzed using the same procedures shows that some of their properties have not evolved strongly with $z$. The absence of trend between  C/$\alpha$  and the metallicity for the pLLSs and LLSs is observed at both high and low $z$. At overlapping metallicities, similar scatter and range of values are observed in C/$\alpha$ at high and low $z$. We show that the fraction of VMP pLLSs/LLSs is 20--47\% (68\% confidence interval) over the redshift interval $z<1$ to $z\sim 4$, i.e., over the last 12 billion years the fraction of VMP pLLSs and LLSs appears to remain relatively constant. The hydrogen densities of the pLLSs and LLSs are also similar at both low and high $z$. 

\item On the other hand, several properties of the pLLSs and LLSs have evolved strongly with $z$. The MDF of the pLLSs and LLSs evolves markedly with $z$, changing from a unimodal distribution at $2.3<z<3.3$ that peaks at  $\xh  \simeq -2.0$ to a bimodal distribution at $z\la 1$ with peaks at $\xh \simeq -1.8$ and $-0.3$. In contrast, the MDF of the DLAs over the same redshift intervals stays unimodal with only an increase of the mean metallicity with decreasing $z$. The ionization parameters, linear scales, and total hydrogen column densities are a factor $\sim 10$ larger on average at $2.3<z<3.3$ than at $z<1$.

\end{enumerate}

These first results from the KODIAQ Z survey already put some strong empirical constraints on the dense ionized gas probed by absorbers with $16 \la \mlnhi \la 18.5$ and their evolution over 12 billion years of cosmic time, before and after the peak of cosmic star formation. However, our sample is still too small to robustly determine if the pLLS and LLS populations at $z>2$ probe similar or widely different physical structures. At $z\la 1$, by doubling the initial sample of pLLSs and LLSs in \citetalias{lehner13}, \citetalias{wotta16} have demonstrated that the MDF of the pLLSs is bimodal, but likely transitions to a unimodal distribution in the LLS regime. Our ongoing KODIAQ Z survey at $z\ga 2$ and COS Legacy survey at $z<1$  will yield much larger samples of pLLSs, LLSs, as well absorbers with $15 \la \mlnhi \la 16$ at both high and low $z$, which will provide new stringent constraints on the properties of the diffuse and dense ionized gas at $0 \la z \la 4$ . 

\section*{Acknowledgements}
Support for this research was partially made by NASA through the Astrophysics Data Analysis Program (ADAP) grant NNX16AF52G. MF acknowledges support by the Science and Technology Facilities Council through grant ST/L00075X/1. Part of this manuscript was written at the winter 2016 retreat hosted by IMPS of UC Santa Cruz Department of Astronomy. We thank the Esalen Institute for its great setting and wonderful hospitality during that retreat. All the data presented in this work were obtained from the Keck Observatory Database of Ionized Absorbers toward QSOs (KODIAQ), which was funded through NASA ADAP grant NNX10AE84G. This research has made use of the Keck Observatory Archive (KOA), which is operated by the W. M. Keck Observatory and the NASA Exoplanet Science Institute (NExScI), under contract with the National Aeronautics and Space Administration.The authors wish to recognize and acknowledge the very significant cultural role and reverence that the summit of Mauna Kea has always had within the indigenous Hawaiian community.

%\bibliographystyle{aasjournal}
%\bibliography{ms} 

%\floattable
\begin{deluxetable*}{lccc}
\tablecaption{Average velocities and column densities of the metal ions \label{t-metal}}
\tablehead{
\colhead{Ions} & \colhead{$[v_1,v_2]$} & \colhead{$v_{\rm a}$}& \colhead{$\log N_a$}  \\
\colhead{} & \colhead{(\km)} & \colhead{(\km)}& \colhead{[cm$^{-2} $]}
}
%\colnumbers
\startdata
\multicolumn{4}{c}{J143316+313126 -- $z_{\rm abs}=2.90116$ -- $\mlnhi = 16.16$} \\
\hline
     \ciit\ $\lambda$1334 & $   -20,   20 $ & $    -2.2  \pm  2.5 $ & $12.44  \pm 0.11  $ \\
    \ciiit\ $\lambda$977  & $   -20,   20 $ & $    -0.3  \pm  0.2 $ & $13.30  \pm 0.01  $ \\
      \civ\ $\lambda$1548 & $	-20,   20 $ & $    -0.4  \pm  2.4 $ & $12.21  \pm 0.12  $ \\ 
     \civt\ $\lambda$1550 & $	-20,   20 $ &		   \nodata  & $ <	12.36	$ \\ 
     \civt		  & $   -20,   20 $ & $    -0.4  \pm  2.4 $ & $12.21  \pm 0.12  $ \\
      \oit\ $\lambda$1302 & $   -20,   20 $ &              \nodata  & $ <       12.58   $ \\
    \siiit\ $\lambda$1260 & $   -20,   20 $ & $    +0.7  \pm  2.7 $ & $11.28  \pm 0.13  $ \\
    \siivt\ $\lambda$1393 & $	-20,   20 $ & $    -1.6  \pm  2.4 $ & $12.04  \pm 0.10  $ \\
    \siivt\ 		  & $	-20,   20 $ & $    -2.6  \pm  1.7 $ & $12.00  \pm 0.07  $ \\
\hline
\multicolumn{4}{c}{J030341--002321 -- $z_{\rm abs}=2.99496$ -- $\mlnhi = 16.17$} \\
\hline
     \civt\ $\lambda$1548 & $   -40,   35 $ & $   -10.3  \pm  0.6 $ & $13.44  \pm 0.01  $ \\
     \civt\ $\lambda$1550 & $   -40,   35 $ & $   -10.4  \pm  2.0 $ & $13.53  \pm 0.05  $ \\
     \civt		  & $   -40,   35 $ & $   -10.4  \pm  0.6 $ & $13.44  \pm 0.01  $ \\
   \siiiit\ $\lambda$1206 & $   -40,   35 $ & $    -3.1  \pm  0.8 $ & $12.58  \pm 0.02  $ \\
    \siivt\ $\lambda$1393 & $   -40,   35 $ & $    -1.6  \pm  1.9 $ & $12.44  \pm 0.04  $ \\
    \siivt\ $\lambda$1402 & $   -40,   35 $ & $   -14.6  \pm  5.7 $ & $12.45  \pm 0.12  $ \\
    \siivt		  & $   -40,   35 $ & $    -1.6  \pm  1.9 $ & $12.44  \pm 0.04  $ \\
\hline
\multicolumn{4}{c}{J014516--094517A -- $z =2.66516$ -- $\mlnhi = 16.17$} \\
\hline
     \ciit\ $\lambda$1334 & $   -25,   20 $ & $    +0.4  \pm  2.9 $ & $12.21  \pm 0.12  $ \\
     \civt\ $\lambda$1548 & $   -25,   20 $ & $    -4.8  \pm  0.3 $ & $13.03  \pm 0.01  $ \\
     \civt\ $\lambda$1550 & $   -25,   20 $ & $    -4.7  \pm  0.5 $ & $13.07  \pm 0.02  $ \\
     \civt	          & $   -25,   20 $ & $    -4.8  \pm  0.3 $ & $13.05  \pm 0.01  $ \\
    \siiit\ $\lambda$1260 & $   -25,   20 $ &              \nodata  & $ <       10.96   $ \\
   \siiiit\ $\lambda$1206 & $   -25,   20 $ & $    -0.6  \pm  0.5 $ & $11.94  \pm 0.02  $ \\
    \siivt\ $\lambda$1393 & $   -25,   20 $ & $    -2.5  \pm  1.7 $ & $11.88  \pm 0.06  $ \\
    \siivt\ $\lambda$1402 & $   -25,   20 $ & $    -2.3  \pm  3.2 $ & $11.93  \pm 0.12  $ \\
    \siivt		  & $   -25,   20 $ & $    -2.4  \pm  1.8 $ & $11.89  \pm 0.05  $ \\
\hline
\multicolumn{4}{c}{J172409+531405 -- $z_{\rm abs}=2.48778$ -- $\mlnhi = 16.20$} \\
\hline
     \ciit\ $\lambda$1334 & $   -28,   28 $ & $    +1.9  \pm  0.2 $ & $\le 14.24  \pm 0.01  $ \\
     \civt\ $\lambda$1548 & $   -28,   28 $ & $    -0.0  \pm  0.4 $ & $13.43  \pm 0.01  $ \\
     \civt\ $\lambda$1550 & $   -28,   28 $ & $    -1.2  \pm  0.4 $ & $13.39  \pm 0.02  $ \\
     \civt		  & $   -28,   28 $ & $    -2.4  \pm  0.6 $ & $13.42  \pm 0.01  $ \\
       \nv\ $\lambda$1242 & $   -28,   28 $ & $    +2.3  \pm  1.7 $ & $13.03  \pm 0.05  $ \\
      \oit\ $\lambda$1302 & $   -28,   28 $ & $    +5.7  \pm  2.2 $ & $13.18  \pm 0.08  $ \\
    \aliit\ $\lambda$1670 & $   -28,   28 $ & $    +0.7  \pm  1.2 $ & $11.92  \pm 0.03  $ \\
    \siiit\ $\lambda$1193 & $   -28,   28 $ & $    -0.7  \pm  0.5 $ & $12.97  \pm 0.02  $ \\
    \siiit\ $\lambda$1304 & $   -28,   28 $ & $    -0.5  \pm  2.4 $ & $12.99  \pm 0.07  $ \\
    \siiit\ $\lambda$1526 & $   -28,   28 $ & $    -1.5  \pm  1.0 $ & $13.02  \pm 0.03  $ \\
    \siiit		  & $   -28,   28 $ & $    -0.9  \pm  0.8 $ & $12.99  \pm 0.02  $ \\
    \siivt\ $\lambda$1402 & $   -28,   28 $ & $    -0.3  \pm  2.0 $ & $12.50  \pm 0.06  $ \\
    \feiit\ $\lambda$1608 & $   -28,   28 $ &              \nodata  & $ <       12.47   $ \\
\hline
\multicolumn{4}{c}{J170100+641209 -- $z_{\rm abs}=2.43307$ -- $\mlnhi = 16.24$} \\
\hline
     \ciit\ $\lambda$1334 & $   -20,   15 $ & $    -1.7  \pm  0.4 $ & $12.83  \pm 0.02  $ \\
     \civt\ $\lambda$1548 & $   -25,   15 $ & $    -3.2  \pm  0.2 $ & $13.29  \pm 0.01  $ \\
    \aliit\ $\lambda$1670 & $   -10,   10 $ & $    -0.4  \pm  0.8 $ & $11.04  \pm 0.07  $ \\
    \siiit\ $\lambda$1193 & $   -10,   10 $ &              \nodata  & $< 11.35	        $ \\
   \siiiit\ $\lambda$1206 & $   -23,   14 $ & $    -0.3  \pm  0.1 $ & $12.97  \pm 0.01  $ \\
    \siivt\ $\lambda$1393 & $   -25,   15 $ & $    -2.4  \pm  0.2 $ & $12.68  \pm 0.01  $ \\
\hline
\multicolumn{4}{c}{J134328+572147 -- $z_{\rm abs}=2.87056$ -- $\mlnhi = 16.30$} \\
\hline
     \civt\ $\lambda$1548 & $   -50,   30 $ & $   -22.2  \pm  6.2 $ & $>13.88  	        $ \\
     \civt\ $\lambda$1550 & $   -50,   30 $ & $   -17.6  \pm  2.0 $ & $13.86  \pm 0.05  $ \\
     \civt	   	  & $   -50,   30 $ & $   -17.6  \pm  2.0 $ & $13.86  \pm 0.05  $ \\
    \siiit\ $\lambda$1526 & $   -50,   30 $ &              \nodata  & $ <       12.73   $ \\
   \siiiit\ $\lambda$1206 & $   -50,   30 $ & $    -5.6  \pm  1.3 $ & $12.71  \pm 0.03  $ \\
    \siivt\ $\lambda$1393 & $   -50,   30 $ & $   -17.9  \pm  1.8 $ & $13.00  \pm 0.04  $ \\
    \siivt\ $\lambda$1402 & $   -50,   30 $ & $   -20.0  \pm  4.4 $ & $13.06  \pm 0.08  $ \\
    \siivt		  & $   -50,   30 $ & $   -19.0  \pm  2.4 $ & $13.01  \pm 0.03  $ \\
\hline
\multicolumn{4}{c}{J012156+144823 -- $z_{\rm abs}=2.66586$ -- $\mlnhi = 16.32$} \\
\hline
     \ciit\ $\lambda$1036 & $   -25,   25 $ & $    +2.4  \pm  1.2 $ & $13.25  \pm 0.04  $ \\
     \ciit\ $\lambda$1334 & $   -25,   25 $ & $    -0.7  \pm  0.3 $ & $13.28  \pm 0.01  $ \\
     \ciit		  & $   -25,   25 $ & $    +0.8  \pm  0.6 $ & $13.27  \pm 0.01    $ \\
    \ciiit\ $\lambda$977  & $   -25,   25 $ & $    +1.2  \pm  1.5 $ & $> 13.87 		$ \\
     \civt\ $\lambda$1548 & $   -25,   25 $ & $    -2.0  \pm  0.1 $ & $13.59  \pm 0.01  $ \\
     \civt\ $\lambda$1550 & $   -25,   25 $ & $    -2.0  \pm  0.3 $ & $13.60  \pm 0.01  $ \\
     \civt		  & $   -25,   25 $ & $    -2.0  \pm  0.1 $ & $13.60  \pm 0.01  $ \\
      \oit\ $\lambda$1302 & $   -25,   25 $ &              \nodata  & $ <       12.23   $ \\
    \aliit\ $\lambda$1670 & $   -25,   25 $ & $    -0.3  \pm  0.9 $ & $11.56  \pm 0.03  $ \\
    \siiit\ $\lambda$1260 & $   -25,   25 $ & $    -2.6  \pm  0.4 $ & $12.28  \pm 0.01  $ \\
    \siiit\ $\lambda$1526 & $   -25,   25 $ & $    +3.4  \pm  2.5 $ & $12.19  \pm 0.14  $ \\
    \siiit		  & $   -25,   25 $ & $    +0.4  \pm  1.3 $ & $12.28  \pm 0.02  $ \\
   \siiiit\ $\lambda$1206 & $   -25,   25 $ & $    +0.8  \pm  0.3 $ & $>13.22 		$ \\
    \siivt\ $\lambda$1393 & $   -25,   25 $ & $    -1.0  \pm  0.1 $ & $13.20  \pm 0.01  $ \\
    \siivt\ $\lambda$1402 & $   -25,   25 $ & $    -0.8  \pm  0.3 $ & $13.22  \pm 0.01  $ \\
    \siivt		  & $   -25,   25 $ & $    -0.8  \pm  0.1 $ & $13.21  \pm 0.01  $ \\
    \feiit\ $\lambda$1608 & $   -25,   25 $ &              \nodata  & $ <       12.31   $ \\
\hline
\multicolumn{4}{c}{J134544+262506 -- $z_{\rm abs}=2.86367$ -- $\mlnhi = 16.36$} \\
\hline
     \ciit\ $\lambda$1036 & $   -20,   20 $ & $    +4.3  \pm  3.8 $ & $12.63  \pm 0.15  $ \\
     \ciit\ $\lambda$1334 & $   -20,   20 $ & $    +3.0  \pm  2.3 $ & $12.65  \pm 0.10  $ \\
     \ciit		& $   -20,   20 $ & $    +3.7  \pm  2.2 $ & $12.65  \pm 0.08  $ \\
     \civt\ $\lambda$1548 & $   -20,   20 $ & $    -0.8  \pm  0.7 $ & $13.09  \pm 0.03  $ \\
     \civt\ $\lambda$1550 & $   -20,   20 $ & $    +2.7  \pm  1.9 $ & $12.99  \pm 0.08  $ \\
     \civt		& $   -20,   20 $ & $    -0.8  \pm  0.7 $ & $13.09  \pm 0.03  $ \\
    \siiit\ $\lambda$1260 & $   -20,   20 $ & $    -1.6  \pm  3.3 $ & $11.39  \pm 0.14  $ \\
   \siiiit\ $\lambda$1206 & $   -20,   25 $ & $    +3.0  \pm  0.2 $ & $13.03 :	      $ \\
    \siivt\ $\lambda$1393 & $   -20,   20 $ & $    -0.0  \pm  0.6 $ & $12.72  \pm 0.02  $ \\
    \siivt\ $\lambda$1402 & $   -20,   20 $ & $    +3.8  \pm  1.6 $ & $12.67  \pm 0.07  $ \\
    \siivt		& $   -20,   20 $ & $    -0.0  \pm  0.6 $ & $12.72  \pm 0.02  $ \\
\hline
\multicolumn{4}{c}{J170100+641209 -- $z_{\rm abs}=2.43359$ -- $\mlnhi = 16.38$} \\
\hline
     \ciit\ $\lambda$1334 & $   -30,   30 $ & $    -5.6  \pm  0.5 $ & $12.81  \pm 0.01  $ \\
     \civt\ $\lambda$1548 & $   -30,   30 $ & $    -8.4  \pm  0.2 $ & $13.12  \pm 0.01  $ \\
    \aliit\ $\lambda$1670 & $   -30,   30 $ & $    -8.9  \pm  2.6 $ & $11.06  \pm 0.06  $ \\
    \siiit\ $\lambda$1260 & $   -30,   30 $ & $    -9.7  \pm  1.2 $ & $11.86  \pm 0.03  $ \\
   \siiiit\ $\lambda$1206 & $   -30,   30 $ & $    -2.9  \pm  0.1 $ & $13.14  \pm 0.01 $ \\
    \siivt\ $\lambda$1393 & $   -30,   30 $ & $    -4.8  \pm  0.2 $ & $12.68  \pm 0.01  $ \\
    \siivt\ $\lambda$1402 & $   -30,   30 $ & $    -3.8  \pm  0.3 $ & $12.70  \pm 0.01  $ \\
    \siivt		  & $   -30,   30 $ & $    -4.3  \pm  0.2 $ & $12.69  \pm 0.01  $ \\
\hline
\multicolumn{4}{c}{J135038--251216 -- $z_{\rm abs}=2.57299$ -- $\mlnhi = 16.39$} \\
\hline
     \ciit\ $\lambda$1334 & $   -30,   25 $ &              \nodata  & $ <       12.04   $ \\
    \ciiit\ $\lambda$977  & $   -40,   40 $ & $    -7.4  \pm  0.7 $ & $13.14  \pm 0.02  $ \\
     \civt\ $\lambda$1548 & $   -40,   25 $ & $   -10.5  \pm  1.8 $ & $12.55  \pm 0.04  $ \\
     \civt\ $\lambda$1550 & $   -40,   25 $ & $   -17.8  \pm  5.7 $ & $12.50  \pm 0.15  $ \\
     \civt\		  & $   -40,   25 $ & $   -17.8  \pm  5.7 $ & $12.55  \pm 0.04  $ \\
    \siiit\ $\lambda$1260 & $   -30,   25 $ &              \nodata  & $ <       11.14   $ \\
   \siiiit\ $\lambda$1206 & $   -40,   40 $ & $    +0.3  \pm  0.9 $ & $12.29  \pm 0.02  $ \\
    \siivt\ $\lambda$1393 & $   -30,   25 $ & $   -11.1  \pm  3.4 $ & $11.88  \pm 0.10  $ \\
    \siivt\ $\lambda$1402 & $   -30,   25 $ &              \nodata  & $ <       11.95   $ \\
\hline
\multicolumn{4}{c}{J130411+295348 -- $z_{\rm abs}=2.82922$ -- $\mlnhi = 16.39$} \\
\hline
     \ciit\ $\lambda$1334 & $   -20,   20 $ &              \nodata  & $ <       12.30   $ \\
      \oit\ $\lambda$1302 & $   -20,   20 $ &              \nodata  & $ <       12.76   $ \\
    \siiit\ $\lambda$1260 & $   -20,   20 $ &              \nodata  & $ <       11.21   $ \\
    \siivt\ $\lambda$1393 & $   -20,   20 $ &              \nodata  & $ <       11.66   $ \\
\hline
\multicolumn{4}{c}{J134544+262506 -- $z_{\rm abs}=2.87630$ -- $\mlnhi = 16.50$} \\
\hline
     \ciit\ $\lambda$1334 & $   -30,   30 $ &              \nodata  & $ <       12.41   $ \\
     \civt\ $\lambda$1548 & $   -30,   30 $ & $    -1.0  \pm  1.2 $ & $13.04  \pm 0.04  $ \\
     \civt\ $\lambda$1550 & $   -30,   30 $ & $    -4.6  \pm  2.3 $ & $13.08  \pm 0.06  $ \\
     \civt 		& $   -30,   30 $ & $    -2.8  \pm  1.3 $ & $13.05  \pm 0.03  $ \\
    \siiit\ $\lambda$1260 & $   -30,   30 $ &              \nodata  & $ <       11.28   $ \\
    \siivt\ $\lambda$1393 & $   -30,   30 $ & $    +1.5  \pm  1.9 $ & $12.44  \pm 0.05  $ \\
\hline
\multicolumn{4}{c}{J212912--153841 -- $z_{\rm abs}=2.90711$ -- $\mlnhi = 16.55$} \\
\hline
     \ciit\ $\lambda$1334 & $   -50,   60 $ & $    -1.0  \pm  2.0 $ & $12.82  \pm 0.03  $ \\
     \civt\ $\lambda$1548 & $   -60,   60 $ & $    +8.6  \pm  0.4 $ & $13.59  \pm 0.01  $ \\
    \siiit\ $\lambda$1260 & $   -50,   50 $ & $    +0.7  \pm  7.0 $ & $12.01  \pm 0.14  $ \\
   \siiiit\ $\lambda$1206 & $   -50,   30 $ & $    -1.2  \pm  0.2 $ & $\ge 13.32  \pm 0.01  $ \\
    \siivt\ $\lambda$1393 & $   -60,   50 $ & $    -1.6  \pm  0.8 $ & $12.97  \pm 0.01  $ \\
    \siivt\ $\lambda$1402 & $   -60,   50 $ & $    +2.2  \pm  1.1 $ & $12.96  \pm 0.01  $ \\
    \siivt		& $   -60,   50 $ & $     +0.3  \pm  0.7 $ & $12.97  \pm 0.01  $ \\
\hline
\multicolumn{4}{c}{J101447+430030 -- $z_{\rm abs}=3.01439$ -- $\mlnhi = 16.63$} \\
\hline
     \ciit\ $\lambda$1334 & $   -20,   20 $ &              \nodata  & $ <       11.94   $ \\
     \civt\ $\lambda$1548 & $   -45,   35 $ & $    -6.5  \pm  0.4 $ & $13.41  \pm 0.01  $ \\
     \civt\ $\lambda$1550 & $   -45,   35 $ & $    -2.9  \pm  0.8 $ & $13.45  \pm 0.02  $ \\
     \civt	  	  & $   -45,   35 $ & $    -4.7  \pm  0.4 $ & $13.42  \pm 0.01  $ \\
     \ovit\ $\lambda$1031 & $   -45,   35 $ & $    -1.7  \pm  0.1 $ & $14.39  \pm 0.01  $ \\
     \ovit\ $\lambda$1037 & $   -45,   35 $ & $    -1.3  \pm  0.2 $ & $14.38  \pm 0.01  $ \\
     \ovit		  & $   -45,   35 $ & $    -1.5  \pm  0.1 $ & $14.39  \pm 0.01  $ \\
    \siiit\ $\lambda$1260 & $   -20,   20 $ &              \nodata  & $ <       10.82   $ \\
    \siivt\ $\lambda$1393 & $   -20,   20 $ &              \nodata  & $ <       11.37   $ \\
\hline
\multicolumn{4}{c}{J131215+423900 -- $z_{\rm abs}=2.48998$ -- $\mlnhi = 16.77$} \\
\hline
     \ciit\ $\lambda$1334 & $   -10,   10 $ & $    +5.3  \pm  2.1 $ & $12.17  \pm 0.15  $ \\
    \ciiit\ $\lambda$977  & $   -30,   30 $ & $    +1.1  \pm  0.6 $ & $>13.61	      $ \\
     \civt\ $\lambda$1548 & $   -15,   15 $ & $    -0.1  \pm  0.8 $ & $13.23  \pm 0.05  $ \\
     \civt\ $\lambda$1550 & $   -15,   15 $ & $    +0.6  \pm  0.9 $ & $13.25  \pm 0.05  $ \\
     \civt		& $   -15,   15 $ & $     +0.3  \pm  0.6 $ & $13.24  \pm 0.04  $ \\
    \aliit\ $\lambda$1670 & $   -10,   10 $ &              \nodata  & $ <       10.77   $ \\
    \siiit\ $\lambda$1260 & $   -10,   10 $ & $    +2.1  \pm  1.6 $ & $11.15  \pm 0.16  $ \\
   \siiiit\ $\lambda$1206 & $   -20,   25 $ & $    +4.6  \pm  0.2 $ & $12.70  \pm 0.01  $ \\
    \siivt\ $\lambda$1393 & $   -20,   30 $ & $    +1.7  \pm  0.8 $ & $12.55  \pm 0.02  $ \\
    \siivt\ $\lambda$1402 & $   -25,   25 $ & $    +0.3  \pm  1.2 $ & $12.59  \pm 0.04  $ \\
    \siivt		& $   -25,   25 $ & $     +1.0  \pm  0.7 $ & $12.57  \pm 0.02  $ \\
\hline
\multicolumn{4}{c}{J144453+291905 -- $z_{\rm abs}=2.46714$ -- $\mlnhi = 16.78$} \\
\hline
     \ciit\ $\lambda$1334 & $   -30,   40 $ & $   +14.8  \pm  1.1 $ & $12.79  \pm 0.02  $ \\
     \civt\ $\lambda$1548 & $   -30,   40 $ & $   +18.1  \pm  0.3 $ & $13.35  \pm 0.01  $ \\
     \civt\ $\lambda$1550 & $   -30,   40 $ & $   +14.9  \pm  0.4 $ & $13.37  \pm 0.01  $ \\
     \civt		& $   -30,   40 $ & $    +16.5  \pm  0.2 $ & $13.36  \pm 0.01  $ \\
    \aliit\ $\lambda$1670 & $   -30,   40 $ &              \nodata  & $ <       10.81   $ \\
    \siiit\ $\lambda$1260 & $   -30,   40 $ & $    +11.4  \pm  3.1 $ & $11.45  \pm 0.10  $ \\
    \siivt\ $\lambda$1393 & $   -30,   40 $ & $    +15.9  \pm  0.4 $ & $12.77  \pm 0.01  $ \\
    \siivt\ $\lambda$1402 & $   -30,   40 $ & $    +16.5  \pm  0.8 $ & $12.72  \pm 0.02  $ \\
    \siivt		& $   -30,   40 $ & $    +16.2  \pm  0.4 $ & $12.75  \pm 0.02  $ \\
\hline
\multicolumn{4}{c}{J020950--000506  -- $z_{\rm abs}=2.57452$ -- $\mlnhi = 16.78$} \\
\hline
     \ciit\ $\lambda$1334 & $   -30,   30 $ & $    +2.2  \pm  0.3 $ & $13.16  \pm 0.01  $ \\
     \civt\ $\lambda$1548 & $   -30,   40 $ & $    -2.7  \pm  0.1 $ & $13.94  \pm 0.01  $ \\
     \civt\ $\lambda$1550 & $   -30,   40 $ & $    -2.9  \pm  0.2 $ & $13.93  \pm 0.01  $ \\
     \civt\tablenotemark{a}		& 		  & $    -1.8  \pm  0.3 $ & $13.90  \pm 0.01  $ \\
     \ovit\ $\lambda$1031 & $   -40,   40 $ & $    +0.7  \pm  0.2 $ & $13.95  \pm 0.01  $ \\
     \ovit\ $\lambda$1037 & $   -40,   40 $ & $    -4.1  \pm  0.4 $ & $13.98  \pm 0.01  $ \\
     \ovit		& $   -40,   40 $ & $    -4.1  \pm  0.4 $ & $13.98  \pm 0.01  $ \\
    \siiit\ $\lambda$1260 & $   -30,   30 $ & $    -0.1  \pm  1.3 $ & $11.83  \pm 0.04  $ \\
    \siivt\ $\lambda$1393 & $   -40,   40 $ & $    +1.1  \pm  0.2 $ & $13.23  \pm 0.01  $ \\
    \siivt\ $\lambda$1402 & $   -40,   40 $ & $    +0.7  \pm  0.3 $ & $13.24  \pm 0.01  $ \\
    \siivt		& $   -40,   40 $ & $     +0.9  \pm  0.2 $ & $13.23  \pm 0.01  $ \\
\hline
\multicolumn{4}{c}{J101723--204658 -- $z_{\rm abs}=2.45053$ -- $\mlnhi = 17.23$} \\
\hline
     \ciit\ $\lambda$1334 & $   -25,   25 $ & $    -3.3  \pm  0.7 $ & $12.98  \pm 0.02  $ \\
    \ciiit\ $\lambda$977  & $   -25,   25 $ & $    +1.2  \pm  2.1 $ & $>13.91	 	$ \\
     \civt\ $\lambda$1548 & $   -25,   25 $ & $    -2.0  \pm  0.4 $ & $13.16  \pm 0.01  $ \\
     \civt\ $\lambda$1550 & $   -25,   25 $ & $    -0.6  \pm  0.8 $ & $13.12  \pm 0.02  $ \\
     \civt		& $   -25,   25 $ & $    -1.3  \pm  0.4 $ & $13.15  \pm 0.01  $ \\
    \aliit\ $\lambda$1670 & $   -25,   25 $ & $    -0.8  \pm  2.7 $ & $11.17  \pm 0.09  $ \\
    \siiit\ $\lambda$1260 & $   -25,   25 $ & $    -5.6  \pm  1.2 $ & $11.71  \pm 0.03  $ \\
   \siiiit\ $\lambda$1206 & $   -25,   25 $ & $    +1.6  \pm  0.6 $ & $>13.21 	  	$ \\
    \siivt\ $\lambda$1393 & $   -25,   25 $ & $    +0.1  \pm  0.3 $ & $12.88  \pm 0.01  $ \\
    \siivt\ $\lambda$1402 & $   -25,   25 $ & $    -0.6  \pm  0.6 $ & $12.91  \pm 0.02  $ \\
    \siivt		& $   -25,   25 $ & $    -0.3  \pm  0.3 $ & $12.89  \pm 0.01  $ \\
\hline
\multicolumn{4}{c}{J025905+001121 -- $z_{\rm abs}=3.08465$ -- $\mlnhi = 17.25$} \\
\hline
    \ciiit\ $\lambda$977 & $   -60,   30 $ & $   -16.3  \pm  2.3 $ & $(>)14.11	      $ \\
     \civt\ $\lambda$1548 & $   -60,   30 $ & $   -15.6  \pm  0.4 $ & $13.66  \pm 0.01  $ \\
     \civt\ $\lambda$1550 & $   -60,   30 $ & $   -18.3  \pm  0.7 $ & $13.65  \pm 0.01  $ \\
     \civt		& $   -60,   30 $ & $   -16.9  \pm  0.4 $ & $13.66  \pm 0.01  $ \\
    \siiit\ $\lambda$1304 & $   -60,   30 $ &              \nodata  & $ <       11.90   $ \\
   \siiiit\ $\lambda$1206 & $   -60,   30 $ & $   -12.9  \pm  1.5 $ & $\le 13.48  \pm 0.03  $ \\
    \siivt\ $\lambda$1393 & $   -60,   30 $ & $   -19.9  \pm  0.5 $ & $13.11  \pm 0.01  $ \\
    \siivt\ $\lambda$1402 & $   -60,   30 $ & $   -18.9  \pm  0.6 $ & $13.12  \pm 0.01  $ \\
    \siivt		& $   -60,   30 $ & $   -19.4  \pm  0.4 $ & $13.12  \pm 0.01  $ \\
\hline
\multicolumn{4}{c}{J132552+663405 -- $z_{\rm abs}=2.38287$ -- $\mlnhi = 17.30$} \\
\hline
     \ciit\ $\lambda$1036 & $   -25,   20 $ &              \nodata  & $ <       12.77   $ \\
    \ciiit\ $\lambda$977  & $   -25,   20 $ & $    -4.0  \pm 17.6 $ & $>13.69   $ \\
     \civt\ $\lambda$1548 & $   -20,   20 $ & $    -3.7  \pm  0.7 $ & $13.29  \pm 0.03  $ \\
     \civt\ $\lambda$1550 & $   -20,   20 $ & $    -4.6  \pm  1.1 $ & $13.32  \pm 0.04  $ \\
     \civt		& $   -20,   20 $ & $    -4.2  \pm  0.6 $ & $13.30  \pm 0.02  $ \\
    \aliit\ $\lambda$1670 & $   -25,   20 $ &              \nodata  & $ <       11.29   $ \\
    \siiit\ $\lambda$1193 & $   -25,   20 $ &              \nodata  & $ <       11.88   $ \\
   \siiiit\ $\lambda$1206 & $   -25,   20 $ & $    -3.2  \pm  0.3 $ & $12.95  \pm 0.02  $ \\
    \siivt\ $\lambda$1393 & $   -25,   20 $ & $    -4.3  \pm  0.8 $ & $12.82  \pm 0.03  $ \\
\hline
\multicolumn{4}{c}{J212912--153841 -- $z_{\rm abs}=2.96755$ -- $\mlnhi = 17.32$} \\
\hline
     \ciit\ $\lambda$1334 & $   -40,   40 $ & $    +1.3  \pm  1.2 $ & $12.98  \pm 0.03  $ \\
     \civt\ $\lambda$1548 & $   -70,   65 $ & $    -1.7  \pm  0.9 $ & $13.34  \pm 0.01  $ \\
     \civt\ $\lambda$1550 & $   -70,   65 $ & $    -2.4  \pm  1.4 $ & $13.37  \pm 0.02  $ \\
     \civt		& $   -70,   65 $ & $    -2.0  \pm  0.8 $ & $13.35  \pm 0.01  $ \\
    \aliit\ $\lambda$1670 & $   -40,   45 $ &              \nodata  & $ <       11.31   $ \\
    \siiit\ $\lambda$1526 & $   -45,   45 $ &              \nodata  & $ <       12.01   $ \\
   \siiiit\ $\lambda$1206 & $   -40,   40 $ & $    +0.7  \pm  0.3 $ & $\le 12.92  \pm 0.01  $ \\
    \siivt\ $\lambda$1393 & $   -45,   45 $ & $    +2.3  \pm  1.5 $ & $12.58  \pm 0.03  $ \\
    \siivt\ $\lambda$1402 & $   -45,   45 $ & $    +2.8  \pm  1.8 $ & $12.58  \pm 0.03  $ \\
    \siivt		& $   -45,   45 $ & $     +2.5  \pm  1.2 $ & $12.58  \pm 0.01  $ \\
\hline
\multicolumn{4}{c}{J095852+120245 -- $z_{\rm abs}=3.22319$ -- $\mlnhi = 17.36$} \\
\hline
     \ciit\ $\lambda$1334 & $   -15,   15 $ &              \nodata  & $ <       12.07   $ \\
     \civt\ $\lambda$1548 & $   -15,   30 $ & $     5.2  \pm  0.3 $ & $13.54  \pm 0.01  $ \\
     \civt\ $\lambda$1550 & $   -15,   30 $ & $     5.7  \pm  0.8 $ & $13.62  \pm 0.07  $ \\
     \civt		& $   -15,   30 $ & $     5.5  \pm  0.4 $ & $13.55  \pm 0.01  $ \\
    \aliit\ $\lambda$1670 & $   -15,   15 $ &              \nodata  & $ <       11.11   $ \\
    \siiit\ $\lambda$1260 & $   -15,   15 $ & $     1.9  \pm  1.6 $ & $\le 11.26  \pm 0.10  $ \\
   \siiiit\ $\lambda$1206 & $   -15,   30 $ & $     4.7  \pm  0.2 $ & $12.85  \pm 0.01  $ \\
    \siivt\ $\lambda$1393 & $   -15,   30 $ & $     2.7  \pm  0.4 $ & $12.87  \pm 0.02  $ \\
    \siivt\ $\lambda$1402 & $   -15,   30 $ & $     4.1  \pm  0.6 $ & $12.86  \pm 0.02  $ \\
    \siivt		& $   -15,   30 $ & $     3.4  \pm  0.4 $ & $12.87  \pm 0.01  $ \\
\hline
\multicolumn{4}{c}{J025905+001121 -- $z_{\rm abs}=3.08204$ -- $\mlnhi = 17.50$} \\
\hline
     \ciit\ $\lambda$1334 & $   -20,   20 $ &              \nodata  & $ <       11.89   $ \\
     \civt\ $\lambda$1548 & $   -20,   20 $ &              \nodata  & $ <       11.98   $ \\
    \siiit\ $\lambda$1304 & $   -20,   20 $ &              \nodata  & $ <       11.68   $ \\
   \siiiit\ $\lambda$1206 & $   -20,   20 $ & $    -0.6  \pm  1.2 $ & $\le 11.80  \pm 0.05  $ \\
    \siivt\ $\lambda$1393 & $   -20,   20 $ &              \nodata  & $ <       11.48   $ \\
\hline
\multicolumn{4}{c}{J162557+264448  -- $z_{\rm abs}=2.55105$ -- $\mlnhi = 17.75$} \\
\hline
    \ciiit\ $\lambda$977  & $   -50,   40 $ & $    -5.7  : 	$ & $>14.14  	      $ \\
     \civt\ $\lambda$1548 & $   -50,   40 $ & $    -9.8  \pm  0.4 $ & $13.66  \pm 0.01  $ \\
     \civt\ $\lambda$1550 & $   -50,   40 $ & $   -11.4  \pm  0.9 $ & $13.63  \pm 0.02  $ \\
     \civt		& $   -50,   40 $ & $   -10.6  \pm  0.5 $ & $13.65  \pm 0.01  $ \\
    \aliit\ $\lambda$1670 & $   -50,   40 $ & $    -6.0  \pm  2.4 $ & $11.96  \pm 0.04  $ \\
    \siiit\ $\lambda$1193 & $   -50,   40 $ & $    +1.0  \pm  5.1 $ & $12.50  \pm 0.10  $ \\
    \siiit\ $\lambda$1260 & $   -50,   40 $ & $    +2.5  \pm  1.2 $ & $12.58  \pm 0.02  $ \\
    \siiit		& $   -50,   40 $ & $     +1.8  \pm  2.5 $ & $12.54  \pm 0.02  $ \\
   \siiiit\ $\lambda$1206 & $   -50,   40 $ & $    -3.3  \pm  0.9 $ & $>13.52  	      $ \\
    \siivt\ $\lambda$1393 & $   -50,   40 $ & $    -4.0  \pm  0.6 $ & $13.44  \pm 0.02  $ \\
    \siivt\ $\lambda$1402 & $   -50,   40 $ & $    -3.3  \pm  0.8 $ & $13.42  \pm 0.02  $ \\
    \siivt		& $   -50,   40 $ & $    -3.6  \pm  0.5 $ & $13.43  \pm 0.01  $ \\
\hline
\multicolumn{4}{c}{J064204+675835 -- $z_{\rm abs}=2.90469$ -- $\mlnhi = 18.42$} \\
\hline
     \ciit\ $\lambda$1334 & $   -60,   30 $ & $   -19.5  \pm 19.9 $ & $>14.62  	      $ \\
     \civt\ $\lambda$1548 & $   -60,   30 $ & $   -14.3  \pm  0.2 $ & $14.29  \pm 0.01  $ \\
     \civt\ $\lambda$1550 & $   -60,   30 $ & $   -11.6  \pm  0.2 $ & $14.31  \pm 0.01  $ \\
     \civt		& $   -60,   30 $ & $   -13.0  \pm  0.2 $ & $14.31  \pm 0.01  $ \\
      \oit\ $\lambda$1039 & $   -60,   30 $ & $   -11.8  \pm  4.3 $ & $14.08  \pm 0.08  $ \\
      \oit\ $\lambda$1302 & $   -60,   30 $ & $   -21.2  \pm  0.3 $ & $13.99  \pm 0.01  $ \\
      \oit		& $   -60,   30 $ & $   -16.5  \pm  2.1 $ & $14.04  \pm 0.03  $ \\
    \aliit\ $\lambda$1670 & $   -60,   30 $ & $   -19.7  \pm  0.3 $ & $12.90  \pm 0.01  $ \\
    \siiit\ $\lambda$1304 & $   -60,   30 $ & $   -24.8  \pm  0.2 $ & $14.08  \pm 0.01  $ \\
    \siiit\ $\lambda$1526 & $   -60,   30 $ & $   -22.8  \pm  0.2 $ & $14.00  \pm 0.01  $ \\
    \siiit		& $   -60,   30 $ & $   -23.8  \pm  0.1 $ & $14.04  \pm 0.04  $ \\
    \siivt\ $\lambda$1393 & $   -60,   30 $ & $   -13.7  \pm  1.7 $ & $>14.01    $ \\
    \siivt\ $\lambda$1402 & $   -60,   30 $ & $   -14.8  \pm  0.2 $ & $14.07  \pm 0.01  $ \\
    \siivt\tablenotemark{b}		& $   -60,   30 $ & $   -14.8  \pm  0.2 $ & $14.13  \pm 0.01  $ \\
    \feiit\ $\lambda$1608 & $   -60,   30 $ & $   -25.1  \pm  1.5 $ & $13.61  \pm 0.03  $ \\
   \feiiit\ $\lambda$1122 & $   -60,   30 $ & $   -20.6  \pm  0.7 $ & $14.02  \pm 0.01  $ \\
\hline
\multicolumn{4}{c}{J030341--002321 -- $z_{\rm abs}=2.94076$ -- $\mlnhi = 18.65$} \\
\hline
     \ciit\ $\lambda$1334 & $   -50,   50 $ & $    +2.3  \pm  0.2 $ & $14.13  \pm 0.01  $ \\
     \civt\ $\lambda$1548 & $   -60,   70 $ & $    -6.4  \pm  0.5 $ & $14.13  \pm 0.01  $ \\
     \civt\ $\lambda$1550 & $   -60,   70 $ & $    -6.8  \pm  0.7 $ & $14.16  \pm 0.01  $ \\
     \civt		& $   -60,   70 $ & $    -6.6  \pm  0.4 $ & $14.14  \pm 0.01  $ \\
      \oit\ $\lambda$1302 & $   -20,   20 $ & $    +2.4  \pm  0.8 $ & $13.15  \pm 0.03  $ \\
    \siiit\ $\lambda$1304 & $   -30,   30 $ & $    +3.7  \pm  1.0 $ & $13.10  \pm 0.03  $ \\
    \siiit\ $\lambda$1526 & $   -30,   30 $ & $    +4.9  \pm  1.2 $ & $13.12  \pm 0.03  $ \\
    \siiit		& $   -30,   30 $ & $     +4.3  \pm  0.8 $ & $13.11  \pm 0.02  $ \\
    \siivt\ $\lambda$1393 & $   -60,   70 $ & $    -6.4  \pm  3.9 $ & $>14.00	      $ \\
    \siivt\ $\lambda$1402 & $   -60,   70 $ & $    -8.3  \pm  0.6 $ & $14.03  \pm 0.01  $ \\
    \siivt\ $^b$		& $   -60,   70 $ & $    -8.3  \pm  0.6 $ & $14.06  \pm 0.02  $ \\
\enddata
\tablecomments{Upper limits (``$<$") are non-detections quoted at the 2$\sigma$ level. Column densities preceded by ``$>$" are lower limits owing to saturation in the absorption. Column densities preceded by ``$\le$" could be somewhat contaminated. For a given atom or ion with more than one transition, we list in the row with no wavelength information the adopted weighted average column densities and velocities.  }
\tablenotetext{a}{\civt\ was fitted with two components to extract reliably the column density of the component directly associated with the LLS.}
\tablenotetext{b}{Corrected from mild saturation (see \citealt{savage91}).}
\end{deluxetable*}

%\floattable
\begin{deluxetable*}{cccccc}
\tablecaption{Summary of the \hit\ parameters and metallicities of the pLLSs and LLSs \label{t-nhi}}
\tablehead{
\colhead{QSO} & \colhead{$z_{\rm abs}$} & \colhead{$\mlnhi$}& \colhead{$b_{\rm HI}$} & \colhead{${\rm [X/H]}$} & \colhead{Notes/} \\
\colhead{} & \colhead{} & \colhead{${\rm [cm^{-2}]}$}& \colhead{(\km)}& \colhead{} & \colhead{References}
}
%\colnumbers
\startdata
\multicolumn{6}{c}{New sample of pLLSs and LLSs} \\
\hline
 J143316$+$313126 & $  2.90116 $ & $  16.16 \pm 0.01	        $ & $    18.1 \pm  0.4  $ & $  -1.80 \pm 0.15  $ & \\
 J030341$-$002321 & $  2.99496 $ & $  16.17 \pm 0.01	        $ & $    37.2 \pm  0.2  $ & $  -1.90 \pm 0.10  $ & \\
 J014516$-$094517A& $  2.66516 $ & $  16.17 \pm 0.01	        $ & $    25.1 \pm  0.2  $ & $  -2.40 \pm 0.20  $ & \\
 J172409$+$531405 & $  2.48778 $ & $  16.20 \pm 0.03	        $ & $    16.1 \pm  0.4  $ & $  +0.20 \pm 0.10  $ & \\
 J170100$+$641209 & $  2.43307 $ & $  16.24 \pm 0.02	        $ & $    24.6 \pm  0.8  $ & $  -1.65 \pm 0.10  $ & \tablenotemark{a}\\
 J134328$+$572147 & $  2.87056 $ & $  16.30 \pm 0.01	        $ & $    35.7 \pm  0.9  $ & $  -1.45 \pm 0.10  $ & \\
 J012156$+$144823 & $  2.66586 $ & $  16.32 \pm 0.01	        $ & $    20.2 \pm  0.2  $ & $  -1.05 \pm 0.10  $ & \\
 J134544$+$262506 & $  2.86367 $ & $  16.36 \pm 0.01	        $ & $    20.1 \pm  0.5  $ & $  -1.65 \pm 0.20  $ & \\
 J170100$+$641209 & $  2.43359 $ & $  16.38 \pm 0.01	        $ & $    23.2 \pm  0.6  $ & $  -1.50 \pm 0.10  $ & \tablenotemark{a}\\
 J135038$-$251216 & $  2.57299 $ & $  16.39 \pm 0.01	        $ & $    36.8 \pm  0.7  $ & $  -2.30 \pm 0.10  $ & \\
 J130411$+$295348 & $  2.82922 $ & $  16.39 \pm 0.01	        $ & $    18.6 \pm  0.2  $ & $ <-1.90 	       $ & \\
 J134544$+$262506 & $  2.87630 $ & $  16.50 \pm 0.04	        $ & $    26.0 \pm  1.2  $ & $  -2.30 \pm 0.10  $ & \\
 J212912$-$153841 & $  2.90711 $ & $  16.55 \,^{+0.15}_{-0.25}  $ & $    27.0 :		$ & $  -1.55 \pm 0.10  $ & \\
 J101447$+$430030 & $  3.01439 $ & $  16.63 \pm 0.01	        $ & $    22.4 \pm  0.2  $ & $ <-2.60 	       $ & \\
 J131215$+$423900 & $  2.48998 $ & $  16.77 \pm 0.01 	        $ & $	 20.0 \pm  0.1  $ & $  -2.50 \pm 0.10  $ & \\
 J144453$+$291905 & $  2.46714 $ & $  16.78 \pm 0.02 	        $ & $	 32.3 \pm  0.5  $ & $  -2.35 \pm 0.15  $ & \tablenotemark{b}\\
 J020950$-$000506 & $  2.57452 $ & $  16.78 \pm 0.03 	        $ & $	 20.0 \pm  1.0  $ & $  -2.00 \pm 0.15  $ & \\
 J101723$-$204658 & $  2.45053 $ & $  17.23 \pm 0.01 	        $ & $	 22.6 \pm  0.1  $ & $  -2.50 \pm 0.15  $ & \\
 J025905$+$001121 & $  3.08465 $ & $  17.25 \pm 0.25 	        $ & $	 18.0 :		$ & $  -2.60 \pm 0.25  $ & \\
 J132552$+$663405 & $  2.38287 $ & $  17.30 \pm 0.30 	        $ & $	 29.0 :		$ & $  -3.00 \pm 0.10  $ & \\
 J212912$-$153841 & $  2.96755 $ & $  17.32 \pm 0.25 	        $ & $	 26.0 :		$ & $ <-2.70 	       $ & \\
 J095852$+$120245 & $  3.22319 $ & $  17.36 \pm 0.05 	        $ & $	 21.0 \pm  1.2  $ & $  -3.35 \pm 0.05  $ & \\
 J025905$+$001121 & $  3.08204 $ & $  17.50 \pm 0.25 	        $ & $	 11.0 :		$ & $ <-2.70 	       $ & \\
 J162557$+$264448 & $  2.55105 $ & $  17.75 \,^{+0.15}_{-0.20}  $ & $    23.0 :		$ & $  -2.25 \pm 0.15  $ & \\
 J064204$+$675835 & $  2.90469 $ & $  18.42 \,^{+0.15}_{-0.30}  $ & $    30.0 :		$ & $  -1.00 \pm 0.20  $ & \\
 J030341$-$002321 & $  2.94076 $ & $  18.65 \,^{+0.15}_{-0.30}  $ & $    23.0 :		$ & $  -2.10 \pm 0.20  $ & \\
\hline
\multicolumn{6}{c}{Sample of pLLSs and LLSs drawn from the literature} \\
\hline
 J144453$+$291905 & $ 2.43886  $ & $  16.43 \pm 0.30 		$ &  \nodata 		  & $  -0.40 \pm 0.30  $ & 1\\	  %Crighton+13
 J044828$-$415728 & $ 2.46416  $ & $  16.94 \pm 0.10 		$ &  \nodata 		  & $  -0.30 \pm 0.11  $ & 2\\	  %Crighton+15
 J101155$+$294141 & $ 2.42901  $ & $  17.75 \pm 0.15 		$ &  \nodata 		  & $  -2.10 \pm 0.20  $ & 3\\
 J134329$+$572148 & $ 2.83437  $ & $  17.78 \pm 0.20 		$ &  \nodata 		  & $  -0.60 \pm 0.20  $ & 3\\
 J143316$+$313126 & $ 2.58615  $ & $  18.15 \pm 0.15 		$ &  \nodata 		  & $ <-2.60             $ & 3 \\
 J121930$+$495054 & $ 2.18076  $ & $  18.60 \pm 0.15 		$ &  \nodata 		  & $ <-1.60 	         $ & 3\\
 J104019$+$572448 & $ 3.26620  $ & $  18.60 \pm 0.20 		$ &  \nodata 		  & $  -1.37\,^{+0.14}_{-0.21}$ & 3\\
\enddata
\tablecomments{The \hit\ absorption was fitted with a single component except otherwise stated. For the new sample: systems with  $b$-values followed by colons were fitted iteratively until a good fit was achieved; systems with errors on the $b$-values were fitted iteratively by hand and automatically by minimizing the reduced-$\chi^2$; both solutions were consistent and we adopted the minimized reduced-$\chi^2$ solution.\\
References: (1) \citealt{crighton13}; (2) \citealt{crighton15}; (3) \citealt{lehner14}. }
\tablenotetext{a}{These two absorbers were analyzed separately and are only separated by $50$ \km. Since they have similar metallicity and likely probing the same structure, we only keep one of these for the metallicity distribution analysis.}
\tablenotetext{b}{This pLLS is best fitted with two components. The total $b$ and \nhi\ are well constrained, but the column densities in each component are not robustly determined. Hence we treat this pLLS as a single absorber. }
\end{deluxetable*}

%\floattable
\begin{deluxetable*}{cccccc}
\tablecaption{Summary of the metallicities for the LYAF, pLLSs, LLSs, and DLAs at $2.3< z< 3.3$ \label{t-metavg}}
\tablehead{
\colhead{Absorbers} & \colhead{Mean\tablenotemark{a}} & \colhead{Median\tablenotemark{b}}& \colhead{SD\tablenotemark{b}} & \colhead{Fraction with} & \colhead{Data} \\
\colhead{} & \colhead{[X/H]} & \colhead{[X/H]}& \colhead{[X/H]}& \colhead{${\rm [X/H]}\le -2.4$\tablenotemark{c}} & \colhead{Source\tablenotemark{d}}
}
%\colnumbers
\startdata
LYAF		& $ -2.85	   $ & $  -2.82    $ & $  \pm 0.75  $ 		& \nodata		& 1 \\
 pLLS		& $ -1.67 \pm 0.18 $ & $  -1.70    $ & $  \pm 0.81  $ 		& 10--27\% (3/18)	& 2 \\
LLS		& $ -2.34 \pm 0.24 $ & $  -2.50    $ & $  \pm 0.80  $ 		& 40--67\% (7/13)	& 2 \\
pLLS+LLS	& $ -2.00 \pm 0.17 $ & $  -2.10    $ & $  \pm 0.84  $ 		& 25--41\% (10/31)	& 2 \\
LLS		& $ -2.08	   $ & $  -2.24    $ & $  +0.50,-0.74  $	& 31\%  (38)$^\dagger$	& 3 \\
SLLS		& $ -1.71	   $ & $  -1.92    $ & $  +0.76,-1.04  $	& 21\% (73)$^\dagger$		& 3 \\
 DLAs		& $ -1.38	   $ & $  -1.39    $ & $  \pm 0.52  $ 		& 1.3--5.0\% (2/80)	& 4 \\
\enddata
\tablenotetext{a}{Mean with error bars are estimated using the KM estimator to account for the upper limits in the sample.}
\tablenotetext{b}{The calculations of the median and standard deviation assume that limits are actual values.}
\tablenotetext{c}{Fraction of VMP absorbers with  ${\rm [X/H]}\le -2.4$ (68\% confidence interval). The numbers between parentheses are the number of absorbers with  ${\rm [X/H]}\le -2.4$ over the sample size, except for $^\dagger$ where it is the probability of finding absorbers lower than the threshold metallicity  (in that case, the number between parentheses is the size sample).}
\tablenotetext{d}{References: 1) \citealt{simcoe04}; this paper; 3) \citetalias{fumagalli16}; 4) \citealt{rafelski12}.}
\end{deluxetable*}

%\floattable
\begin{deluxetable*}{lccccc}
\tablecaption{Comparison between the high and low $z$ pLLS/LLS samples \label{t-cloudyavg}}
\tablehead{
\colhead{$z$} & \colhead{Mean} & \colhead{Median} & \colhead{St. Dev.} & \colhead{Min.} & \colhead{Max.}
}
%\colnumbers
\startdata
\multicolumn{6}{c}{$\mlnhi$ $[{\rm cm^{-2}}]$} \\
\hline
low		& $ 16.75	$ & $  16.48	$ & $  0.67  $	& $16.11$	& $18.40$ \\
 high		& $ 17.06	$ & $  16.78	$ & $  0.81  $	& $16.16$	& $18.65$ \\
\hline
\multicolumn{6}{c}{$\log U$} \\
\hline
low		& $ -3.2	$ & $  -3.1	$ & $  0.5  $	& $-4.0$	& $-2.0$ \\
 high		& $ -2.4	$ & $  -2.3	$ & $  0.7  $	& $-4.0$	& $-1.5$ \\
\hline
\multicolumn{6}{c}{$\log n_{\rm H}$ $[{\rm cm^{-3}}]$} \\
\hline
low		& $ -2.3	$ & $  -2.4	$ & $  0.6  $	& $>-4.0$	& $-1.2$ \\
 high		& $ -2.3	$ & $  -2.4	$ & $  0.7  $	& $-3.3$	& $-0.7$ \\
\hline
\multicolumn{6}{c}{$\log N_{\rm H}$ $[{\rm cm^{-2}}]$} \\
\hline
low		& $ 18.9	$ & $  18.9	$ & $  0.7  $	& $17.7$	& $20.0$ \\
 high		& $ 20.0	$ & $  20.1	$ & $  0.9  $	& $17.3$	& $21.5$ \\
\hline
\multicolumn{6}{c}{$\log l$ [pc]} \\
\hline
low		& $ 2.8 	$ & $  3.0	$ & $  1.1  $	& $0.5$		& $4.6$ \\
 high		& $ 3.8 	$ & $  4.0	$ & $  1.4  $	& $-0.5$	& $6.3$ \\
\hline
\multicolumn{6}{c}{$T$ ($10^4$\,K)} \\
\hline
low		& $ 1.4		$ & $  1.4	$ & $  0.5  $	& $0.4$ 	& $2.7 $ \\
 high		& $ 2.1		$ & $  2.1	$ & $  0.7  $	& $0.6	$	& $3.5 $ \\
\enddata
\tablecomments{
Low and high $z$ in column (1) correspond to $z<1$ and $2.3<z<3.3$, respectively. Lower or upper limits were treated as values to calculate the mean, median, and standard deviation (note that the two absorbers at  $2.3<z<3.3$ where only a lower limit on $\log U\ge -4$ was set by hand are not included to calculate these numbers, but their inclusion would not change these values significantly). Values for the low $z$ samples are from \citetalias{lehner13} and references therein and for the high $z$ sample from this work and adapted from \citet{crighton13,crighton15,lehner14}. The \nhi\ values were estimated from the spectra; all the other values were obtained from the Cloudy models.
}
\end{deluxetable*}

%\floattable
\begin{deluxetable*}{lccccc}
\tablecaption{Comparison of  ${\rm [C/\alpha]}$ estimated from the Cloudy models and directly from the data \label{t-calpha}}
\tablehead{
\colhead{QSO} & \colhead{$z_{\rm abs}$} & \colhead{$\log N_{\rm HI}$} &  \colhead{${\rm [X/H]}$} & \colhead{${\rm [C/\alpha]_{\rm Cloudy}}$} & \colhead{${\rm [C/\alpha]_{\rm data}}$} \\
\colhead{}&\colhead{}&  \colhead{$[{\rm cm}^{-2}]$}&\colhead{}&\colhead{}&\colhead{}
} 
%\colnumbers
\startdata
\multicolumn{6}{c}{L13 sample} \\
\hline
	PG1338+416     & $   0.3488 $ & $   16.30 \pm 0.13 $ & $ -0.75 \pm 0.15  $ & $ +0.15 \pm 0.15 $ & $ +0.09 \pm 0.10 $ \\
	J1419+4207     & $   0.2889 $ & $   16.40 \pm 0.06 $ & $ -0.65 \pm 0.15  $ & $ -0.15 \pm 0.15 $ & $ \ge -0.14 	   $\tablenotemark{a}\\ 
        PG1216+069     & $   0.2823 $ & $   16.40 \pm 0.05 $ & $ <-1.65 	 $ & $ +0.00 : 	      $ & $ >-0.28  	   $ \\
        J1619+3342     & $   0.2694 $ & $   16.48 \pm 0.05 $ & $ -1.60 \pm 0.10  $ & $ -0.10 \pm 0.10 $ & $ -0.23 \pm 0.05 $ \\
        J1435+3604     & $   0.3730 $ & $   16.65 \pm 0.07 $ & $  -1.85 \pm 0.10 $ & $ -0.15 \pm 0.20 $ & $ >-0.65    	   $\tablenotemark{b}  \\
        PKS0552-640    & $   0.3451 $ & $   16.90 \pm 0.08 $ & $  <-1.50	 $ & $ -0.15 \pm 0.15 $ & $ -0.36 \pm 0.10 $\tablenotemark{b}  \\
\hline
\multicolumn{6}{c}{This paper} \\
\hline
  J012156$+$144823  & $2.66586 $ & $ 16.32 \pm 0.01 $  & $ -1.05 \pm 0.10 $ & $ +0.05 \pm 0.10 $ & $>-0.33: $\tablenotemark{c}  \\
  J135038$-$251216  & $2.57299 $ & $ 16.39 \pm 0.01 $  & $ -2.30 \pm 0.10 $ & $ -0.05 \pm 0.10 $ & $ -0.15 \pm 0.10 $  \\
  J131215$+$423900  & $2.48998 $ & $ 16.77 \pm 0.01 $  & $ -2.50 \pm 0.10 $ & $ -0.55 \pm 0.10 $ & $ >-0.09$  \\
  J101723$-$204658  & $2.45053 $ & $ 17.23 \pm 0.01 $  & $ -2.50 \pm 0.15 $ & $ +0.10 \pm 0.15 $ & $ >-0.28: $\tablenotemark{c} \\
  J132552$+$663405  & $2.38287 $ & $ 17.30 \pm 0.30 $  & $ -3.00 \pm 0.10 $ & $ -0.20 \pm 0.10 $ & $>-0.26 $  \\
\enddata
\tablecomments{We only consider here systems for which we can estimate directly from the observations $(N_{\rm CII} + N_{\rm CIII} +N_{\rm CIV})/(N_{\rm SiII} + N_{\rm SiIII} +N_{\rm SiIV})$ at $2.3<z<3.3$  and  $(N_{\rm CII} + N_{\rm CIII})/(N_{\rm SiII} + N_{\rm SiIII} )$ at $z<1$.}
\tablenotetext{a}{Assuming that any possible levels of saturation in \ciiit\ and \siiiit\  are mild.}
\tablenotetext{b}{\ciit\ is not available for that absorber and its column is assumed negligible relative to \ciiit\ based on other absorbers. }
\tablenotetext{c}{Assuming that the saturation in  \siiiit\ is mild (one pixel reaches zero flux level); the colon emphasizes that this result is more uncertain. }
\end{deluxetable*}

\appendix

\section{Description of the absorbers and Cloudy analysis in the new sample}

In this Appendix, we provide more details for the new sample of pLLSs and LLSs, in particular about their velocity profiles and column densities, and the Cloudy photoionization models that were used to determine the metallicity of these absorbers. The redshift of each absorber is defined based on the strongest \hi\ component. For each Cloudy simulation run, we set the \hi\ column density to the value summarized in Table~\ref{t-nhi} and consider the errors on \nhi\ and on the metal lines to determine the errors on the metallicity and the ionization parameter. For each absorber, we vary the ionization parameter $U$ and the metallicity to search for models that are consistent with the constraints set by the column densities determined from the observations (also check \citetalias{lehner13} for more information regarding the methodology to estimate the metallicity and for comparison with the low redshift sample). In the figures that follow, we show the normalized profiles of most of the metal lines observed for each absorber by Keck HIRES as a function of the restframe velocity.  Finally, in Table~\ref{t-cloudy}, we tabulate the relevant properties of the pLLSs and LLSs directly derived from the absorption profiles (redshift and \nhi) and inferred from the estimated column densities of metals and \hi\ using Cloudy (the metallicity $\xh$; the relative abundance of carbon relative to $\alpha$ elements [$\alpha/$C]; the total H column density, $N_{\rm H}$; the ionization parameter $U$; the ionization fraction $N_{\rm HII}/N_{\rm H}$; the density $n_{\rm H}$; and the linear-scale of the absorber $l$).

\begin{figure}
\epsscale{1} 
\plotone{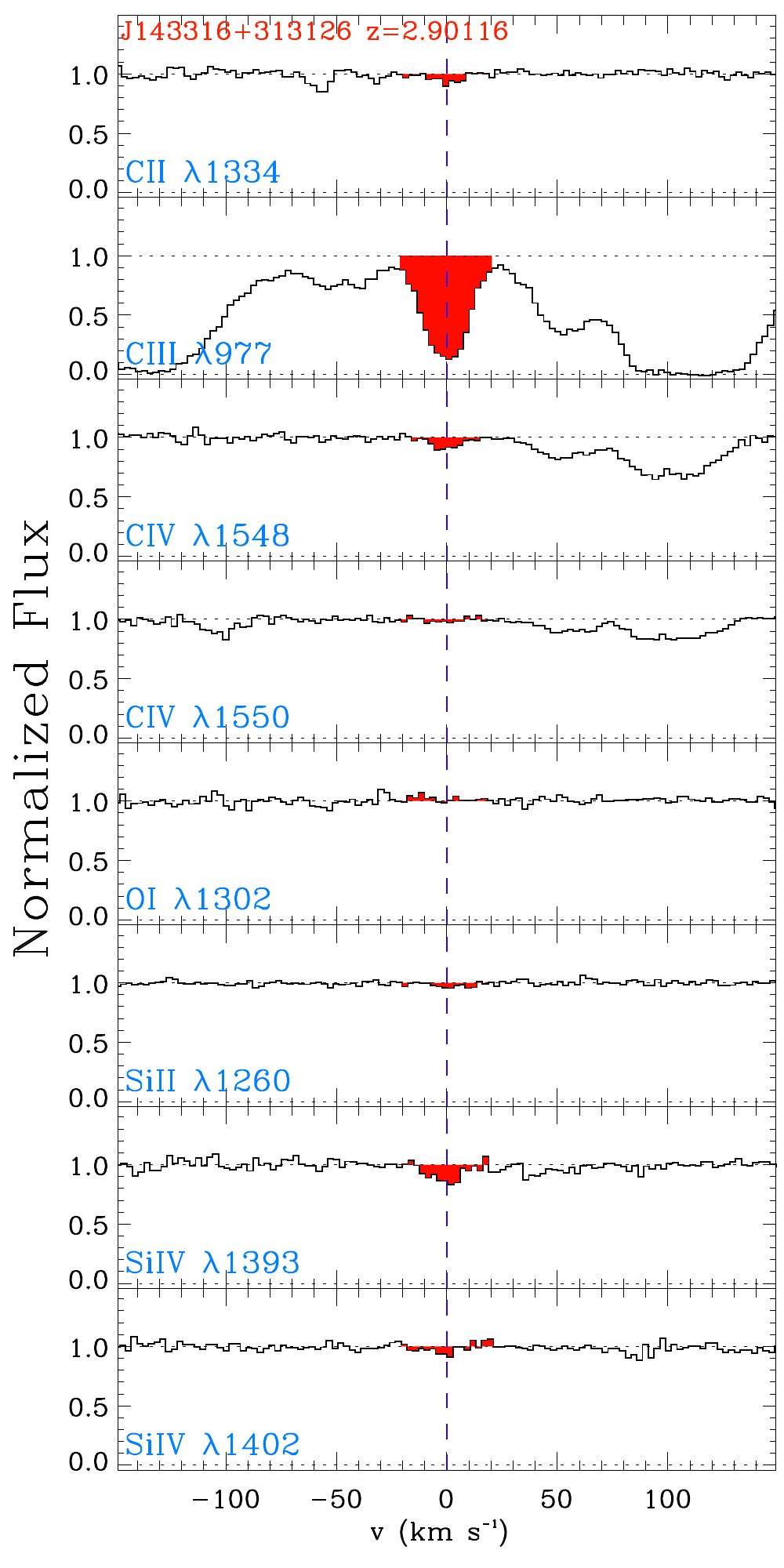}
 \caption{Normalized profiles of the metal absorption lines as a function of velocity centered on the absorber at $z=2.90116$ observed toward J143316+313126. The red portion in each profile shows the approximate velocity range of the absorption associated with the pLLS.  The reader should refer to Table~\ref{t-metal} for the exact integration velocity intervals. The vertical dashed lines mark the zero velocity.}
 \label{f-J143316}
\end{figure}

\noindent
{\it -- J143316+313126 -- $z=2.90116$ -- $\mlnhi = 16.16$}: For this pLLS, the following ions are available (see Fig.~\ref{f-J143316}): \cii, \ciii, \civ, \siii, \siiv. Although the absorption is weak for each transition of the \civ\ and \siiv\ doublets (and not detected at the 3$\sigma$ level for \civ\ $\lambda$1550), the column densities and the limit are consistent between the weak and strong transitions, respectively. \ciii\ is blended but the component associated with the pLLS appears free of blend. \siii\ and \cii\ are very weak, but detected at the $3.9$ and $4.6\sigma$ level, respectively.

For this absorber, the observations (detections of \siii, \cii, \ciii, and non-detections of \siiv\ and \civ) are well constrained with a Cloudy model with  $\xh  = -1.80 \pm 0.15 $, $\log U = -2.80 \pm 0.15$,  and $[{\rm C}/\alpha]= 0.00 \pm 0.15$. If the metallicity is lower, then $U$ must increase to match some of the observables, but in this case there is no adequate solution that fits simultaneously the \siii/\siiv\ and \cii/\civ\ ratios.  The metallicity cannot be much higher because otherwise too much \siii\ would be produced over the $\log U$ interval satisfying the \cii/\ciii\ ratio. We therefore adopt this solution for this pLLS.

\begin{figure}
\epsscale{1} 
\plotone{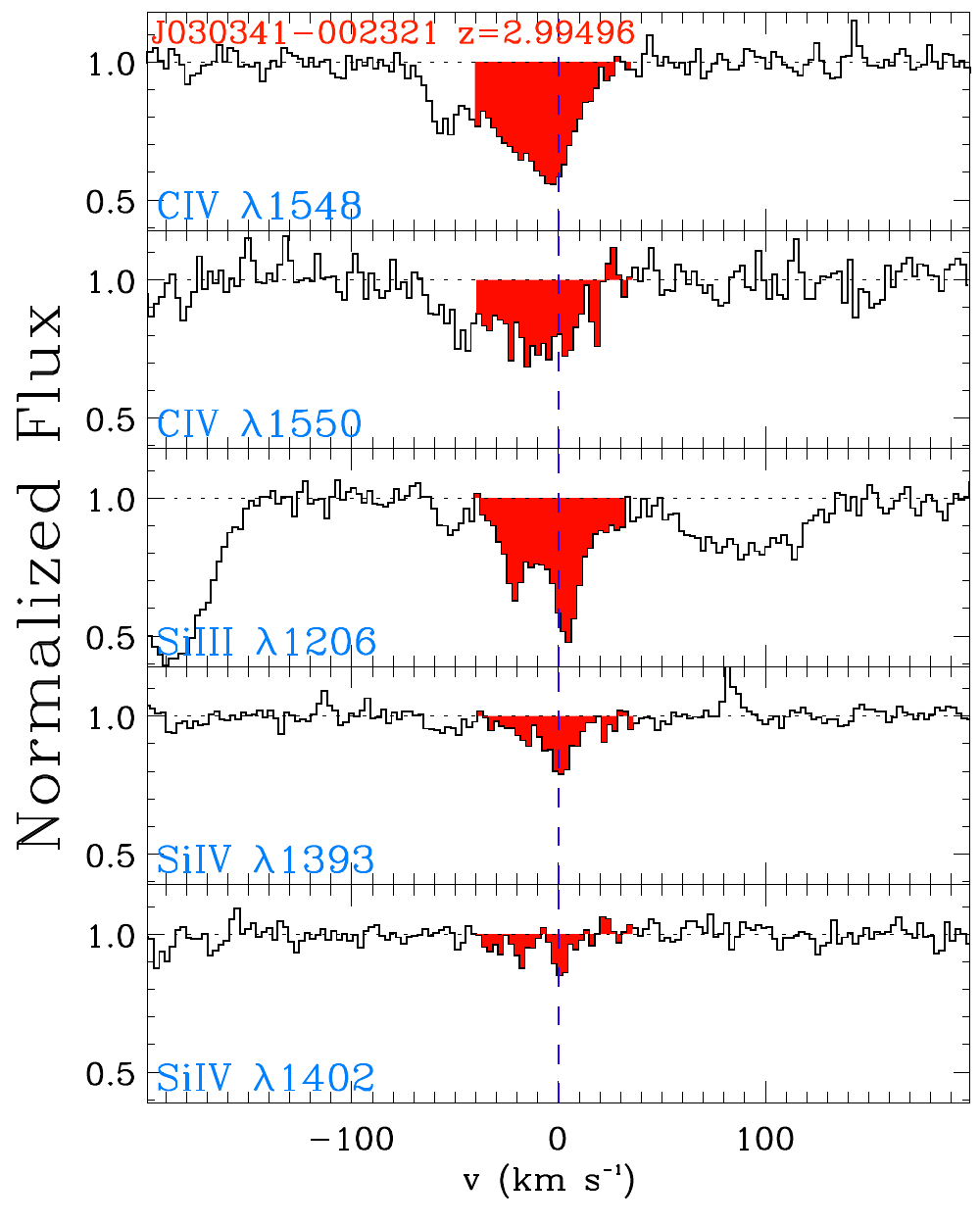}
 \caption{Same as Fig.~\ref{f-J143316}, but for another absorber.}
 \label{f-J030341a}
\end{figure}

\noindent
{\it -- J030341$-$002321 -- $z=2.99496$ -- $\mlnhi = 16.17$}:
For this pLLS, the following ions are available (see Fig.~\ref{f-J030341a}): \civ, \siiii, and \siiv. Both transitions of the \siiv\ doublet give similar column densities, but the strong transition is more securely detected and we adopted $N$ from the stronger transition. For \civ, there is some mild contamination in the weak transition of the doublet based on the comparison of the AOD profiles, and therefore we adopted $N$ from \civ\ $\lambda$1548. \siiii\ $\lambda$1206 appears uncontaminated based on its similar velocity profile to that of \siiv. The profiles of \siiii\ and \siiv\ have two components at about $+4$ and $-22$ \km, the positive velocity one being much stronger (see Fig.~\ref{f-J030341a}). For \civ, there are also two components, but at $-55$ and $-2.7$ \km, the latter associated with the pLLS being much stronger (as well as broader than observed in \siiii\ and \siiv). Unfortunately for this pLLS, all the useful \cii\ and \siii\ transitions are contaminated. 

To constrain the Cloudy photoionization model, we first use \siiii\ and \siiv. To match the amount of \siiv\ for this pLLSs, the metallicity needs to be at least $-1.95$ dex for any $U$. For this metallicity, a model with  $\log U = -1.85$ and  $[{\rm C}/\alpha] \simeq -0.25$ would simultaneously match the column densities of \siiii, \siiv, and \civ\ within 1$\sigma$. If the metallicity increases, $U$ must increase to match the  \siiii/\siiv\ ratio, but if $\log U \ga -1.5$ and $\xh  \ga -1.70$, the model would fail to match this ratio. For this pLLS, we therefore adopt $\xh  = -1.90 \pm 0.10 $, $\log U = -1.70 \pm 0.10$,  and $0\la [{\rm C}/\alpha]\la -0.7$.

\begin{figure}
\epsscale{1} 
\plotone{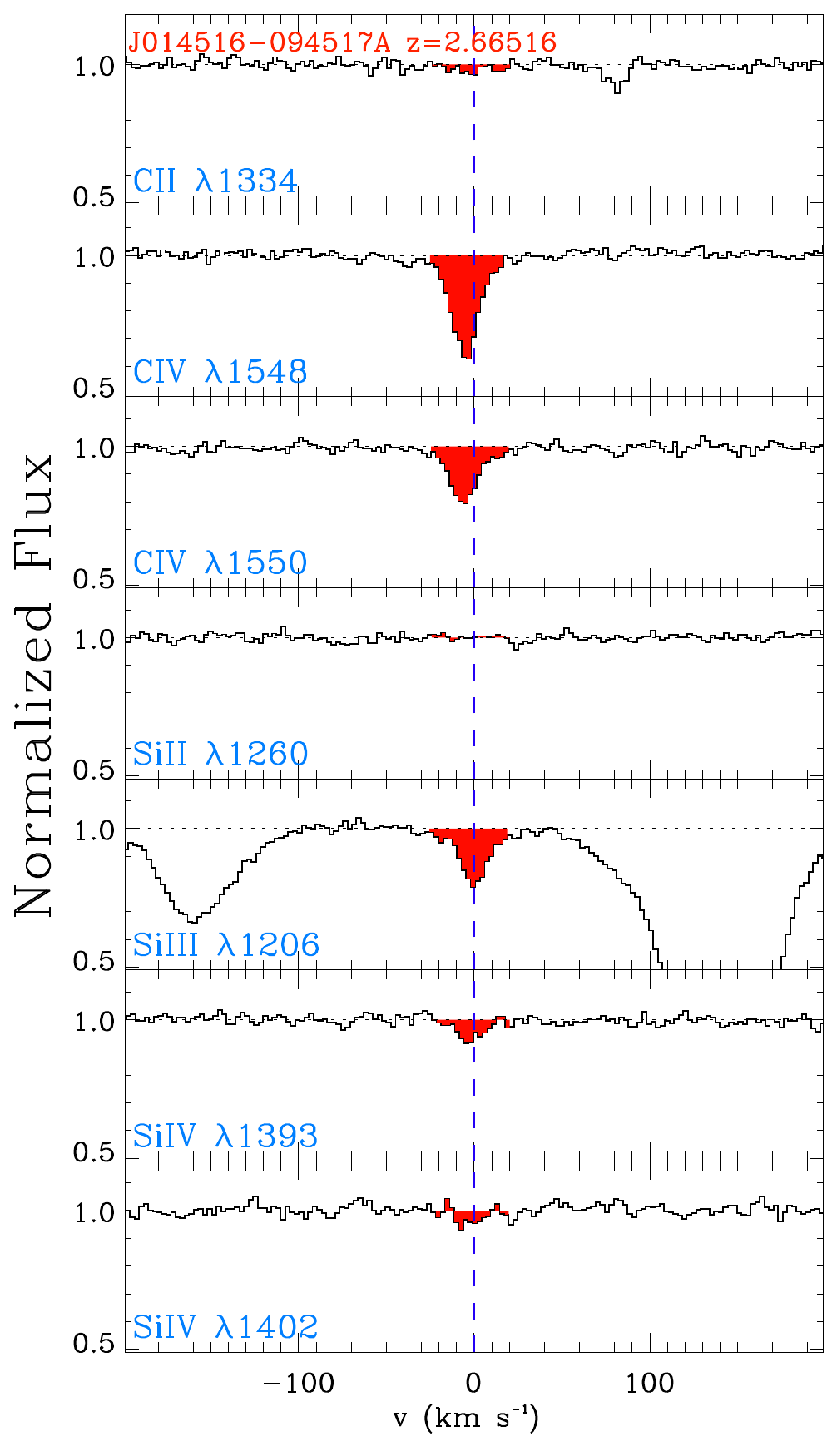}
 \caption{Same as Fig.~\ref{f-J143316}, but for another absorber.}
 \label{f-J014516}
\end{figure}

\noindent
{\it -- J014516$-$094517A -- $z =2.66516$ -- $\mlnhi = 16.17$}: 
For this pLLS, the following ions are available (see Fig.~\ref{f-J014516}): \cii, \civ, \siii, \siiii, and \siiv. Both transitions of the doublets of \civ\ and \siiv\ are detected and their column densities agree within 1$\sigma$. \siii\ $\lambda$1260 is not detected but a sensitive upper limit on $N$ can be derived thanks to the excellent S/N in the spectrum. \cii\ $\lambda$1334 is detected at 4$\sigma$ ($W_\lambda=3.25 \pm 0.78$ m\AA). \siiii\ $\lambda$1206 is well detected ($W_\lambda=24.4 \pm 0.70$ m\AA) and its velocity profile is similar to that of \siiv, implying it is unlikely contaminated. We note that there is no evidence of multiple components in this absorber, but there is a small velocity shift between \civ\ and  the other ions (about 4 \km, see Fig.~\ref{f-J014516} and Table~\ref{t-metal}). 

To constrain the Cloudy photoionization model, we first rely on the unambiguously detected \siiv\ and non detected \siii. The metallicity must be at least $-2.4$ dex to yield the observed  $N_{\rm SiIV}$. For this metallicity, $\log U = -1.85 $, and $[{\rm C}/\alpha] = 0$, the model is in agreement with the limit on \siii\ and  correctly predicts the column densities of \siiii, \siiv, \cii, and \civ. Taking into account the 1$\sigma$ uncertainty on \siiv, the metallicity cannot be much lower than $-2.55$ dex. If the metallicity increases to $-2.3$, this would imply $\log  U = -1.6$ to fit \siii, \siiii, and \siiv, and $[{\rm C}/\alpha] = -0.5$ to match the column density of \civ. However, in that case, the predicted amount of \cii\ would be too small by a factor $\sim 4$, which could imply that the observed absorbing feature is not \cii. The metallicity and $ U$ cannot be much higher than $-2.20$ and $-1.5$ dex, respectively, in order to match the \siiv/\siiii\ column density ratio. For this pLLS, we therefore adopt $\xh  = -2.40 \pm 0.20 $, $\log U = -1.85 \,^{+0.35}_{-0.10}$, and $[{\rm C}/\alpha] = 0.0  \,^{+0.10}_{-0.50}$.

\begin{figure}
\epsscale{1} 
\plotone{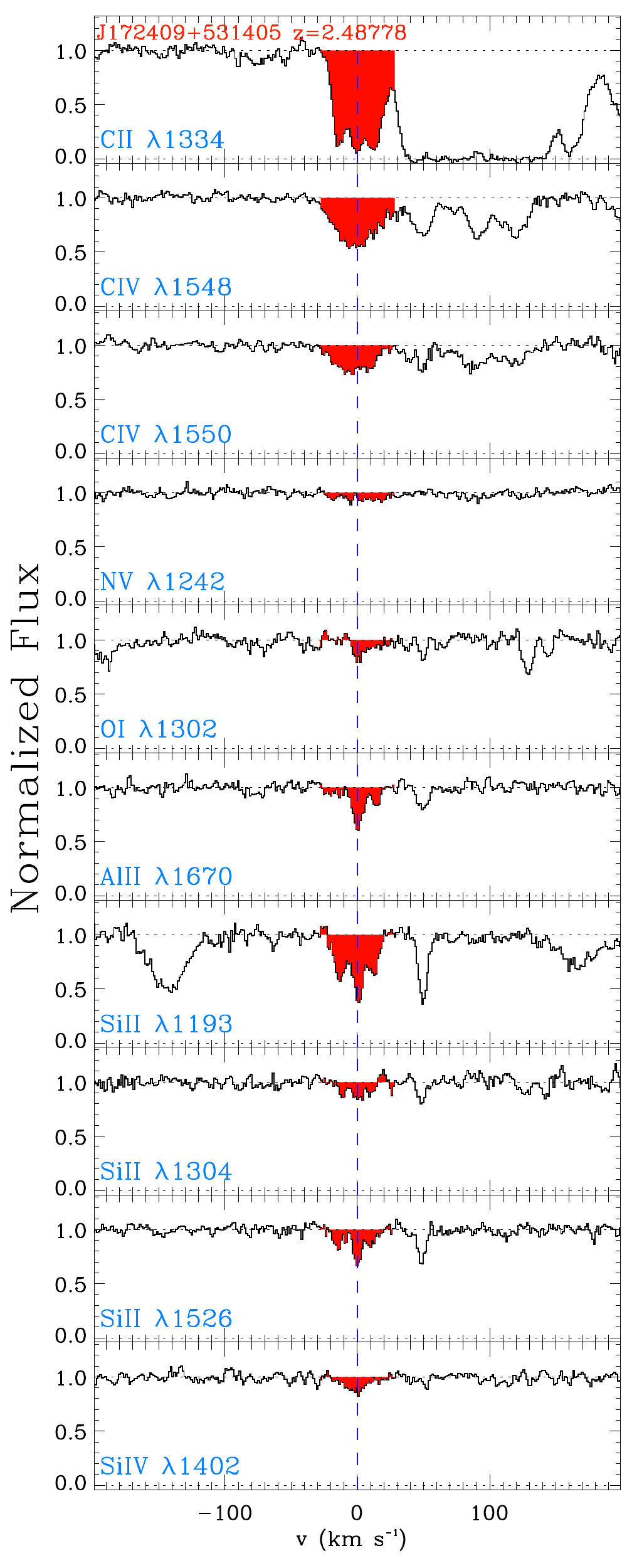}
 \caption{Same as Fig.~\ref{f-J143316}, but for another absorber.}
 \label{f-J172409}
\end{figure}
\noindent
{\it -- J172409+531405 -- $z=2.48778$ -- $\mlnhi = 16.20$}: For this pLLS, the following ions are available (see Fig.~\ref{f-J172409}): \cii, \civ, \nv, \oi, \alii, \siii, \siiv, and \feii. Several transitions of \siii\ are detected, giving consistent results for the column density. The velocity profile of \alii\  is similar to that of \siii, with 3 distinct components between $-28$ and $+28$ \km\ (as well as an additional one at $+50$ \km, which is not considered for the metallicity estimate since this absorption is related to the \hi\ absorption with $\mlnhi \simeq 15.87$). \civ, \siiv, and \nv\ are also detected, but their profiles have only a single broad component, implying that the gas is multiphase (and indeed the high ions cannot be reproduced by the photoionization model described below). The \cii\ absorption is very strong, and the absorption between $+10$ and $+30$ \km\ is most likely contaminated. \feii\ is not detected at the 3$\sigma$ level, while \oi\ is detected but only in the stronger absorption near $0$ \km. \siiii\ and \ciii\ are both detected, but are contaminated to some levels in view of the differences in their profiles compared to the other ions. Near this pLLS, there are other lower \hi\ column density absorbers at $+50$ \km\ seen in singly and highly ionized species and between $+50$ and $+130$ \km. For the metallicity estimate, we only consider the absorption between  $-28$ and $+28$ \km\ where \nhi\ of the pLLS was estimated. 

\begin{figure}
\epsscale{1} 
\plotone{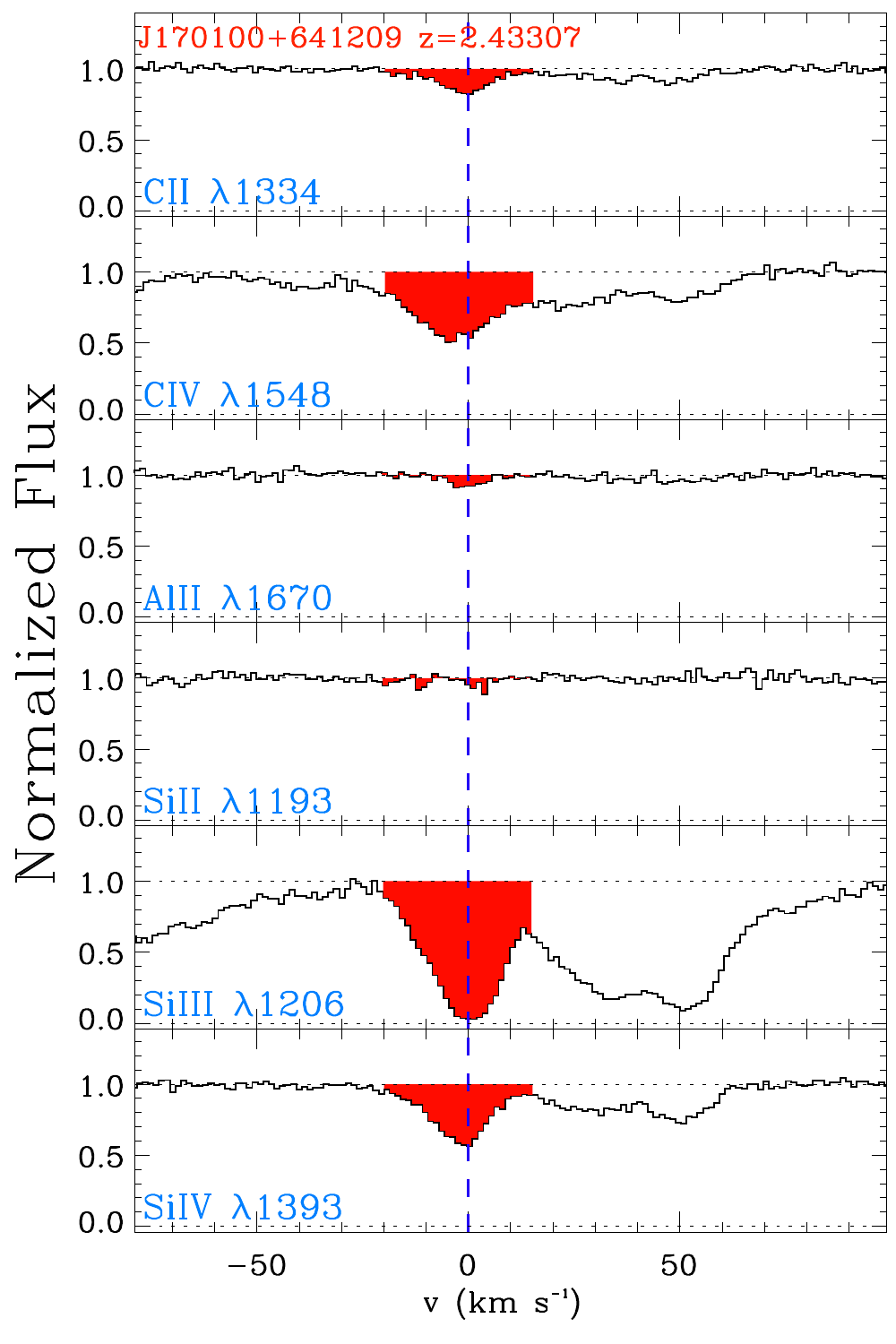}
 \caption{Same as Fig.~\ref{f-J143316}, but for another absorber.}
 \label{f-JJ170100}
\end{figure}

For the Cloudy model, we first consider \siii, \alii, and \oi\ since they are reliably detected and trace the narrow \hi\ component associated with the pLLS. A model with $\xh  = +0.20 \pm 0.10 $ and $\log U = -4.00 \pm 0.10$ matches well the column densities of \siii, \alii, and \oi. The predicted column density of \cii\ is about 13.7 dex, consistent with the lower limit on the \cii\ column density. For this model, the high ions are not reproduced by several orders of magnitude. \feii\ is overproduced by at least $+0.3$ dex, suggesting that some $\alpha/{\rm Fe}$ enhancement (or possibly some dust depletion) is present in this high metallicity pLLS. A higher/lower metallicity would produce too much/little \oi. Although the \oi\ velocity component aligns well with the main component of \siii\ and \alii, we note the presence of several unidentified features near the \oi\ absorption (see Fig.~\ref{f-J172409}). It is therefore plausible that \oi\ may not be real, but even in this case, the metallicity would not change much since the lowest metallicity would need to be higher than $-0.2$ dex to produce enough \alii\ and match the \siii/\alii\ ratio. The metallicity cannot be much higher either in order to satisfy $N_{\rm OI}$ (or its limit if contaminated) and the \siii/\alii\ ratio. We therefore adopt the model that matches simultaneously the constraints from  \siii, \alii, and \oi:  $\xh  = +0.20 \pm 0.10 $ and $\log U = -4.00 \pm 0.10$.

\noindent
{\it -- J170100+641209 -- $z=2.43307$ -- $\mlnhi = 16.24$}: For this pLLS, several ions are detected but each one in a single transition only: \cii, \civ, \alii, \siiii, \siiv; \siii\ $\lambda$1193 is not detected (see Fig.~\ref{f-JJ170100}). This pLLS is blended with another one at $z=2.43359$ (about 50 \km\ in the restframe; see below) that we treat separately since in this case we could reliably determine \nhi\ in each component. Since each absorption profile reveals a single component structure, we suspect little contamination by unrelated absorbers. 

For this absorber, the observations are well constrained with a cloudy model with  $\xh  = -1.65 \pm 0.10 $, $\log U = -2.25 \pm 0.15$,  and $[{\rm C}/\alpha]= +0.20 \pm 0.10$. This model matches well the \siiii/\siiv\ and \cii/\civ\ ratios (as well as \alii/\siii). The metallicity cannot be higher because otherwise too much \alii\ and \siii\ would be produced and cannot be lower because too little \siiv\ would be otherwise produced. We therefore adopt the solution above for this pLLS.

\begin{figure}
\epsscale{1} 
\plotone{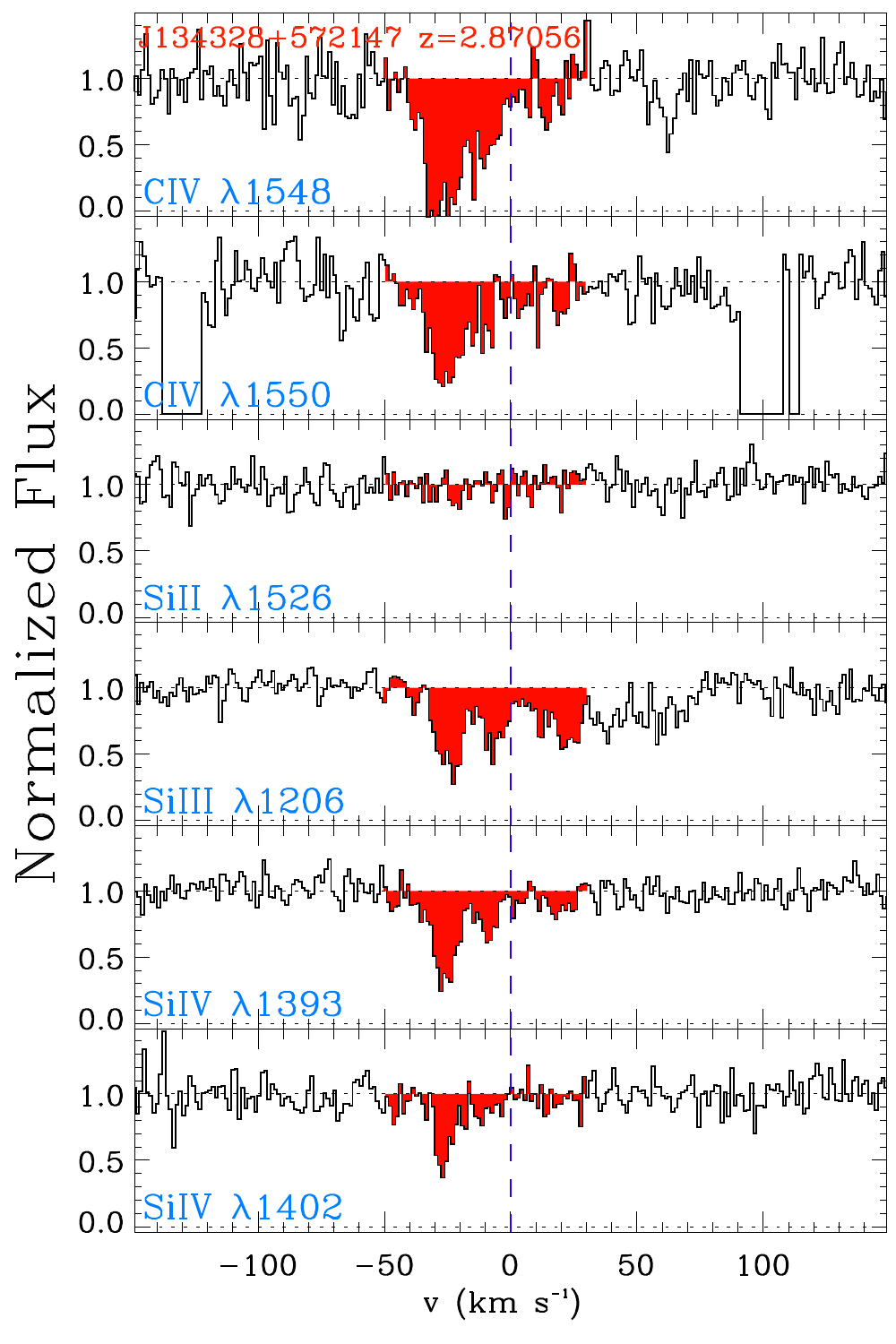}
 \caption{Same as Fig.~\ref{f-J143316}, but for another absorber.}
 \label{f-J134328}
\end{figure}

\noindent
{\it -- J134328+572147 -- $z=2.87056$ -- $\mlnhi = 16.30$}: For this pLLS, the following ions are available (see Fig.~\ref{f-J134328}): \civ, \siiii, \siiv. For \siii, only the transitions at 1304 and 1526 \AA\ are not contaminated, but neither gives a sensitive limit on the column density of \siii. There 
is evidence that \civ\ $\lambda$1548 is contaminated in the $-24$ \km\ velocity component when comparing the apparent column density profiles of the \civ\ doublet; at other velocities, the \civ\ profiles are identical. We therefore adopt $N_{\rm CIV}$ from the weak transition. The two transitions of the \siiv\ doublet yield consistent column densities within less than 1$\sigma$. We note that the profiles have 3 components at about $-24, -4, +24$ \km, with the stronger component being at $-24$ \km. The \hi\ profiles at 915 to 926 \AA\ are too noisy, and, at longer wavelengths, too saturated to discern any velocity component structure. We therefore estimated the column densities by integrating the profiles from $-50$ to $+30$ \km.

For this pLLS, the metallicity cannot be lower than $-1.6$ dex to produce the observed amount of \siiv\ for any $U$. A Cloudy model with  $\xh  = -1.45 \pm 0.10 $, $\log U = -1.55 \pm 0.10$,  and $[{\rm C}/\alpha]= -0.70 \pm 0.10$ reproduces well the observables. The metallicity and $U$ could be higher if $[{\rm C}/\alpha] \ll -0.70$, which is not very likely based on the ${\rm C}/\alpha$ nucleosynthesis history (see Fig.~\ref{f-calpha}). We therefore adopt this solution for this pLLS.

\begin{figure*}
\epsscale{1} 
\plotone{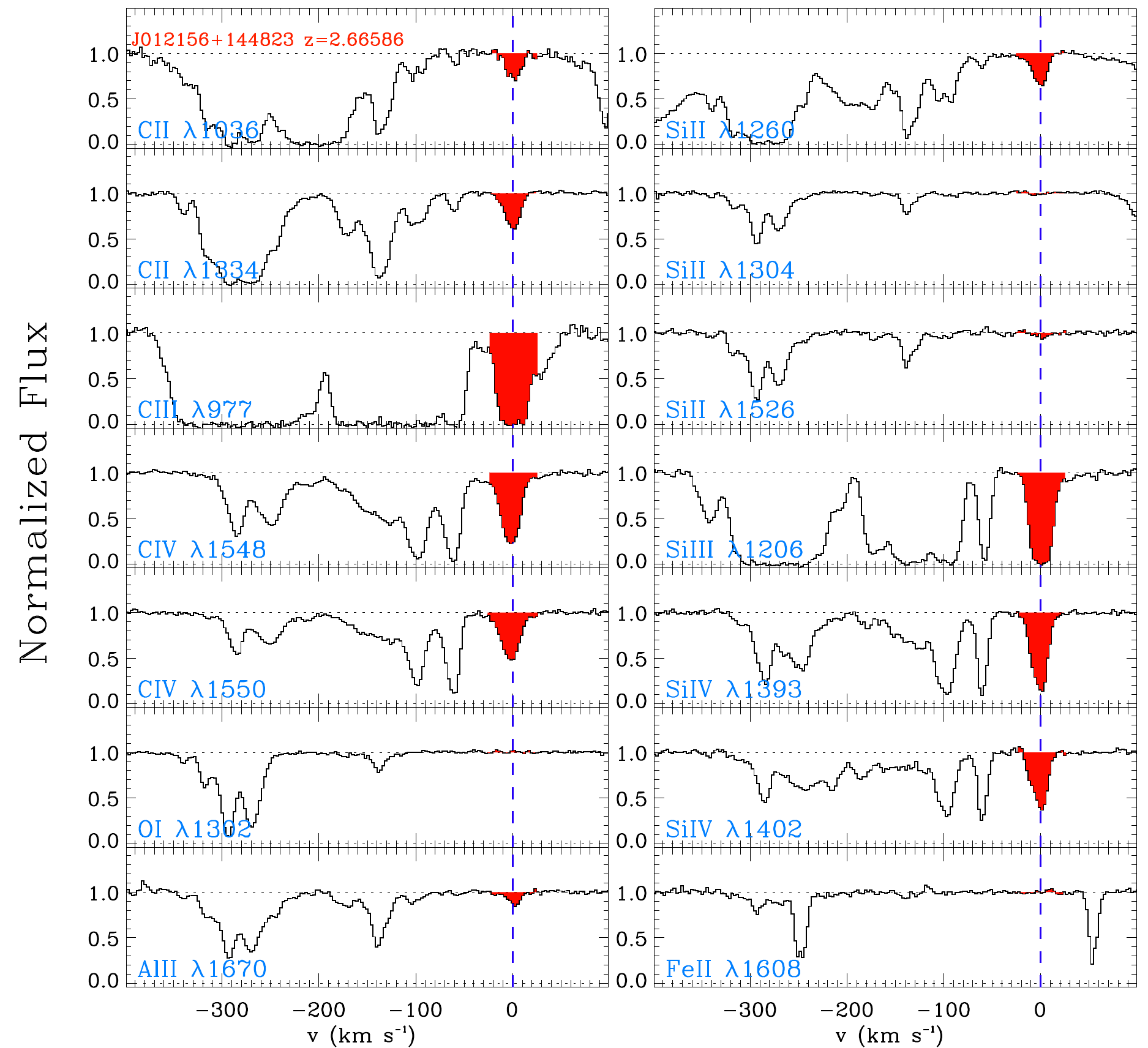}
 \caption{Same as Fig.~\ref{f-J143316}, but for another absorber.}
 \label{f-J012156}
\end{figure*}

\noindent
{\it -- J012156+144823 -- $z=2.66586$ -- $\mlnhi = 16.32$}: For this pLLS, the following ions are available (see Fig.~\ref{f-J012156}): \cii, \ciii, \civ, \oi, \alii, \siii, \siiii, \siiv, and \feii. \cii, \civ, \siii, and \siiv\ have each two transitions that yield very similar column densities. \alii\ and \siiii\ have very similar velocity profiles, suggesting that they are not contaminated. \ciii\ $\lambda$977 could be partially contaminated based on the extra absorption observed at $v\ga +25$ \km. \oi\ and \feii\ are not detected at the $2\sigma$ level. In this case, the absorption in all the observed ions has a single component between $-25$ and $+25$ \km\ associated with the pLLS. This absorber is associated to a SLLS ($\mlnhi \simeq 19.05$) and LLS/SLLS ($\mlnhi \ga 18.45$) at $z=2.66245$ and 2.66415 as it can be seen in the metal lines where the absorption extends to about $-350$ \km. 

For this absorber, the observations are well constrained with a cloudy model with  $\xh  = -1.00 \pm 0.10 $, $\log U = -2.40 \pm 0.10$,  and $[{\rm C}/\alpha]= +0.00 \pm 0.10$. This is because over $-4 \la \log U \la -2$ and for this \nhi, there is only a small interval of $U$ where $N_{\rm SiII}$ and $N_{\rm SiIV}$ overlap. This model  also matches \cii/\civ\ and predicts the observed $N_{\rm AlII}$ within 1$\sigma$ and is in agreement with the lower limits on \siiii\ and \ciii. We therefore adopt that solution for this pLLS.

\begin{figure}
\epsscale{1} 
\plotone{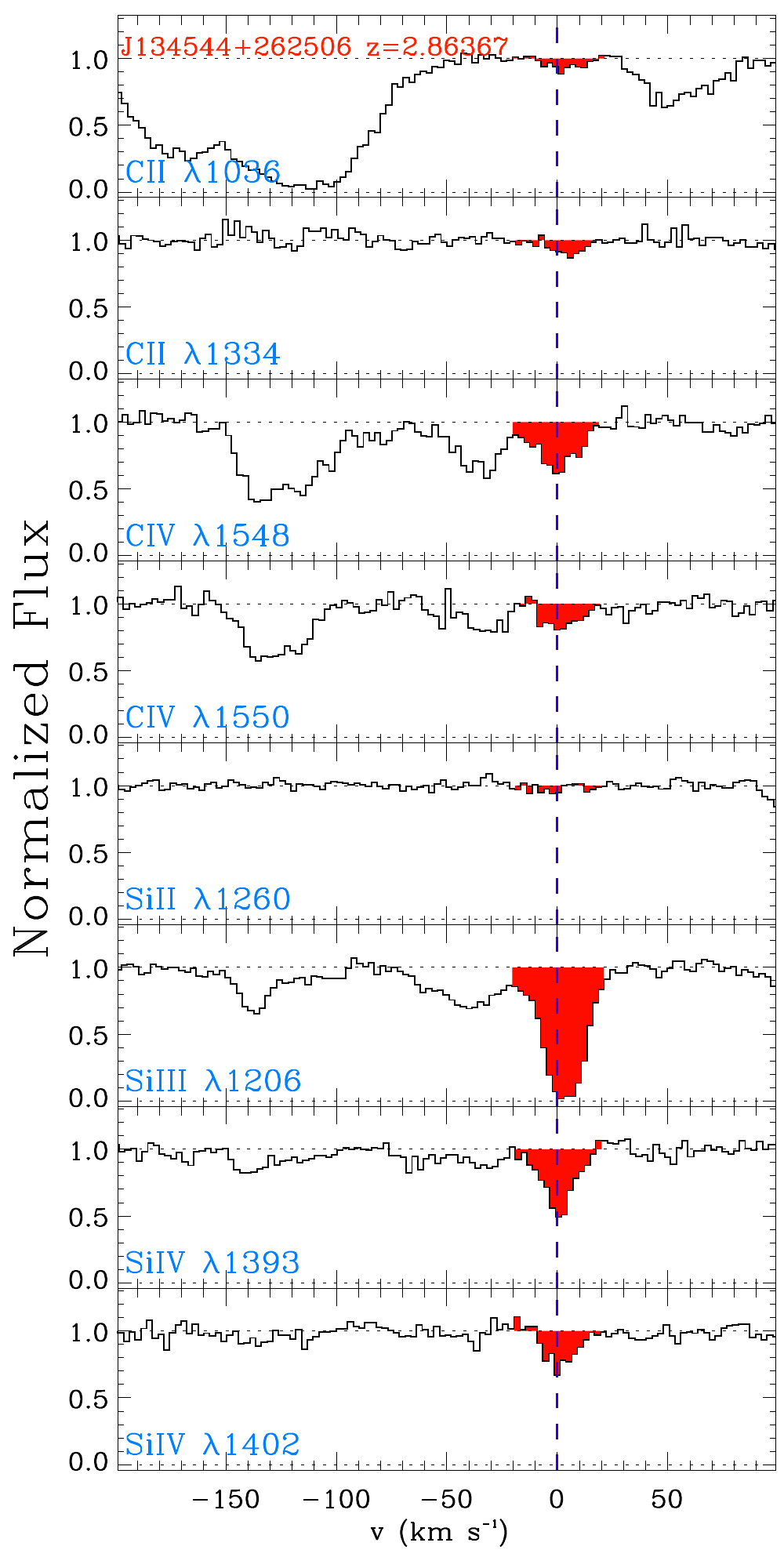}
 \caption{Same as Fig.~\ref{f-J143316}, but for another absorber.}
 \label{f-J134544}
\end{figure}

\noindent
{\it -- J134544+262506 -- $z=2.86367$ -- $\mlnhi = 16.36$}: For this pLLS, the following ions are available (see Fig.~\ref{f-J134544}): \cii, \civ, \siii, \siiii, and \siiv. Both \cii\ and \siii\ are weak but detected above the $3\sigma$ level. In particular, \cii\ $\lambda$$\lambda$1334, 1036 give consistent column densities. Both transitions of the doublets of \civ\ and \siiv\ are detected and their column densities agree within 1$\sigma$.  \siiii\ is strong and could be saturated or contaminated (see below). Near this pLLS, there is clearly evidence of additional absorption components at about $-125$ and $-35$ \km\ in the metal ions. These components are also identified in the \hi\ absorption and closely match the \hi\ fit model even though the metal and \hi\ lines were independently fitted (see Table~\ref{t-nhi}). We only consider the absorption between $-20$ and $+20$ \km, which is directly associated with the pLLS. 

For this absorber, it was difficult to match all the observables within 1$\sigma$ on the column densities of the metal lines. Within about 2$\sigma$, a cloudy model with  $\xh  = -1.65 \pm 0.20 $, $\log U = -2.10 \pm 0.20$,  and $[{\rm C}/\alpha]= -0.10 \pm 0.20$ would match the column densities of \cii, \civ, \siii, \siiii, and \siiv\ derived from the observations. A higher/lower metallicity and $U$ would fail to match appropriately the \siii/\siiv\ and \cii/\civ\ ratios. We therefore adopt this solution.

\begin{figure}
\epsscale{1} 
\plotone{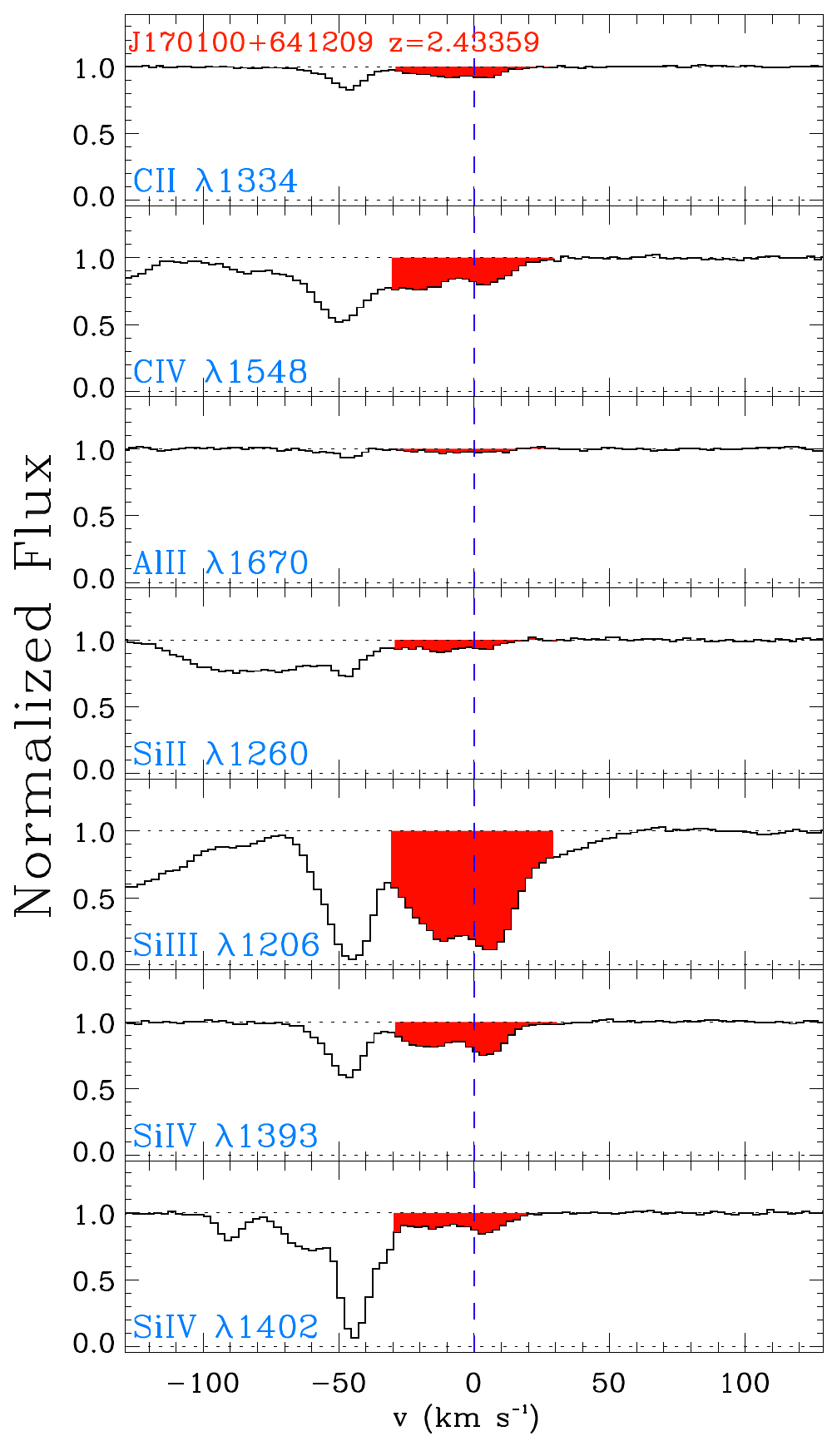}
 \caption{Same as Fig.~\ref{f-J143316}, but for another absorber.}
 \label{f-JJ170100a}
\end{figure}

\noindent
{\it -- J170100+641209 -- $z=2.43359$ -- $\mlnhi = 16.38$}: For this pLLS, several ions are detected: \cii, \civ, \alii, \siii, \siiii, \siiv\ (both transitions of the doublet) (see Fig.~\ref{f-JJ170100a}). This pLLS is blended with another one at $z= 2.43307 $ (about $-50$ \km\ in their restframe, see above). As all the profiles reveal a similar velocity structure with two main components, we suspect little contamination by unrelated absorbers.

For this absorber, the observations are well constrained with a Cloudy model with  $\xh  = -1.50 \pm 0.10$, $\log U = -2.35 \pm 0.15$,  and $[{\rm C}/\alpha]= -0.05 \pm 0.15$. This model matches well the  \siii/\siiv, \siiii/\siiv, and \cii/\civ\ ratios (as well as \alii/\siii). The metallicity cannot change much either way because otherwise \siii/\siiv\ and \siiii/\siiv\ would not be matched by the model for any $U$. We therefore adopt the solution above for this pLLS.

\begin{figure}
\epsscale{1} 
\plotone{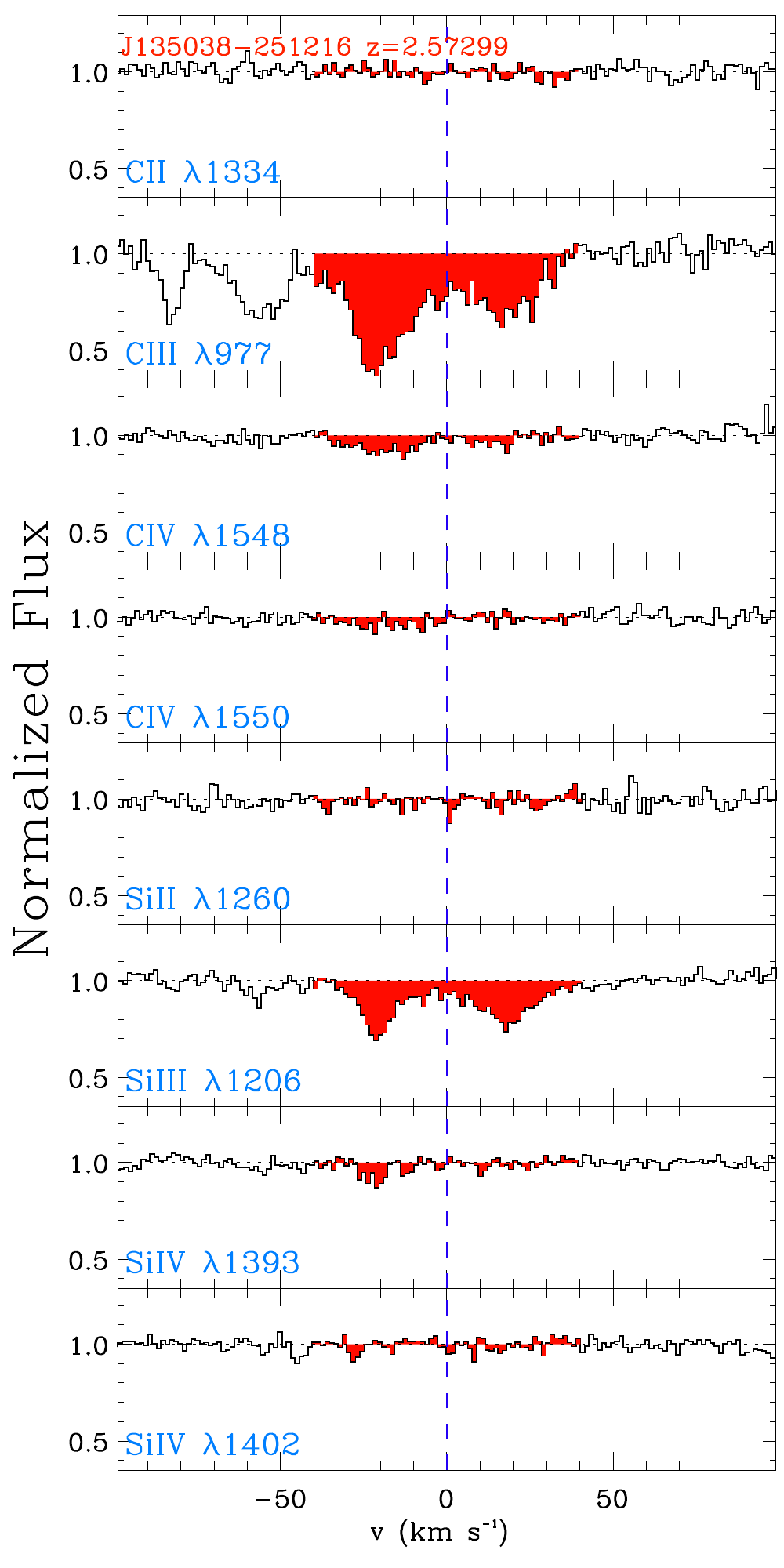}
 \caption{Same as Fig.~\ref{f-J143316}, but for another absorber.}
 \label{f-J135038}
\end{figure}

\noindent
{\it -- J135038$-$251216 -- $z=2.57299$ -- $\mlnhi = 16.39$}: For this pLLS, the following ions are available (see Fig.~\ref{f-J135038}): \cii, \ciii, \civ, \siii, \siiii, and \siiv. \cii\ and \siii\ are not detected and only the stronger transitions of the \civ\  and \siiv\ doublets are detected above the $3\sigma$ level. There is, however, an overall good agreement in the structure of the velocity profiles between \ciii, \siiii, and \civ, giving us confidence that these lines are not contaminated and these ions probe the same gas-phase gas. At the velocities over which the absorption of the pLLS is observed, there are two velocity components at $-20$ and $+20$ \km\ observed in the metal ionic lines. 

For this absorber, the observations are well constrained with a cloudy model with $\xh  = -2.30 \pm 0.10 $, $\log U = -2.45 \pm 0.10$,  and $[{\rm C}/\alpha]= -0.05 \pm 0.10$. This model matches well the \siiii/\siiv\ and \ciii/\civ\ column density ratios as well as the limits on the column densities of \cii\ and \siii. The metallicity cannot be higher because otherwise too much \cii\ and \siii\ would be produced and cannot be lower because too little \siiii\ would be otherwise produced.

\begin{figure}
\epsscale{1} 
\plotone{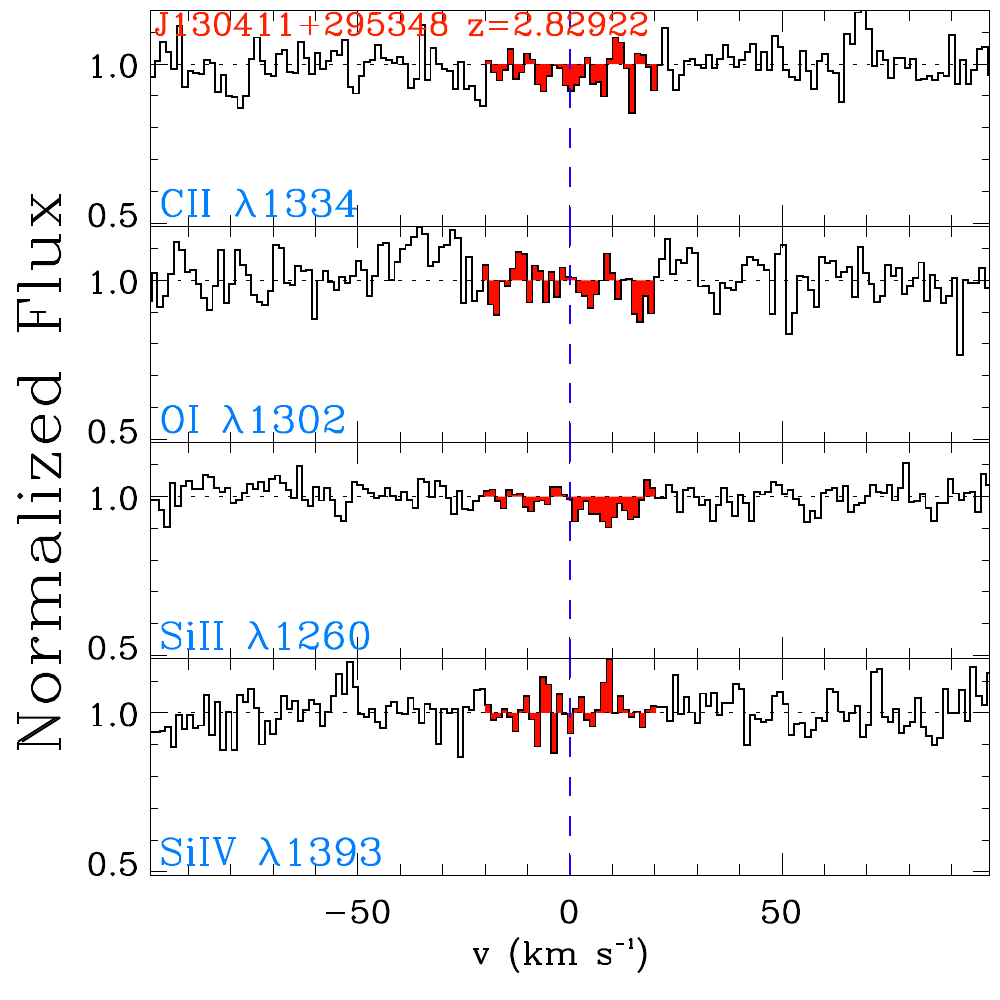}
 \caption{Same as Fig.~\ref{f-J143316}, but for another absorber.}
 \label{f-J130411}
\end{figure}

\noindent
{\it -- J130411+295348 -- $z=2.82922$ -- $\mlnhi = 16.39$}: For this pLLS, no metal lines are detected (see Fig.~\ref{f-J130411}). \ciii\ and \siiii\ are both contaminated. The \civ\ doublet is not covered. So we have to rely only \cii\ $\lambda$1334, \siii\ $\lambda$1260, and \siiv\ $\lambda$1393 to constrain the ionization model. 

For this LLS, we have to make the assumption that $\log U\ge -4$ (based on the other models and see Fig.~\ref{f-udist}) to be able to constrain the ionization model. We can only place an upper limit on the metallicity of  $\xh  <-1.7$ and $\log U \ge -4$. This model satisfies all the limits, but we cannot constrain better the metallicities with the current observables. We note that if we use instead the mean $\langle \log U \rangle = -2.4$ derived from the $\log U$ distribution for our sample of pLLSs and LLSs (see Fig.~\ref{f-udist}), then $\xh \le -2.80$. To be conservative, we adopt the former value.

\begin{figure}
\epsscale{1} 
\plotone{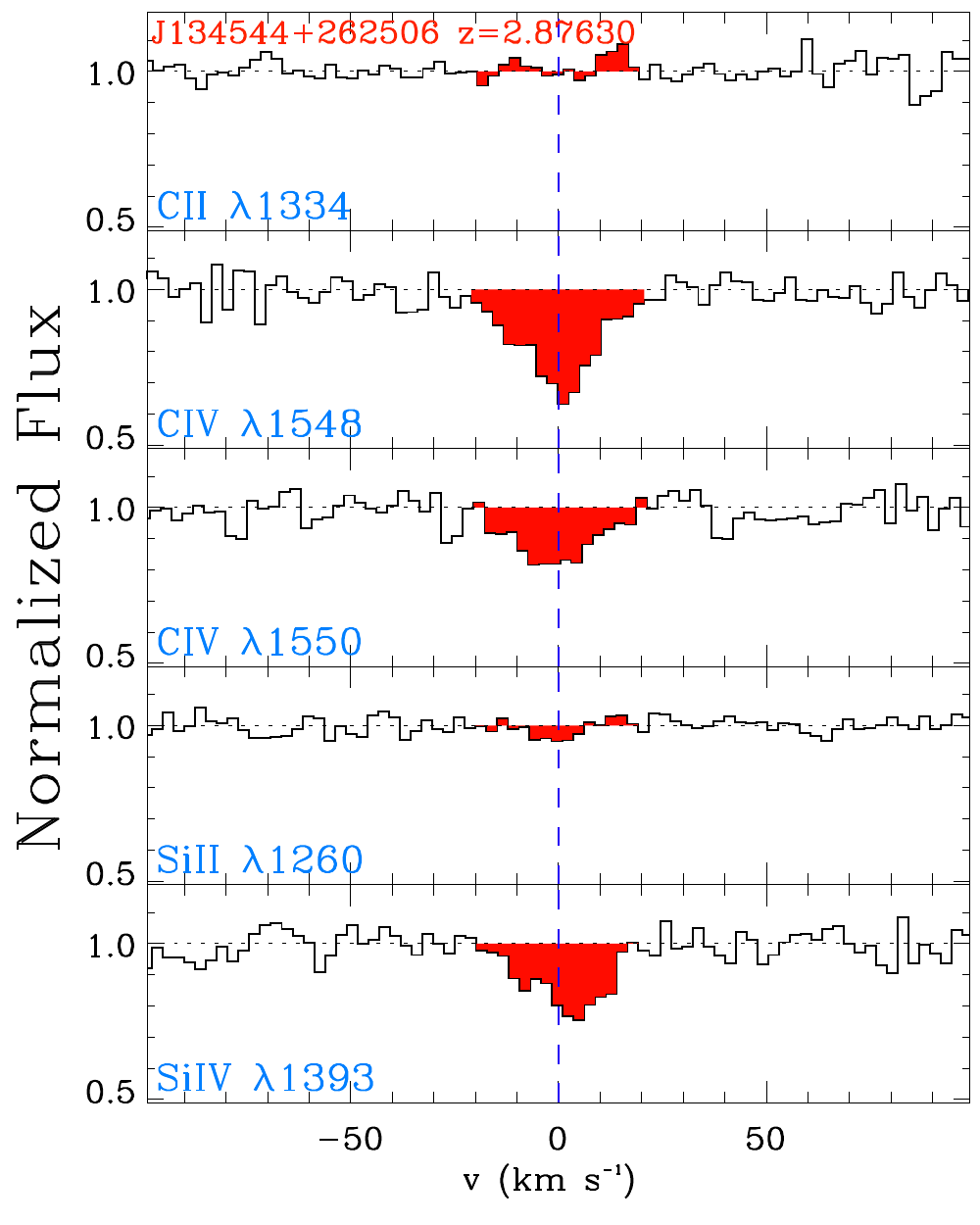}
 \caption{Same as Fig.~\ref{f-J143316}, but for another absorber.}
 \label{f-J134544a}
\end{figure}

\noindent
{\it -- J134544+262506 -- $z=2.87630$ -- $\mlnhi = 16.50$}: For this pLLS, the following ions are available (see Fig.~\ref{f-J134544a}): \cii, \civ, \siii, and \siiv. \cii\ and \siii\ are not detected at the $3\sigma$ level. The wavelength coverage of the observations did not cover \siiv\ $\lambda$1402. The two transitions of the \civ\ doublet give similar $N$ within $1\sigma$. The velocity profiles of \civ\ and \siiv\ are dominated by a single component. 

In order to  match the amount of \siiv\ and the upper limit on \siii, the metallicity needs to be at least $-2.40$ dex; a lower metallicity would produce too little \siiv. The metallicity and $U$ cannot be much higher either because otherwise it would violate the upper limit on \siii/\siiv\ and would require  $[{\rm C}/\alpha]\ll -0.5$. For this pLLS, we therefore adopt $\xh  = -2.30 \pm 0.10 $, $\log U = -1.80 \pm 0.10$,  and $[{\rm C}/\alpha]= -0.50 \pm 0.10$.

\begin{figure}
\epsscale{1} 
\plotone{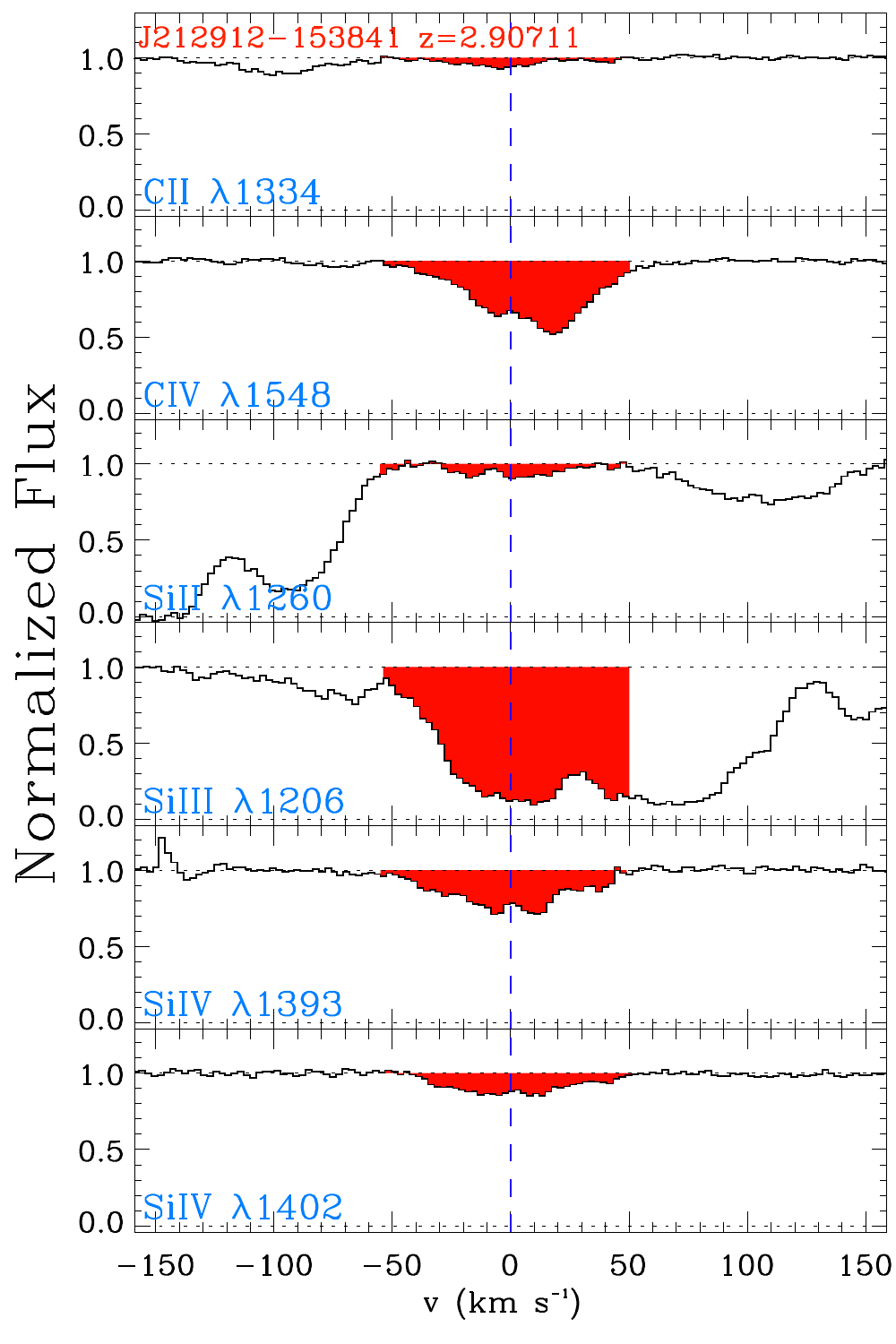}
 \caption{Same as Fig.~\ref{f-J143316}, but for another absorber.  }
 \label{f-J212912}
\end{figure}

\noindent
{\it -- J212912$-$153841 -- $z=2.90711$ -- $\mlnhi = 16.55$}: For this pLLS, several ions are detected: \cii, \civ, \siii, \siiii, \siiv\ (see Fig.~\ref{f-J212912}). The weak transition of the \civ\ doublet is contaminated, but the two transitions of the \siiv\ doublet give essentially the same column density.  \siiii\ $\lambda$1206 is partially blended and we only integrate the profiles to $+20$ \km; despite this contamination, \siiii\ provides a stringent lower limit on the amount of \siiii\ in this pLLS. The metal lines are dominated by two components, which are not resolved in the \hi\ lines.  As the ions all reveal similar absorption profiles, we suspect little contamination by unrelated absorbers. 

For this pLLS, the observations are well constrained with a cloudy model with  $\xh  = -1.55 \pm 0.10$, $\log U = -2.30 \pm 0.10$,  and $[{\rm C}/\alpha]= -0.20 \pm 0.10$. This model matches the \siii/\siiv, \siiii/\siiv, and \cii/\civ\ column density ratios. The metallicity cannot be much lower because otherwise not enough \siiii\ would be produced and cannot be much higher because otherwise too much \siii\ would be produced for any values of $U$. 

\begin{figure}
\epsscale{1} 
\plotone{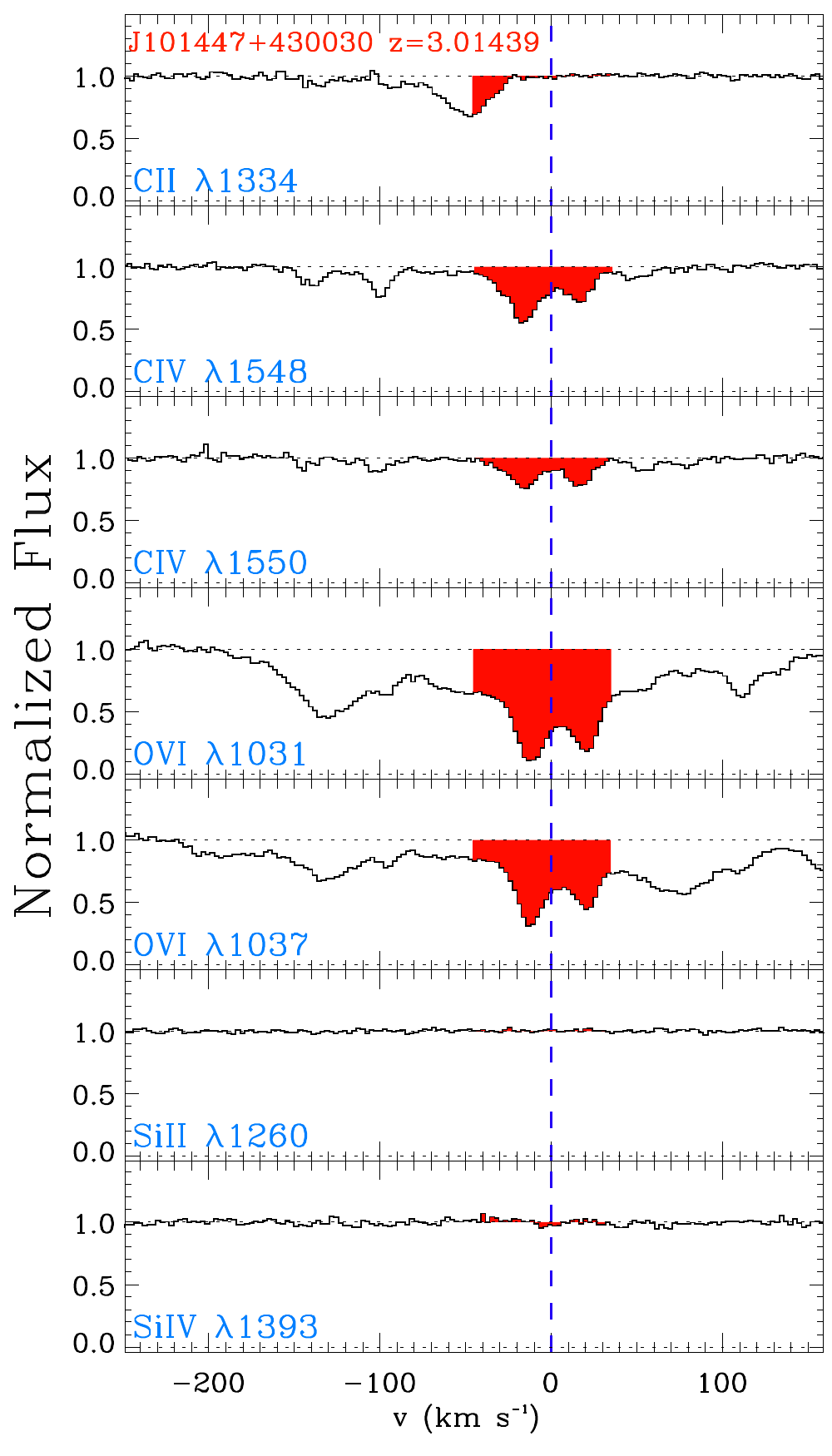}
 \caption{Same as Fig.~\ref{f-J143316}, but for another absorber.}
 \label{f-J101447}
\end{figure}

\noindent
{\it -- J101447+430030 -- $z=3.01439$ -- $\mlnhi = 16.63$}: For this pLLS, only \civ\ and \ovi\ are detected; \siii, \siiv\, and \cii\ are not detected at the $3\sigma$ level despite the high S/N level (see Fig.~\ref{f-J101447}, note that part of the \cii\ profile is contaminated). 

For this pLLS, \civ\ and \ovi\ must trace a different gas-phase since there is no valid photoionization solution for that \nhi\ that would fit simultaneously the column densities of \hi, \civ, \ovi, and  column-density limits on \cii, \siii, and \siiv. We therefore can only use the non-detections to constrain our models, but these are not sufficient to constrain reliably $U$. We therefore make the assumption we have already made that $\log U \ge -4$. For that value, the metallicity must be $\xh \le -2.60$, which also satisfies the limits on \cii\ and \siiv\ for that value of $U$. If $U$ increases, the metallicity must decrease, and in particular if we use instead the mean  $\langle \log U \rangle = -2.4$ derived from the $\log U$ distribution for our sample of pLLSs and LLSs, then $\xh <-3.40$. To be conservative, we adopt here $\xh <-2.60$ and $\log U \ge -4$.

\begin{figure}
\epsscale{1} 
\plotone{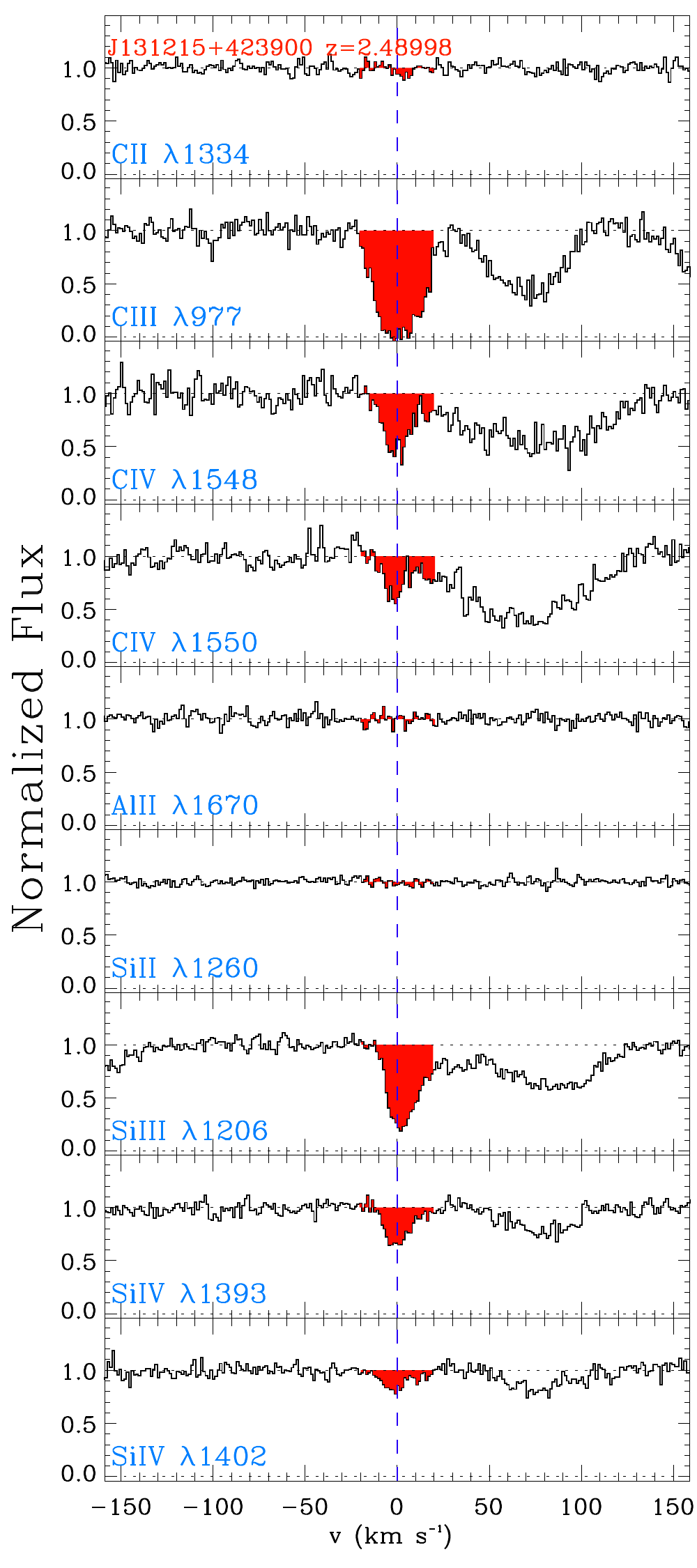}
 \caption{Same as Fig.~\ref{f-J143316}, but for another absorber.}
 \label{f-J131215}
\end{figure}

\noindent
{\it -- J131215+423900 -- $z=2.48998$ -- $\mlnhi = 16.77$}: For this pLLS, \ciii, \civ, \siiii, \siiv\ are well detected, while \siii\ and \cii\ are very weak and \alii\ is not detected at the $3\sigma$ level (see Fig.~\ref{f-J131215}). Owing to the weakness of the \siii\ $\lambda$1260 absorption, the column density is not well constrained, but it cannot be larger than the value quoted in Table~\ref{t-metal}. Both transitions of the \civ\ and \siiv\ doublets are detected with an excellent agreement for the column densities, respectively. There is evidence for a single component between about $-25$ and $+25$ \km\ as observed in the absorption of the \hi\ transitions. We note some broad absorption centered at $+75$ \km\ ($z=2.49089$) in \civ, \ciii, \siiii, and \siiv. This broad absorption is only observed in the \hi\ transitions at 972, 1025, and 1215 \AA. 

For this pLLS, the  \siii/\siiv\ and \siiii/\siiv\ ratios constrain well the photoionization model with  $\xh  = -2.50 \pm 0.10$ and $\log U = -1.60 \pm 0.10$. To match the \cii/\civ\ ratio, we derive $[{\rm C}/\alpha]= -0.55 \pm 0.10$; this is also consistent with the lower limit on $N_{\rm CIII}$. The metallicity cannot be much higher or lower, otherwise it would not match the column densities of  \siiii\ and \siii.

\begin{figure}
\epsscale{1} 
\plotone{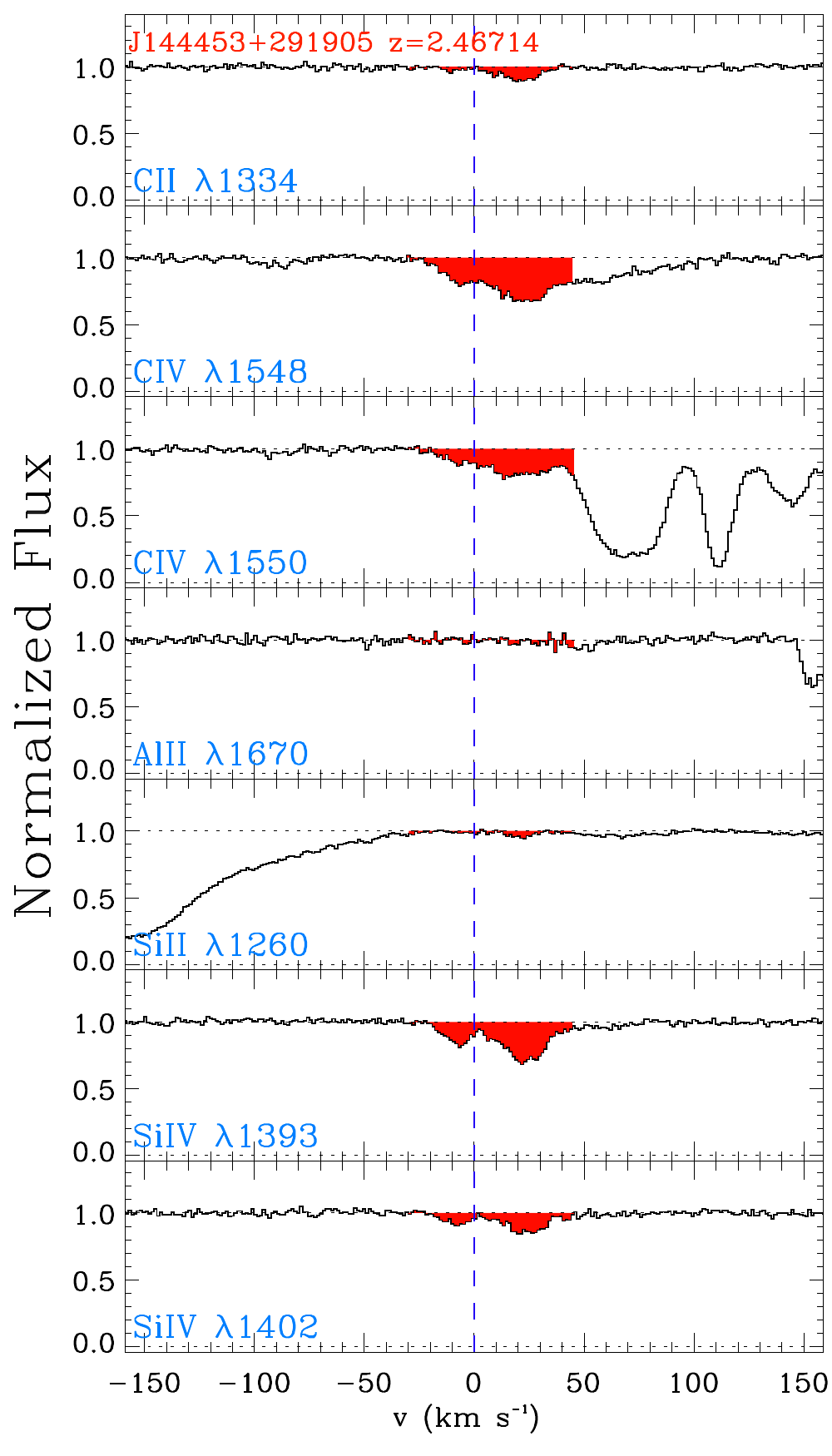}
 \caption{Same as Fig.~\ref{f-J143316}, but for another absorber.}
 \label{f-J144453}
\end{figure}

\noindent
{\it -- J144453+291905 -- $z=2.46714$ -- $\mlnhi = 16.78$}: For this pLLS, \cii, \civ, \siii, \siiv\ are all detected, while \alii\ is not detected at the $3\sigma$ level (see Fig.~\ref{f-J144453}). Both transitions of the \civ\ and \siiv\ doublets are detected with an excellent agreement for the column densities, respectively, although we note that both \civ\ transitions are contaminated at $v>+40$ \km\ (it is, however, unlikely that the column density of \civ\ could be increased by more than $\sim$0.2 dex). There are two main components in this pLLS, with the component at $+20$ \km\ being the strongest. The fit to the \hi\ transitions also requires two components, but the two components are too blended in the \hi\ transitions to robustly separate them (and indeed the central velocities of the independent \hi\ fit are quite different from that of the metal ions). We therefore treat these two components as a single pLLS. 

For this absorber, the  \siii/\siiv\ and \cii/\civ\ ratios simultaneously constrain the photoionization model with  $\xh  = -2.30 \pm 0.15$ and $\log U = -1.90 \pm 0.20$ and $[{\rm C}/\alpha]= 0.00 \pm 0.10$. The metallicity cannot be much higher or lower, or otherwise the model would produce too much or too little \siii\ relative to \siiv. This solution also matches the non-detection of \alii. 

\begin{figure}
\epsscale{1} 
\plotone{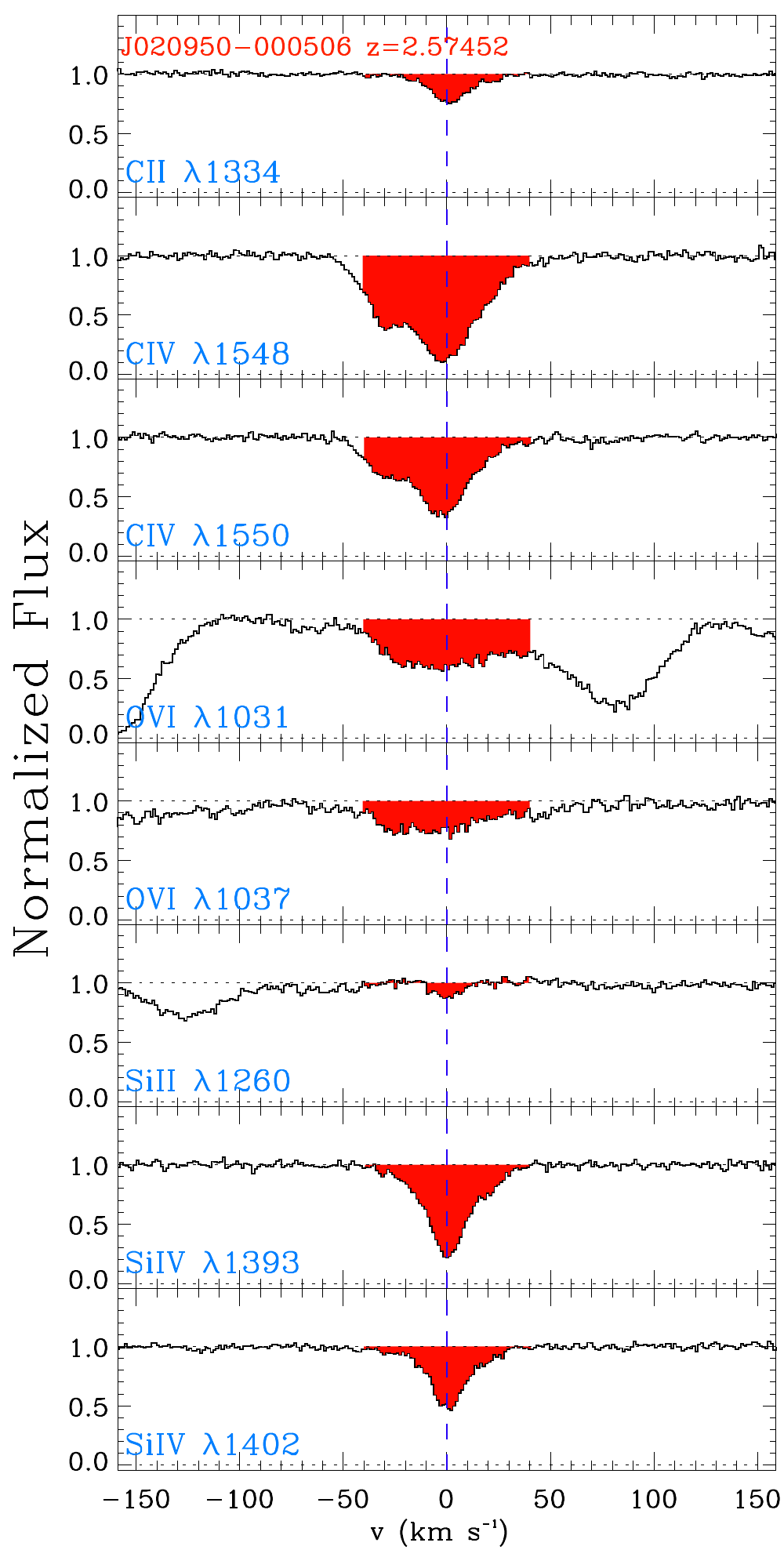}
 \caption{Same as Fig.~\ref{f-J143316}, but for another absorber.}
 \label{f-J020950}
\end{figure}

\noindent
{\it -- J020950$-$000506  -- $z=2.57452$ -- $\mlnhi = 16.78$}: For this pLLS, \cii, \civ, \siii, \siiv, and \ovi\ are all detected (see Fig.~\ref{f-J020950}). Both transitions of the \civ\ and \siiv\ doublets are detected with an excellent agreement for the column densities, respectively. The absorption in \siiv\ and lower ions is dominated by single component at the same redshift as the \hi\ absorption. For \civ, there are two components, while for \ovi, there is a very broad component. Because of the close blending \civ, we undertook a profile fit of the \civ\ to determine the column density in the component at $0$ \km.

For this absorber, the  \siii/\siiv\ and \cii/\civ\ ratios simultaneously constrain the photoionization model with  $\xh  = -2.00 \pm 0.15$ and $\log U = -1.90 \pm 0.15$ and $[{\rm C}/\alpha]= +0.15 \pm 0.15$. The metallicity cannot be much higher or lower, or otherwise the model would produce too much or too little \siii\ relative to \siiv. 

\begin{figure}
\epsscale{1} 
\plotone{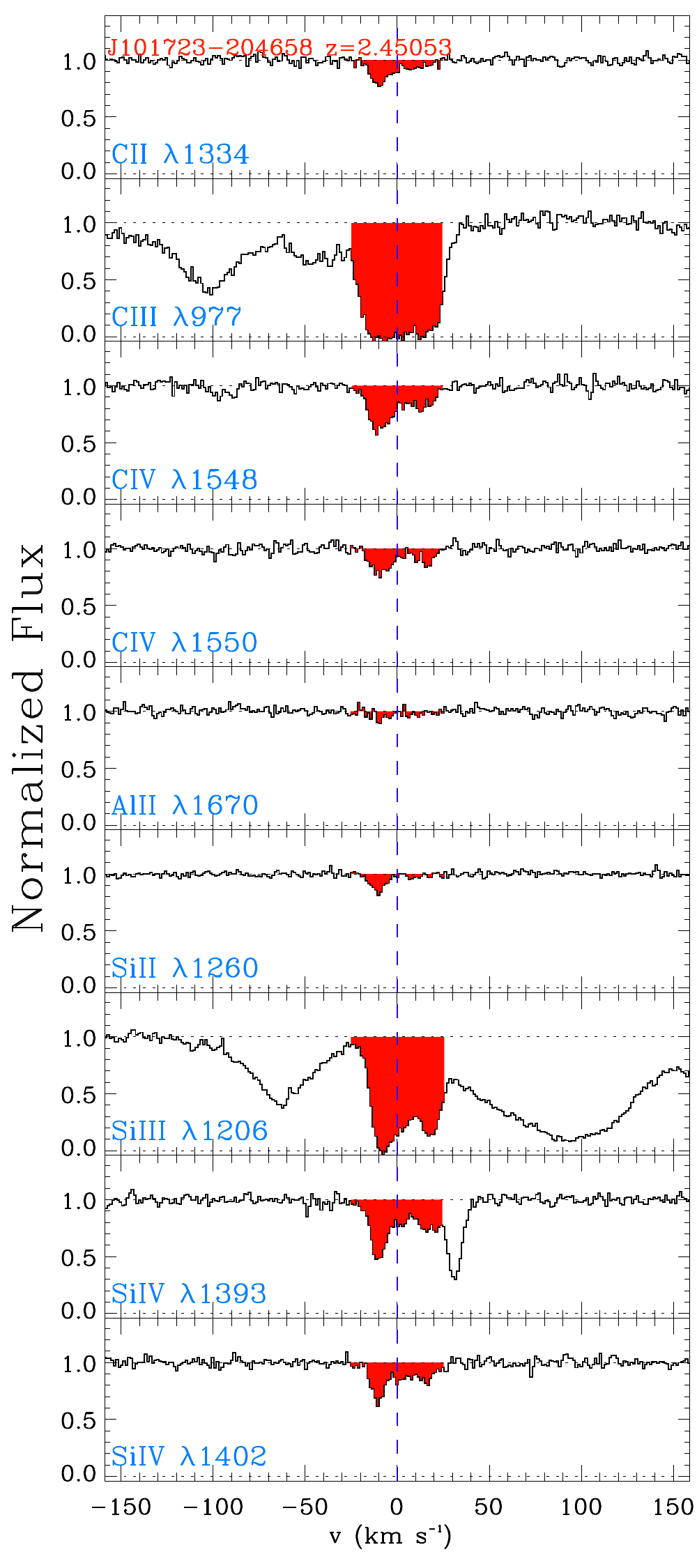}
 \caption{Same as Fig.~\ref{f-J143316}, but for another absorber.}
 \label{f-J101723}
\end{figure}

\noindent
{\it -- J101723$-$204658 -- $z=2.45053$ -- $\mlnhi = 17.23$}: For this LLS, \cii, \ciii, \civ, \siii, \siiii, \siiv, and \alii\ are all detected (see Fig.~\ref{f-J101723}). Both transitions of the \civ\ and \siiv\ doublets are detected with an excellent agreement for the column densities, respectively. Both \ciii\ and \siiii\ are strong, saturated, and quite possibly blended with unrelated absorbers. The absorption in the low, intermediate, and high ions all follows a similar velocity structure, with two components at $-10$ and $+15$ \km, the negative velocity component being the strongest. 

For this absorber, we use the  \siii/\siiv\ and \cii/\civ\ ratios to simultaneously constrain the photoionization model:  a solution  with $\xh  = -2.50 \pm 0.15$ and $\log U = -2.30 \pm 0.15$ and $[{\rm C}/\alpha]= +0.10 \pm 0.15$ are in agreement with these observed ratios as well as the limits on \siiii\ and \ciii. This model also predicts $N_{\rm AlII}$ within about $2\sigma$ of the observed value. 

\begin{figure}
\epsscale{1} 
\plotone{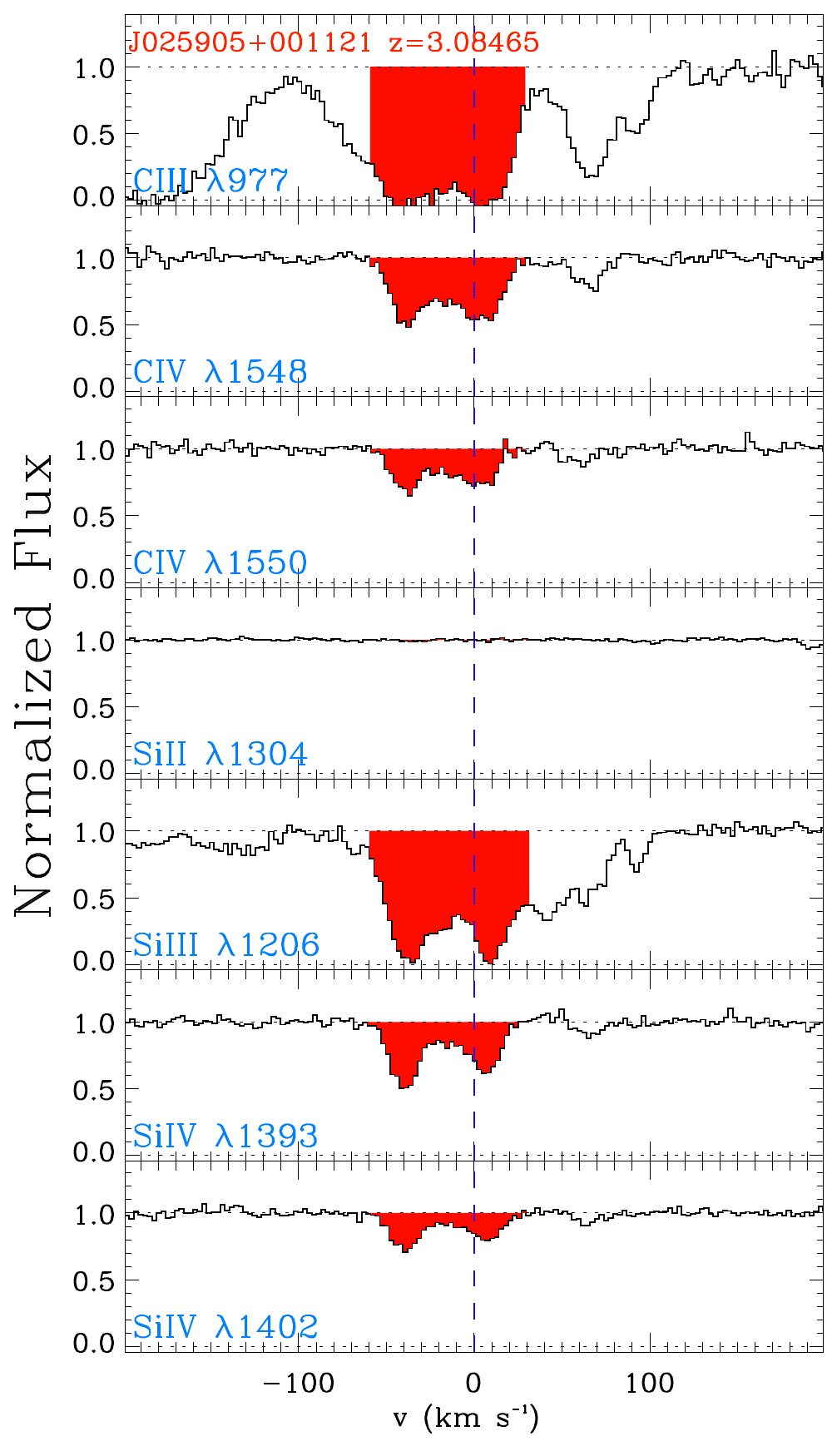}
 \caption{Same as Fig.~\ref{f-J143316}, but for another absorber.}
 \label{f-J025905a}
\end{figure}

\noindent
{\it -- J025905+001121 -- $z=3.08465$ -- $\mlnhi = 17.25$}: For this LLS, there is a detection of \ciii, \civ, \siiii, and \siiv, but no detection of \siii\ at the $3\sigma$ level  (see Fig.~\ref{f-J025905a}). Both \ciii\ and \siiii\ are likely contaminated and saturated to some levels. However, based on the similarity in the velocity profiles between \siiii\ and \siiv, it is unlikely that  $N_{\rm SiIII}$ is overestimated by more than 0.2--0.3 dex. For this LLS, there are two components of about similar strength observed in all the ions near $+5$ and $-35$ \km; an additional weak absorption is observed at $+50$ \km, but not included in the integration of the column densities. 

For this absorber, we use the \siiii/\siiv\ ratio as well as the limit on \siii/\siiv\ to simultaneously constrain the photoionization model. There is some tension between the non-detection of \siii\ and \siiii, but allowing for a contamination of about 0.2 dex in the \siiii\ absorption, a model with  $\xh  = -2.60 \pm 0.25$ and $\log U = -1.90 \pm 0.25$ and $[{\rm C}/\alpha]= -0.20 \pm 0.25$ satisfies all the observational constraints. 

\begin{figure}
\epsscale{1} 
\plotone{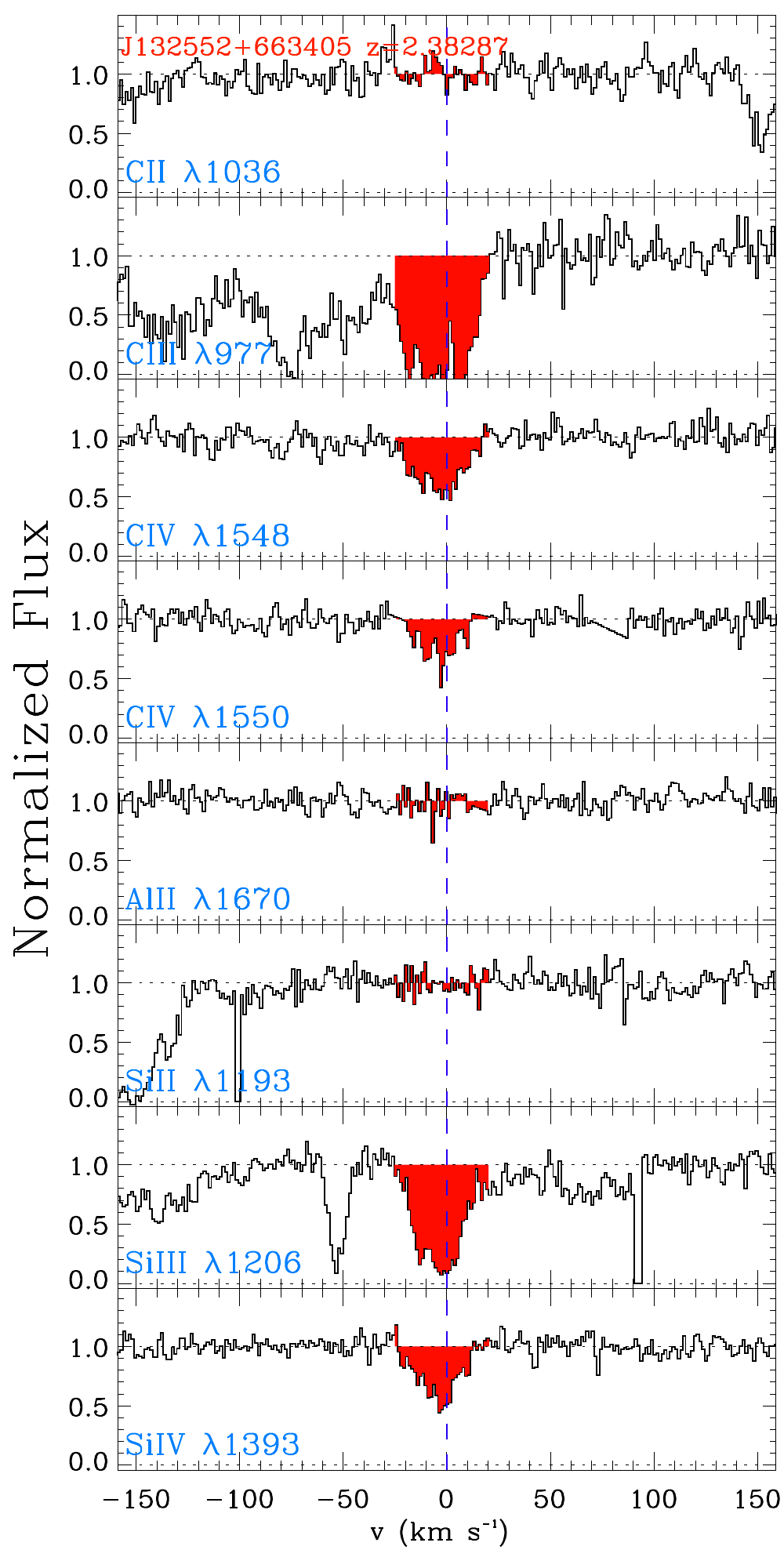}
 \caption{Same as Fig.~\ref{f-J143316}, but for another absorber.}
 \label{f-J132552}
\end{figure}

\noindent
{\it -- J132552+663405 -- $z=2.38287$ -- $\mlnhi = 17.30$}: For this LLS, \ciii, \civ, \siiii,  and \siiv\ are detected, while \cii, \siii, and \alii\ are not at the $3\sigma$ level (see Fig.~\ref{f-J132552}). Both transitions of the \civ\ doublets are detected, with an excellent agreement in the derived column densities. Both \ciii\ and \siiii\ are strong (\ciii\ is saturated and could be partially contaminated). The absorption in all the ions is dominated by a single component. 

For this absorber, we use the \siiii/\siiv\ ratio as well as limits on \siii, \cii, \alii, and \ciii\  to simultaneously constrain the photoionization model:  a solution with $\xh  = -3.00 \pm 0.10$ and $\log U = -1.90 \pm 0.15$ and $[{\rm C}/\alpha]= -0.20 \pm 0.20$ satisfies these observational constraints.

\begin{figure}
\epsscale{1} 
\plotone{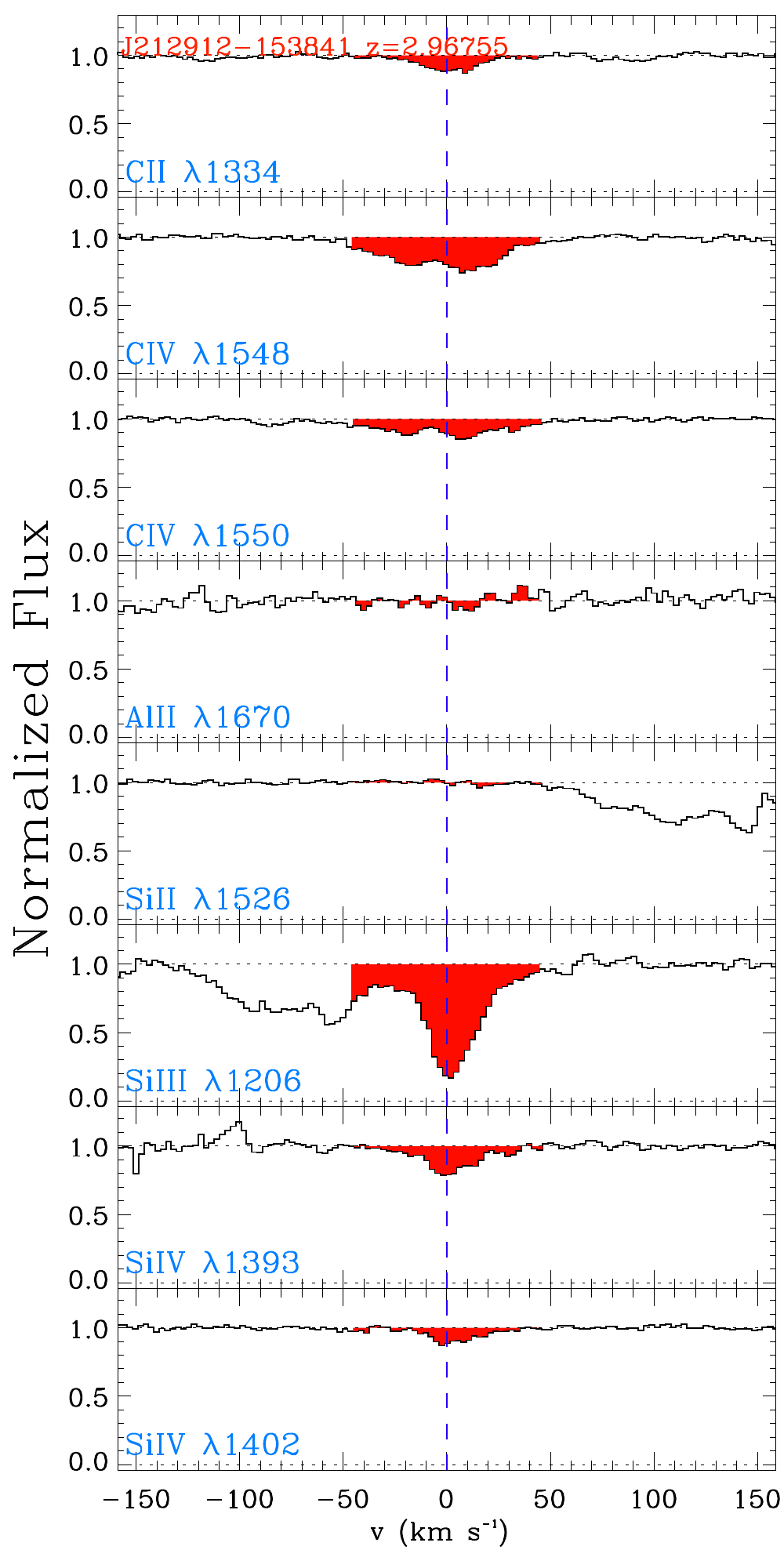}
 \caption{Same as Fig.~\ref{f-J143316}, but for another absorber.}
 \label{f-J212912a}
\end{figure}

\noindent
{\it -- J212912-153841  -- $z=2.96755$ -- $\mlnhi = 17.32$}: For this LLS, \cii, \civ, \siiii, \siiv\ are detected while \siii\ and \alii\ are not (see Fig.~\ref{f-J212912a}). Both transitions of the \civ\ and \siiv\ doublets are detected with excellent agreement for the column densities, respectively. The \siiv\ profiles  are dominated by a single component, while the \civ\ profiles have two main components and are more extended, suggesting that the bulk or \civ\ and \siiv\ may not trace the same gas. \siiii\ $\lambda$1206 is partially blended, and could be partially contaminated; \siiii\ provides an upper limit on the amount of \siiii\ in this LLS. 

For this absorber, we can only place an upper limit on the metallicity $\xh  \le -2.70$, $\log U \ge -2.30 $ based on the limits on \siii/\siiv\ and \siiii/\siiv. This limit is consistent with the non-detection of \alii\  and would imply $[{\rm C}/\alpha]\ga +0.40$ for \cii\ and \civ.

\begin{figure}
\epsscale{1} 
\plotone{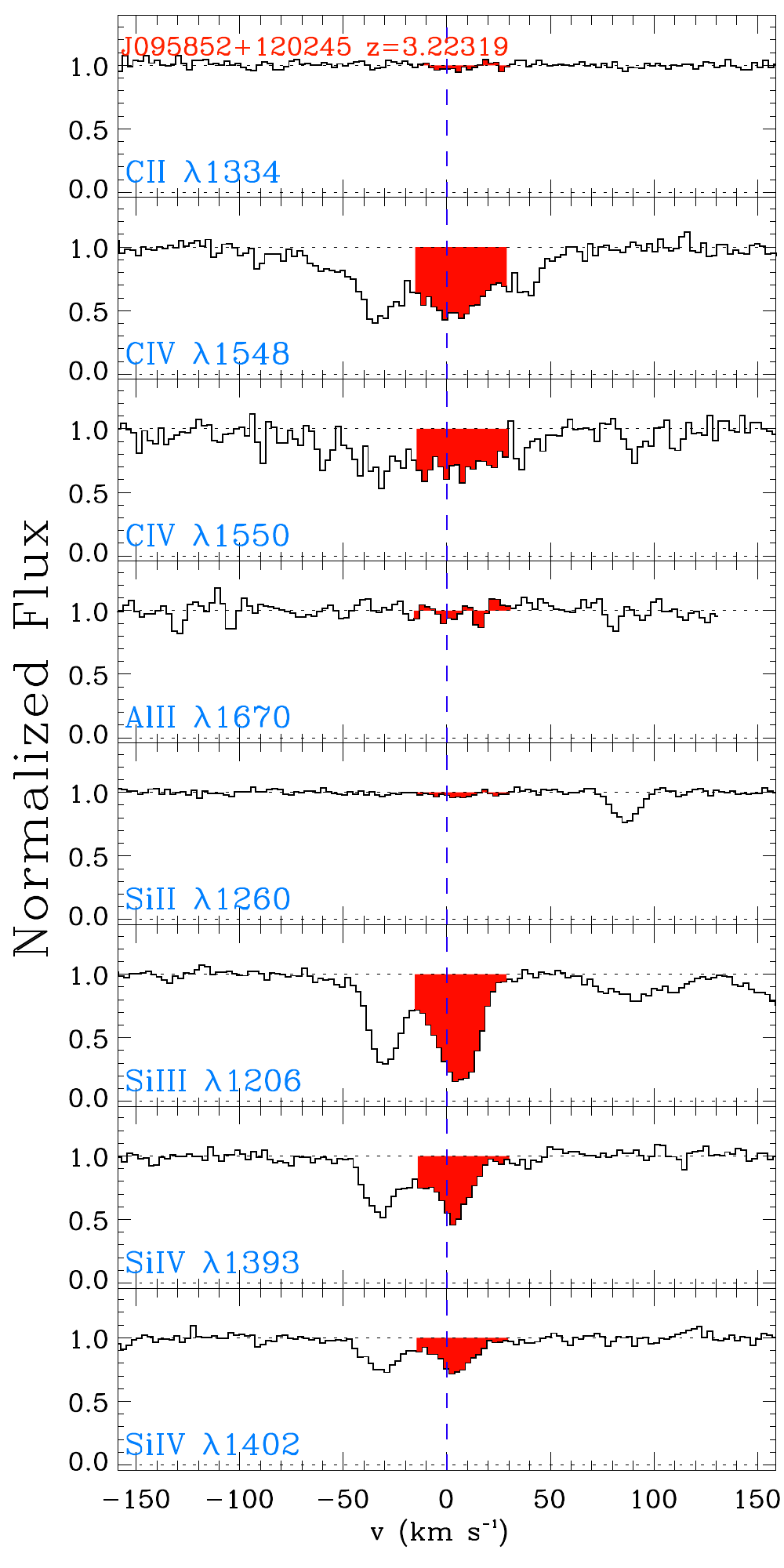}
 \caption{Same as Fig.~\ref{f-J143316}, but for another absorber.}
 \label{f-J095852}
\end{figure}

\noindent
{\it -- J095852+120245 -- $z=3.22319$ -- $\mlnhi = 17.36$}: For this LLS, \civ, \siii, \siiii, and \siiv\ are detected, while \cii\ and \alii\ are not at the $3\sigma$ level (see Fig.~\ref{f-J095852}). Both transitions of the \siiv\ and \civ\ doublets are detected, with an excellent agreement for the column densities, respectively. \siiii\ and \siiv\ have very similar velocity profiles, implying that \siiii\ is unlikely to be contaminated. For this LLS, the metals have two components about 0 and $-30$ \km\ (and possibly additional ones in \civ). However, the component at $-30$ \km\ is only seen in the strong \hi\ transition, not in the weaker transitions where a single component fits extremely the weak Lyman series transitions. Therefore we only integrate the profiles of the metal lines to estimate the column density in the stronger component near 0 \km\ (see Fig.~\ref{f-J095852}).  

For this absorber, we use the \siiii/\siiv\ ratio as well as the limit on \siii/\siiv\ to simultaneously constrain the photoionization model, which lead to a solution with $\xh  = -3.35 \pm 0.05$ and $\log U = -1.50 \pm 0.10$. For that LLS, using \civ, we find $[{\rm C}/\alpha]= -0.20 \pm 0.10$, which is also consistent with the limit on \cii.

\begin{figure}
\epsscale{1} 
\plotone{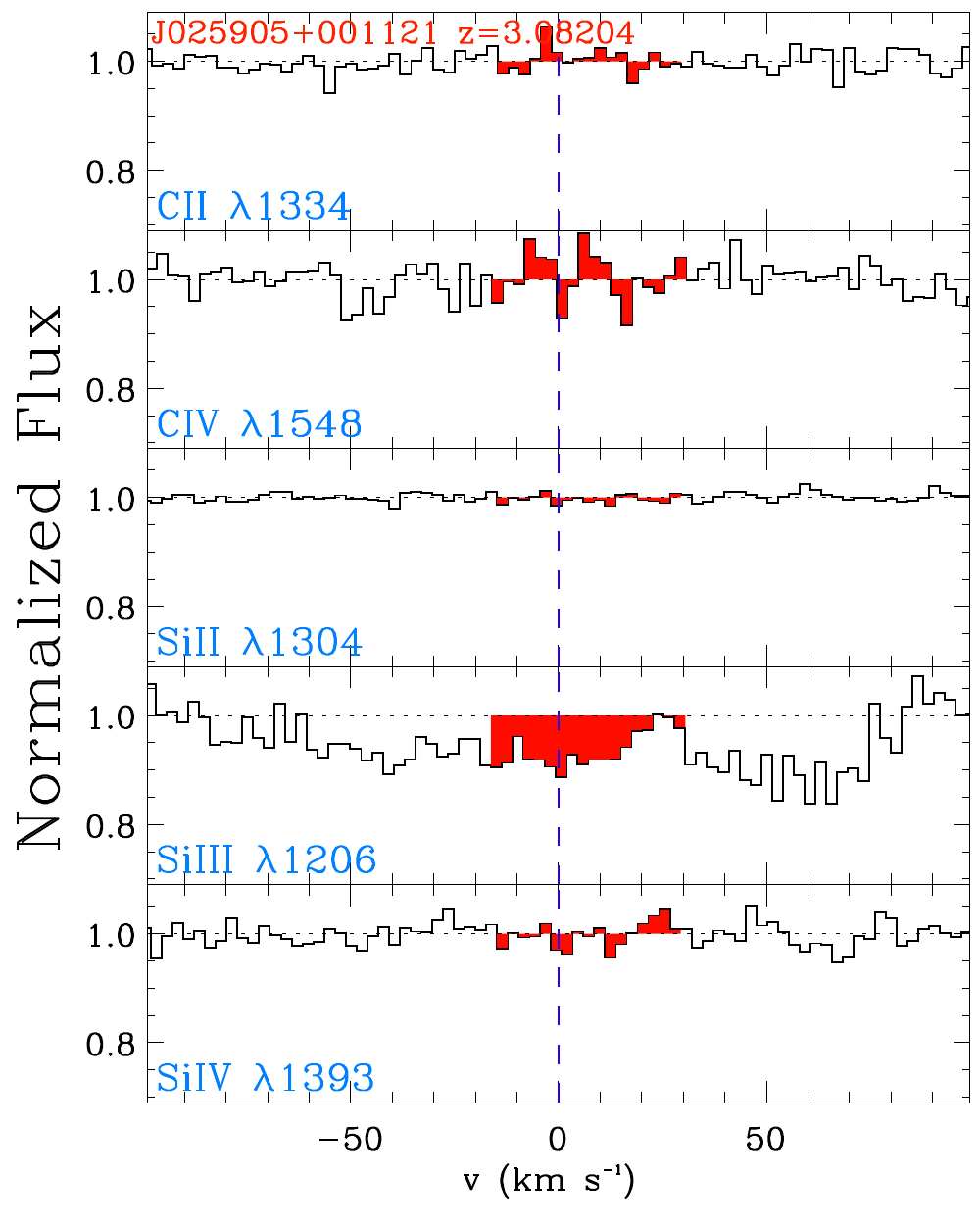}
 \caption{Same as Fig.~\ref{f-J143316}, but for another absorber.}
 \label{f-J025905b}
\end{figure}

\noindent
{\it -- J025905+001121 -- $z=3.08204$ -- $\mlnhi = 17.50$}: For this LLS, there is no detection of \cii, \civ, \siii, and \siiv\ at the $3\sigma$ level  (see Fig.~\ref{f-J025905b}). There is absorption near \siiii, but it is likely contaminated by other absorbers in view of the relatively broad absorption, the absence of such absorption in the higher ions, and other absorption features near this redshift. We therefore treat the absorption of \siiii\ as an upper limit.  

While this absorber is reminiscent of a pristine LLS, the contamination of \siiii\ (and \siii\ $\lambda$1260) implies that we can only place the following limits on $\xh <-2.70$, $\log U > -3.60$, and $[{\rm C}/\alpha]>-0.60$. The metallicity cannot be higher than this limit for this $\log U$ because otherwise too much \siii\ would be produced relative to \siiii. If we use instead the mean  $\langle \log U \rangle = -2.4$ derived from the $\log U$ distribution for our sample of pLLSs and LLSs (see Fig.~\ref{f-udist}), then $\xh \le -4.10$ based on the \siiii/\siiv\ ratio. To be conservative, we, however, adopt here $\xh <-2.7$ and $\log U \ge -3.6$. 

\begin{figure}
\epsscale{1} 
\plotone{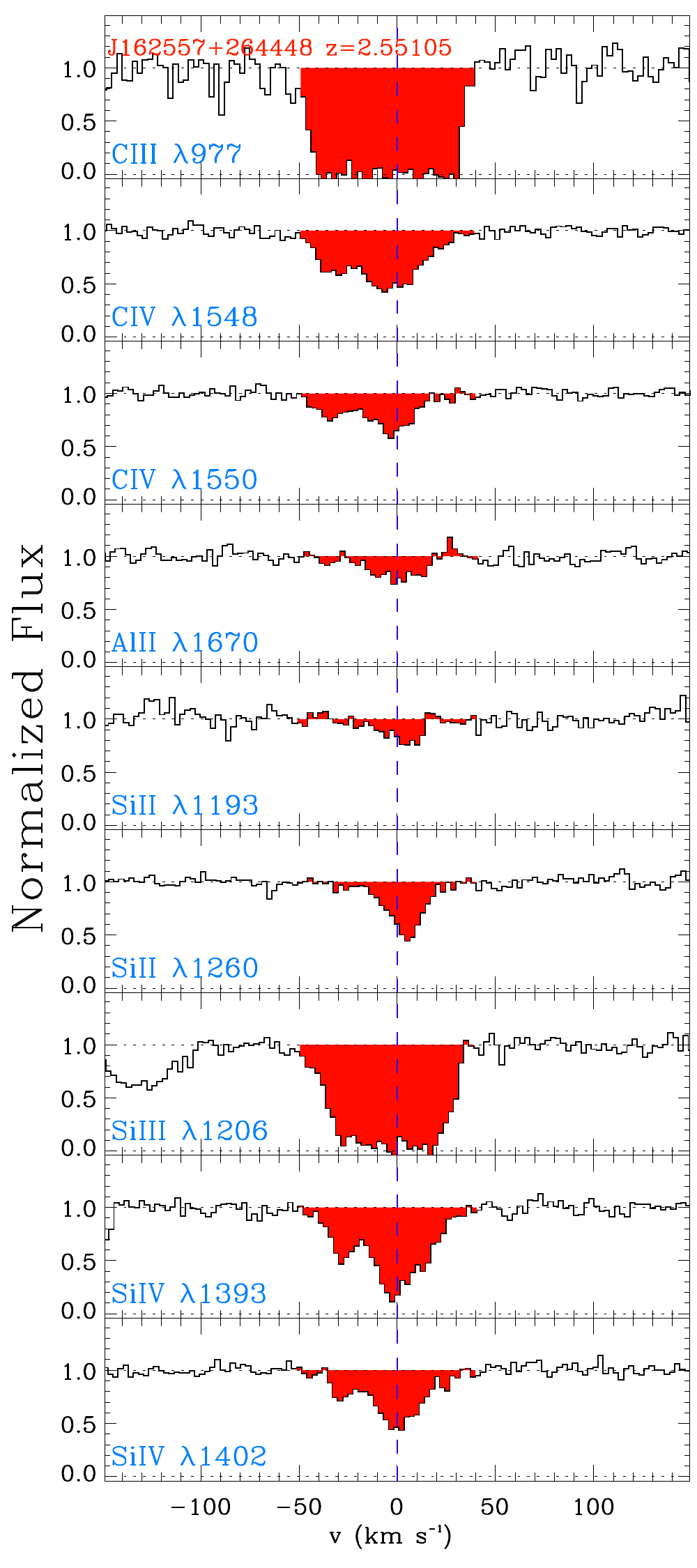}
 \caption{Same as Fig.~\ref{f-J143316}, but for another absorber.}
 \label{f-J025905}
\end{figure}

\noindent
{\it -- J162557+264448  -- $z=2.55105$ -- $\mlnhi = 17.75$}: For this LLS, there are detections of \ciii, \civ, \siii, \siiii, \siiv, and \alii\  (see Fig.~\ref{f-J025905}). Both transitions of the \siiv\ and \civ\ doublets are detected, with an excellent agreement for the column densities, respectively. Both \ciii\ and \siiii\ are saturated. Within $1\sigma$, there is a good agreement for $N$ between \siii\ $\lambda$1193 and $\lambda$1260. The \alii\ velocity profile is similar to that of \siii, implying there is no evidence of contamination for that transition. There are several components observed in the velocity profiles, but the absorption is dominated by the component at 0 \km. 

For this absorber, we use the \siii/\siiv\ ratio as well as the limit on \siiii/\siiv\ to constrain the photoionization model. The model is well constrained with  $\xh  = -2.25 \pm 0.25$ and $\log U = -2.20 \pm 0.15$ and $[{\rm C}/\alpha]= -0.30 \pm 0.15$ by the observations. This solution requires $[$Al/Si$]=-0.20 \pm 0.15$. 

\begin{figure*}
\epsscale{1} 
\plotone{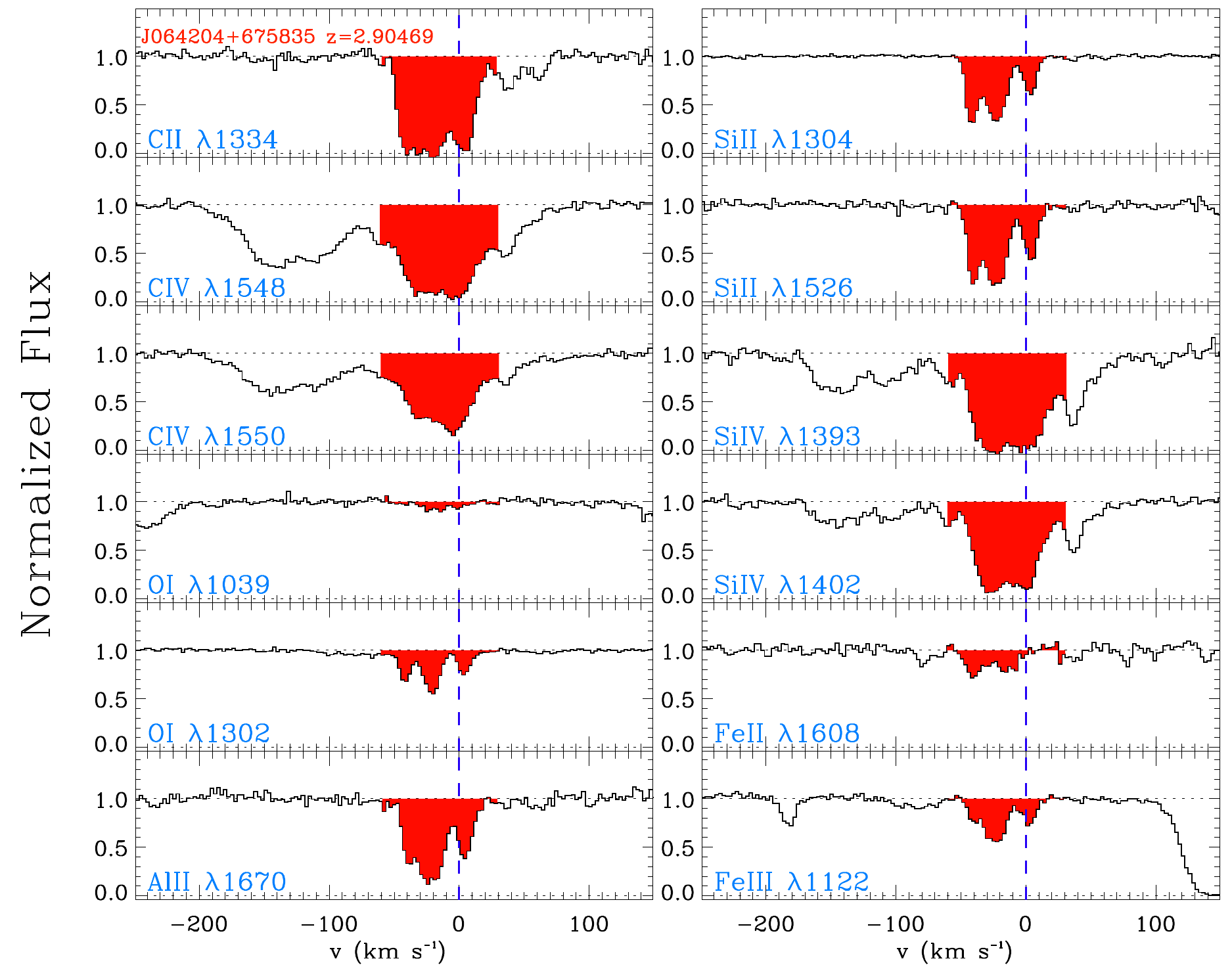}
 \caption{Same as Fig.~\ref{f-J143316}, but for another absorber.}
 \label{f-J064204}
\end{figure*}

\noindent
{\it -- J064204+675835 -- $z=2.90469$ -- $\mlnhi = 18.42$}: This is the second strongest LLS in our new sample, with detections of \oi, \cii, \alii, \siii, \feii, and \feiii\ (\siiii\ and \ciii\ are also detected but saturated and most likely contaminated) (see Fig.~\ref{f-J064204}). The high ions \civ\ and \siiv\ are also detected but have a different velocity structure than the low ions and extend over much larger velocities. We therefore use \oi\ and the low ions to constrain the photoionization model. We integrate the velocity profiles over the 3 observed components that spread between $-60$ and $+30$ \km\ since there is not enough information from the \hi\ profiles to determine which component is the most likely associated with the LLS. 

For this strong LLS, the \oi/\siii\ ratio constrains the photoionization model with $\xh  = -1.00 \pm 0.20$ and $\log U = -3.00 \pm 0.15$. For that solution, we find $[{\rm C}/\alpha]\ge 0$. There is some tension for \alii\ (overproduced by about 0.3) and \feii/\feiii\ (\feii\ is underproduced by about 0.2 dex, while \feiii\ is overproduced by 0.15 dex). However, since $N_{\rm OI} \simeq N_{\rm SiII}$, the metallicity cannot change by a large amount for this \nhi\ value. This model also implies that \civ\ and \siiv\ are underproduced by about 1 and 0.5 dex, respectively, which is consistent with the different velocity profiles between the high and low ions.

\begin{figure}
\epsscale{1} 
\plotone{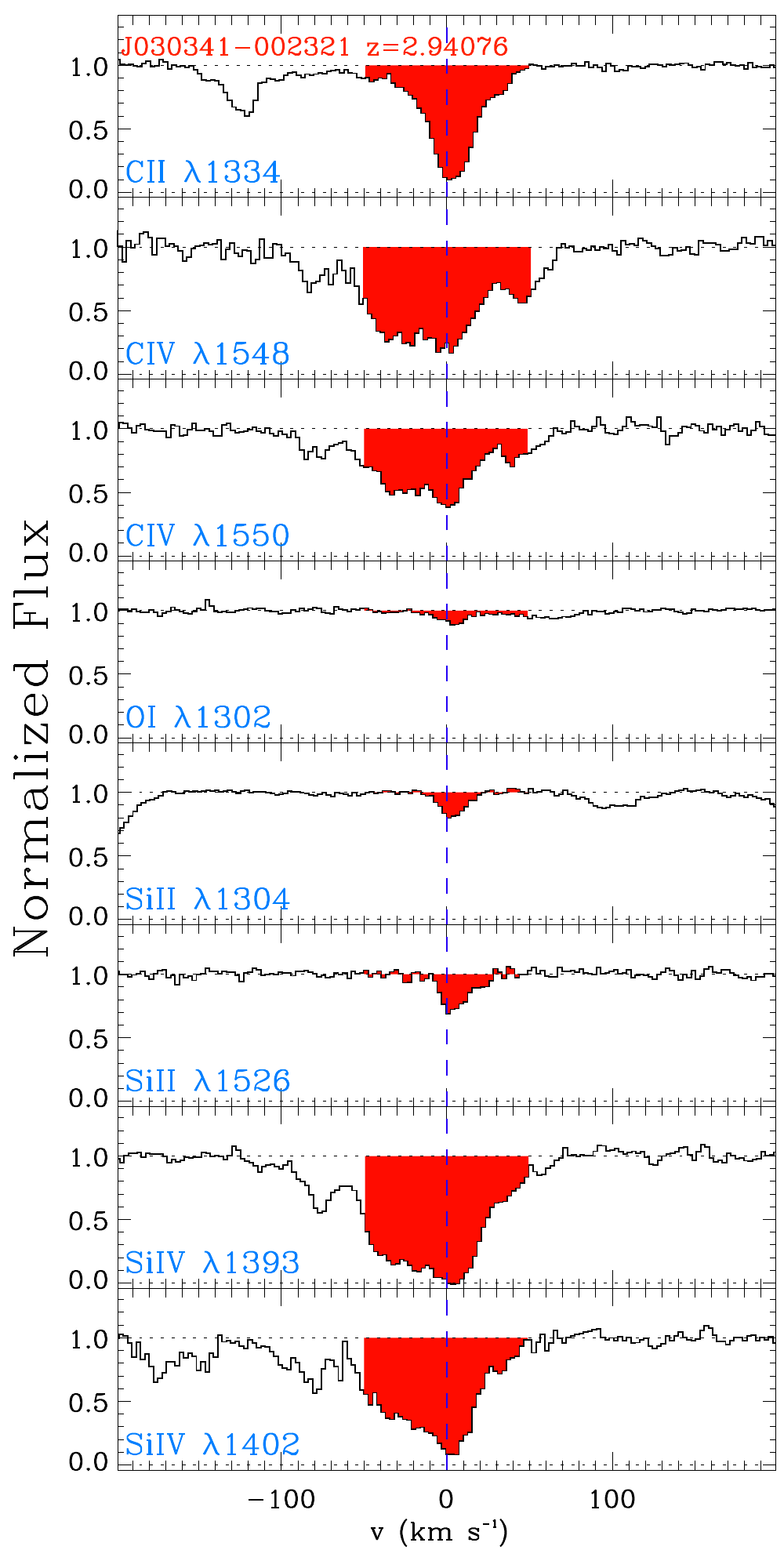}
 \caption{Same as Fig.~\ref{f-J143316}, but for another absorber.}
 \label{f-J030341}
\end{figure}

\noindent
{\it -- J030341$-$002321 -- $z=2.94076$ -- $\mlnhi = 18.65$}: This is the strongest LLS in our new sample, with a detection of \oi, \cii, and \siii. The high ions \civ\ and \siiv\ are also detected but have quite different velocity structure than the low ions that are dominated by a single velocity component (see Fig.~\ref{f-J030341}). 

For this strong LLS, the \oi/\siii\ ratio constrains well the photoionization model with  $\xh  = -2.10 \pm 0.20$ and $\log U = -2.70 \pm 0.15$. For that solution, we find $[{\rm C}/\alpha]= +0.30 \pm 0.20$ . This solution implies that \civ\ and \siiv\ are underproduced by about 1 dex, which is consistent with the detection of \oi\ and the very different velocity profiles between the high and low ions.

\clearpage
%\floattable
\begin{deluxetable*}{ccccccccccr}
\tablecaption{Cloudy results \label{t-cloudy}}
\tablehead{
\colhead{QSO} & \colhead{$z_{\rm abs}$} & \colhead{$\log N_{\rm HI}$}& \colhead{$\log N_{\rm H}$} &  \colhead{${\rm [X/H]}$} & \colhead{${\rm [C/\alpha]}$} & \colhead{$\log U$} & \colhead{\nhii/\nh} & \colhead{$T$} & \colhead{$\log n_{\rm H}$} & \colhead{$\log l$} \\
\colhead{}&\colhead{}&  \colhead{$[{\rm cm}^{-2}]$}&  \colhead{$[{\rm cm}^{-2}]$}&\colhead{}&\colhead{}&\colhead{}&  \colhead{(\%)} &  \colhead{($10^4$\,K)}&  \colhead{$[{\rm cm}^{-3}]$} &  \colhead{[pc]}
}
%\colnumbers
\startdata
  J143316$+$313126  & $2.90116 $ & $ 16.16 $ & $ 18.92 $ & $ -1.80 $ & $ -0.15 $ & $ -2.65 $ & $  99.8 $ & $ 1.9 $ & $ -2.19 $ & $  1.85$ \\
  J030341$-$002321  & $2.99496 $ & $ 16.17 $ & $ 19.96 $ & $ -1.90 $ & $ -0.40 $ & $ -1.75 $ & $ 100.0 $ & $ 2.9 $ & $ -3.02 $ & $  4.49$ \\
  J014516$-$094517A & $2.66516 $ & $ 16.17 $ & $ 19.86 $ & $ -2.40 $ & $ -0.10 $ & $ -1.85 $ & $ 100.0 $ & $ 2.9 $ & $ -2.87 $ & $  4.24$ \\
  J172409$+$531405  & $2.48778 $ & $ 16.20 $ & $ 17.34 $ & $ +0.20 $ & \nodata   & $ -4.00 $ & $  92.7 $ & $ 0.6 $ & $ -0.70 $ & $ -0.45$ \\
  J134544$+$262506  & $2.86367 $ & $ 16.20 $ & $ 19.60 $ & $ -1.65 $ & $ -0.10 $ & $ -2.10 $ & $ 100.0 $ & $ 2.4 $ & $ -2.65 $ & $  3.76$ \\
  J170100$+$641209  & $2.43307 $ & $ 16.24 $ & $ 19.47 $ & $ -1.65 $ & $ +0.20 $ & $ -2.25 $ & $  99.9 $ & $ 2.3 $ & $ -2.44 $ & $  3.43$ \\
  J134328$+$572147  & $2.87056 $ & $ 16.30 $ & $ 20.36 $ & $ -1.45 $ & $ -0.70 $ & $ -1.55 $ & $ 100.0 $ & $ 3.2 $ & $ -3.20 $ & $  5.06$ \\
  J012156$+$144823  & $2.66586 $ & $ 16.32 $ & $ 19.32 $ & $ -1.00 $ & $ +0.05 $ & $ -2.40 $ & $  99.9 $ & $ 1.8 $ & $ -2.32 $ & $  3.15$ \\
  J170100$+$641209  & $2.43359 $ & $ 16.38 $ & $ 19.49 $ & $ -1.50 $ & $ -0.05 $ & $ -2.35 $ & $  99.9 $ & $ 2.1 $ & $ -2.34 $ & $  3.35$ \\
  J135038$-$251216  & $2.57299 $ & $ 16.39 $ & $ 19.34 $ & $ -2.30 $ & $ -0.05 $ & $ -2.50 $ & $  99.9 $ & $ 2.0 $ & $ -2.21 $ & $  3.06$ \\
  J130411$+$295348  & $2.82922 $ & $ 16.39 $ & $>17.73 $ & $<-1.70 $ & \nodata   &$\ge -4.00$& $> 95.2 $ & $>1.2 $ & $<-0.74 $ & $>-0.02$ \\
  J134544$+$262506  & $2.87630 $ & $ 16.50 $ & $ 20.31 $ & $ -2.30 $ & $ -0.50 $ & $ -1.80 $ & $ 100.0 $ & $ 3.1 $ & $ -2.95 $ & $  4.77$ \\
  J212912$-$153841  & $2.90711 $ & $ 16.55 $ & $ 19.73 $ & $ -1.55 $ & $ -0.20 $ & $ -2.30 $ & $  99.9 $ & $ 2.1 $ & $ -2.46 $ & $  3.69$ \\
  J101447$+$430030  & $3.01439 $ & $ 16.63 $ & $>17.97 $ & $<-2.60 $ & \nodata   &$\ge -4.00$& $> 95.3 $ & $>1.2 $ & $<-0.77 $ & $> 0.25$ \\
  J131215$+$423900  & $2.48998 $ & $ 16.77 $ & $ 20.73 $ & $ -2.50 $ & $ -0.55 $ & $ -1.70 $ & $ 100.0 $ & $ 3.3 $ & $ -3.00 $ & $  5.23$ \\
  J144453$+$291905  & $2.46714 $ & $ 16.78 $ & $ 20.48 $ & $ -2.30 $ & $ +0.00 $ & $ -1.90 $ & $ 100.0 $ & $ 2.8 $ & $ -2.80 $ & $  4.78$ \\  %
  J020950$-$000506  & $2.57452 $ & $ 16.78 $ & $ 20.47 $ & $ -2.05 $ & $  0.15 $ & $ -1.90 $ & $ 100.0 $ & $ 2.8 $ & $ -2.81 $ & $  4.79$ \\
  J101723$-$204658  & $2.45053 $ & $ 17.23 $ & $ 20.43 $ & $ -2.50 $ & $ +0.10 $ & $ -2.30 $ & $  99.9 $ & $ 2.1 $ & $ -2.40 $ & $  4.34$ \\
  J025905$+$001121  & $3.08465 $ & $ 17.25 $ & $ 20.89 $ & $ -2.60 $ & $ -0.20 $ & $ -1.90 $ & $ 100.0 $ & $ 2.5 $ & $ -2.88 $ & $  5.29$ \\
  J132552$+$663405  & $2.38287 $ & $ 17.30 $ & $ 20.95 $ & $ -3.00 $ & $ -0.20 $ & $ -1.90 $ & $ 100.0 $ & $ 2.6 $ & $ -2.79 $ & $  5.25$ \\
  J212912$-$153841  & $2.96755 $ & $ 17.32 $ & $ 20.48 $ & $<-2.70 $ & $>+0.40 $ &$\ge-2.30$ & $ >99.9 $ & $>2.1 $ & $<-2.46 $ & $> 4.45$ \\
  J095852$+$120245  & $3.22319 $ & $ 17.36 $ & $ 21.46 $ & $ -3.35 $ & $ -0.20 $ & $ -1.50 $ & $ 100.0 $ & $ 3.0 $ & $ -3.30 $ & $  6.27$ \\
  J025905$+$001121  & $3.08204 $ & $ 17.50 $ & $>19.17 $ & $<-2.70 $ & $>-0.60 $ & $>-3.60 $ & $> 97.8 $ & $>1.3 $ & $<-1.18 $ & $> 1.86$ \\
  J162557$+$264448  & $2.55105 $ & $ 17.75 $ & $ 20.86 $ & $ -2.25 $ & $ -0.30 $ & $ -2.20 $ & $  99.9 $ & $ 2.1 $ & $ -2.51 $ & $  4.88$ \\
  J064204$+$675835  & $2.90469 $ & $ 18.42 $ & $ 20.08 $ & $ -1.00 $ & \nodata   & $ -3.00 $ & $  97.8 $ & $ 1.4 $ & $ -1.76 $ & $  3.35$ \\
  J030341$-$002321  & $2.94076 $ & $ 18.65 $ & $ 20.45 $ & $ -2.10 $ & $ +0.30 $ & $ -2.70 $ & $  98.4 $ & $ 1.6 $ & $ -2.06 $ & $  4.02$ \\
\enddata
\end{deluxetable*}

\end{document}